\documentclass[a4paper,fleqn,usenatbib]{mnras}
\usepackage[T1]{fontenc}
\usepackage{lmodern}
\usepackage{newtxtext,newtxmath}

\usepackage[T1]{fontenc}
\usepackage{ae,aecompl}

\usepackage{graphicx}	
\usepackage{amsmath}	
\usepackage{amssymb}	

\title[Young stellar discs at the Galactic Centre]{Star formation at the Galactic Centre: coevolution of multiple young stellar discs}

\author[Mastrobuono-Battisti et al.]{
Alessandra Mastrobuono-Battisti,$^{1}$\thanks{mastrobuono@mpia.de}
Hagai B. Perets,$^{2}$ Alessia Gualandris,$^{3}$
 \newauthor
Nadine Neumayer$^{1}$ and Anna C. Sippel$^{1}$\\
$^{1}$Max Planck Institute for Astronomy, K\"onigstuhl 17, D69117, Heidelberg, Germany\\
$^{2}$Physics Department, Technion - Israel Institute of Technology, Haifa 3200004, Israel\\
$^{3}$Department of Physics, Faculty of Engineering and Physical Sciences, University of Surrey, Guildford GU2 7XH, UK
}

\date{Accepted XXX. Received YYY; in original form ZZZ}

\pubyear{2019}

\begin{document}
\label{firstpage}
\pagerange{\pageref{firstpage}--\pageref{lastpage}}
\maketitle

\begin{abstract}
Studies of the Galactic Centre suggest that in-situ star formation may have given rise to the observed stellar population near the central supermassive black hole (SMBH). Direct evidence for a recent starburst is provided by the currently observed young stellar disc ($2$-$7$\,Myr) in the central $0.5$\,pc of the Galaxy. This result suggests that star formation in galactic nuclei may occur close to the SMBH and produce initially flattened stellar discs. Here we explore the possible build-up and evolution of nuclear stellar clusters near SMBHs through in-situ star formation producing stellar discs similar to those observed in the Galactic Centre and other nuclei. We make use of $N$-body simulations to model the evolution of multiple young stellar discs, and explore the potential observable signatures imprinted by such processes. Each of the five simulated discs is evolved for $100$\,Myr before the next one is introduced in the system. We find that populations born at different epochs show different morphologies and kinematics. Older and presumably more metal poor populations are more relaxed and extended, while younger populations show a larger amount of rotation and flattening. We conclude that star formation in central discs can reproduce the observed properties of multiple stellar populations in galactic nuclei differing in age, metallicity and kinematic properties. 
\end{abstract}



\begin{figure*}
\includegraphics[width=3.68cm, trim = 0.4cm 2cm 0.4cm 0.5cm, clip=true]{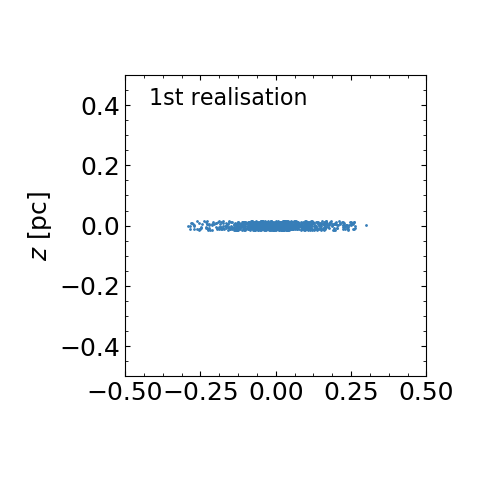}
\includegraphics[width=3.32cm, trim = 0.4cm 1.5cm 0.4cm 0.4cm, clip=true]{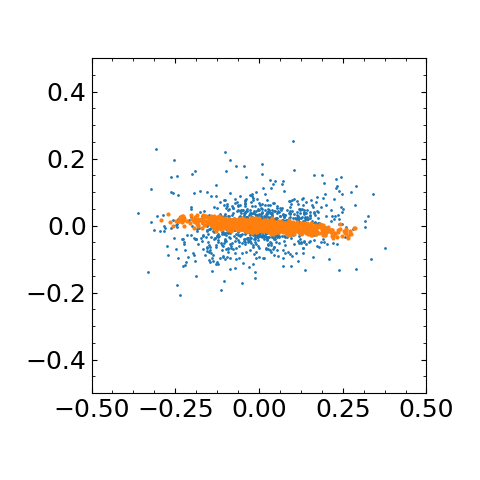}
\includegraphics[width=3.32cm, trim = 0.4cm 1.5cm 0.4cm 0.4cm, clip=true]{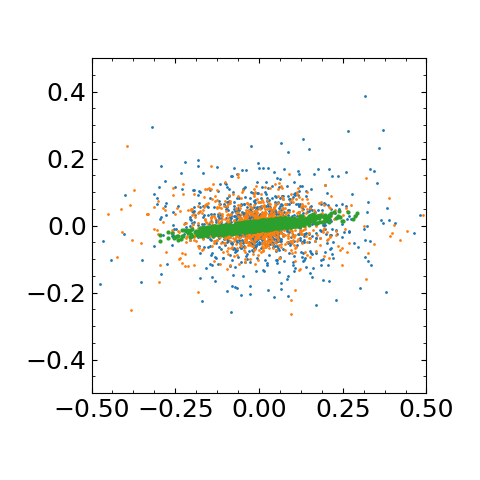}
\includegraphics[width=3.32cm, trim = 0.4cm 1.5cm 0.4cm 0.4cm, clip=true]{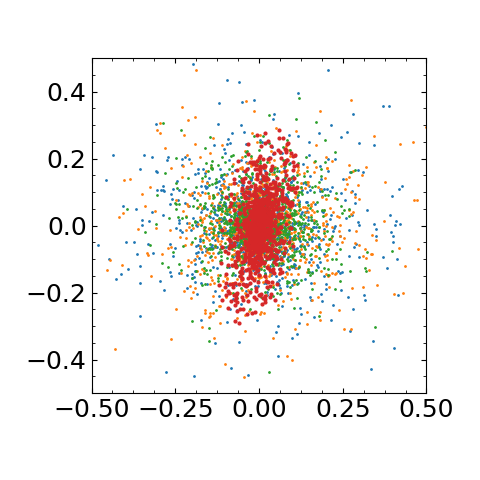}
\includegraphics[width=3.32cm, trim = 0.4cm 1.5cm 0.4cm 0.4cm, clip=true]{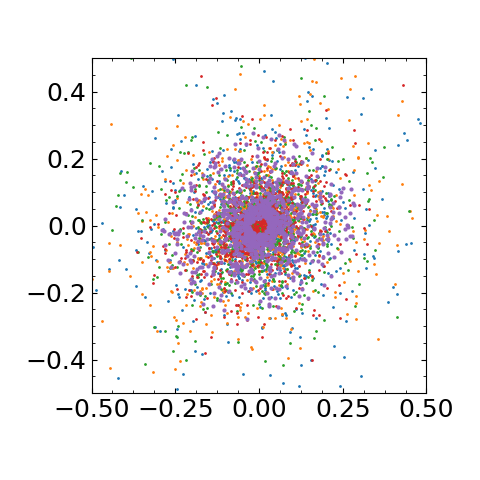}

\includegraphics[width=3.68cm, trim = 0.4cm 1.1cm 0.4cm 0.5cm, clip=true]{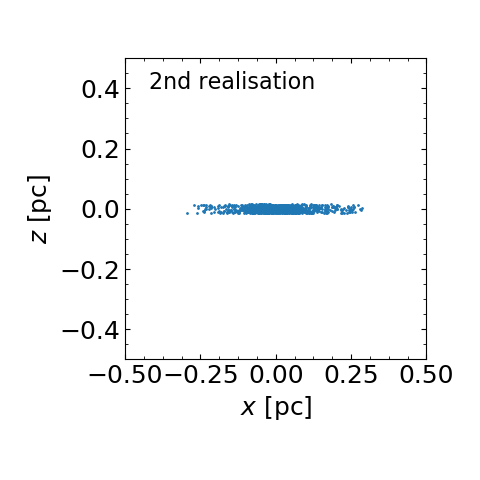}
\includegraphics[width=3.32cm, trim = 0.4cm 0.4cm 0.4cm 0.4cm, clip=true]{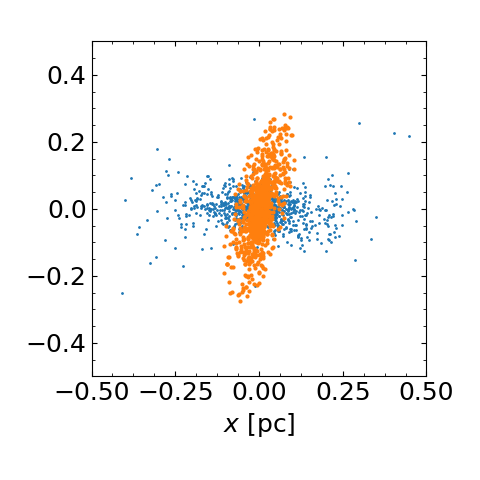}
\includegraphics[width=3.32cm, trim = 0.4cm 0.4cm 0.4cm 0.4cm, clip=true]{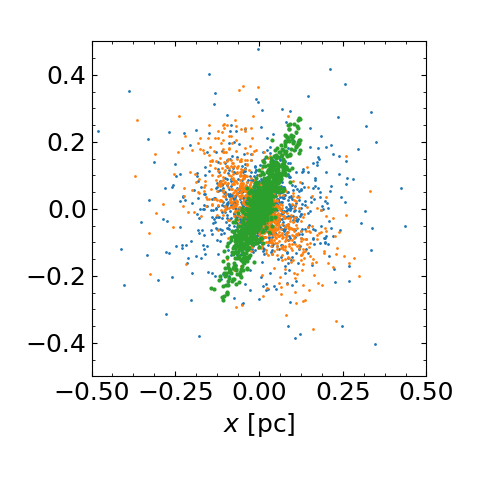}
\includegraphics[width=3.32cm, trim = 0.4cm 0.4cm 0.4cm 0.4cm, clip=true]{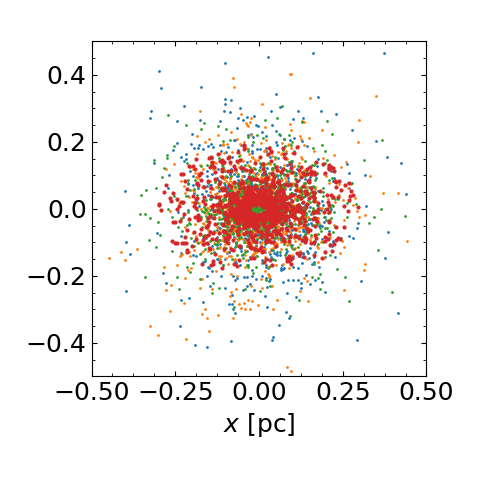}
\includegraphics[width=3.32cm, trim = 0.4cm 0.4cm 0.4cm 0.4cm, clip=true]{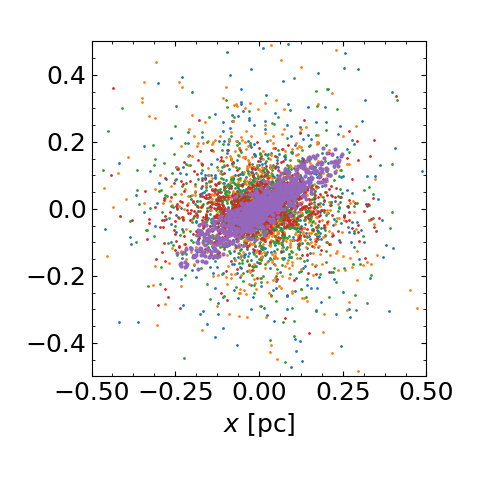}
\caption{Snapshots of the simulated stars in the xz plane taken when each disc is introduced in the system. Different colours represent stars in different discs and the top panels are for the first realisation while the bottom panels are for the second realisation. The discs are inclined differently and the third disc in the first realisation and the second one in the second realisation are retrograde.}\label{fig:snaps}
\end{figure*}
\begin{figure}
\centering
\includegraphics[width=0.45\textwidth]{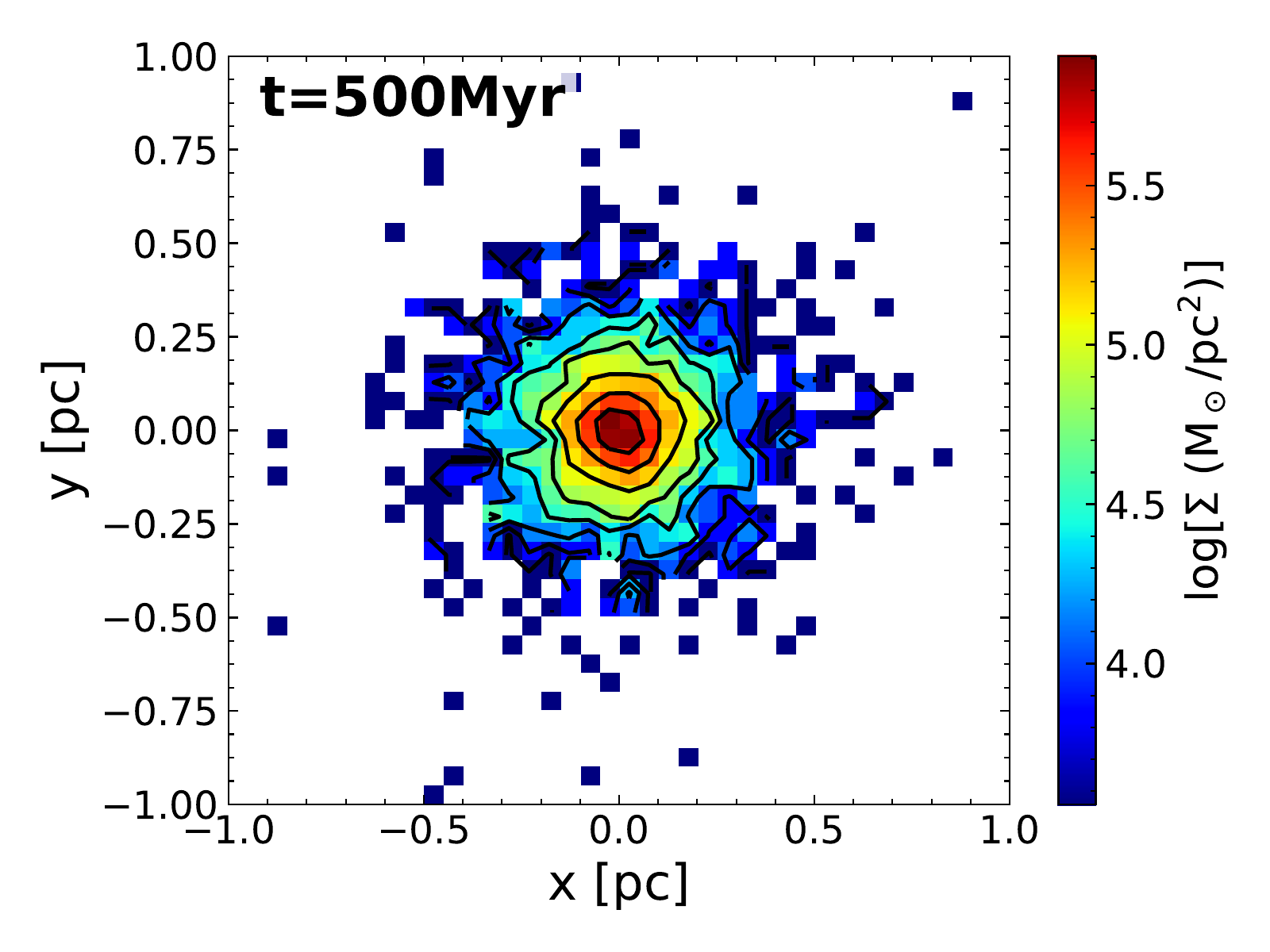}
\includegraphics[width=0.45\textwidth]{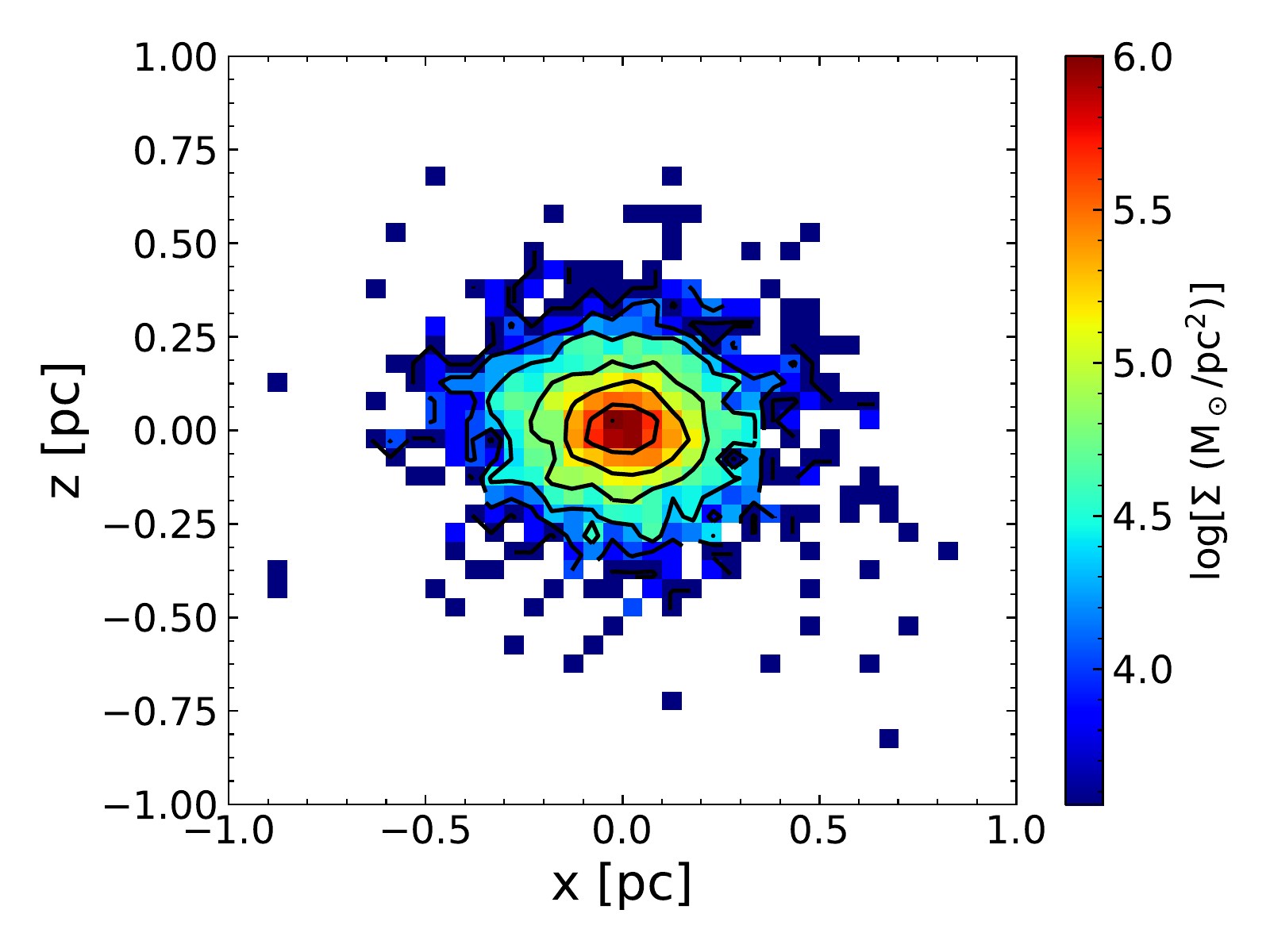}
\includegraphics[width=0.45\textwidth]{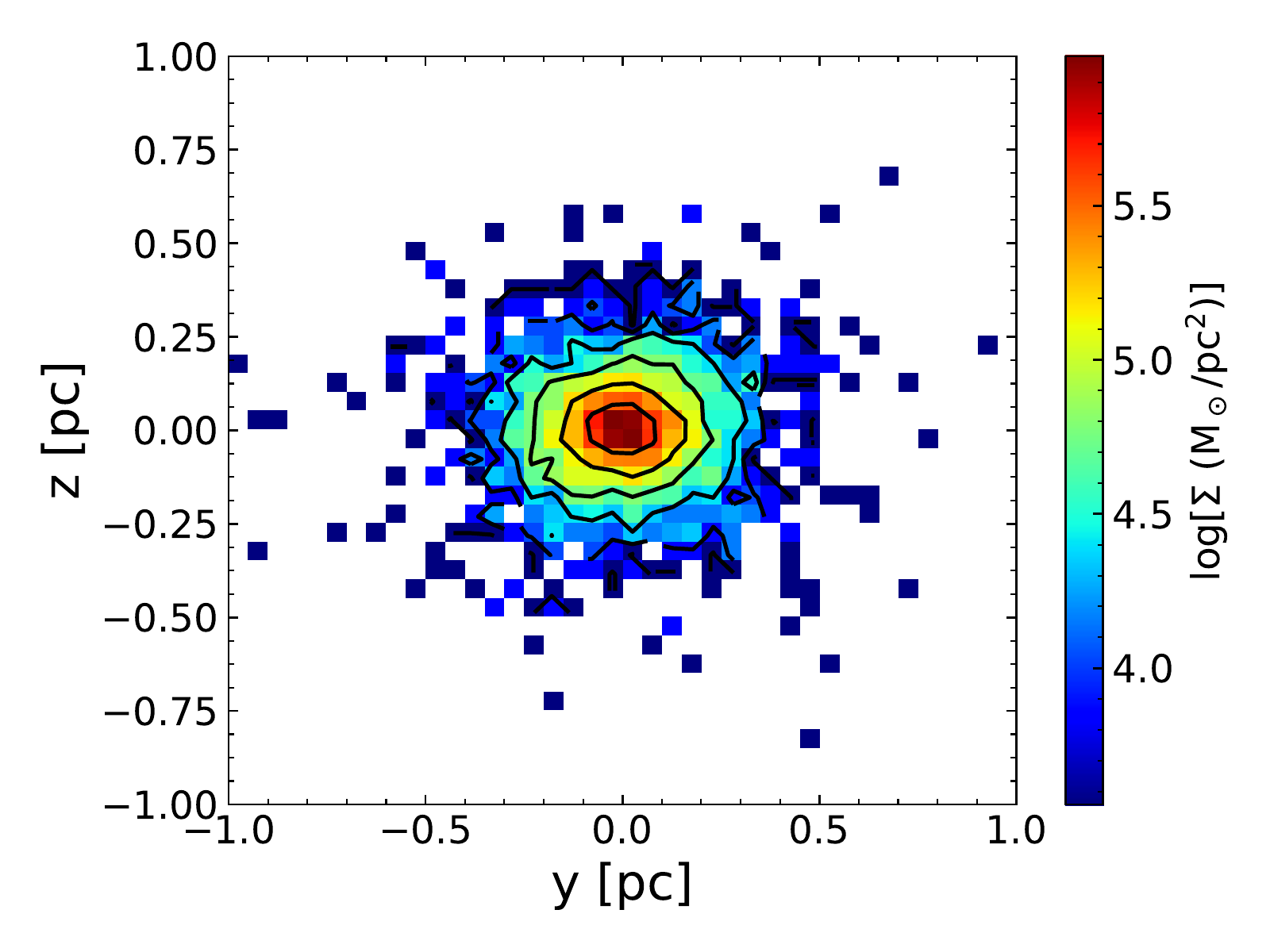}
\caption{Contour density maps of the five simulated discs { in the first realisation} considered together after 500\,Myr of evolution. The system is seen face-on (top panel) and edge-on (middle and bottom panels).}\label{fig:maps}
\end{figure}
\section{Introduction}
The Galactic Centre (GC) hosts a relatively old (older than few Gyr) $\sim2.5\times10^7$M$_\odot$ nuclear star cluster (NSC) with an effective radius of $4.2$\,pc \citep{FN14, S14}.
At the centre of this cluster lies a supermassive black hole (SMBH), Sgr A$^*$, of mass $4.3\times10^6$M$_\odot$ \citep{GH98, EI05, GI09, BGS16, GP17}.
Mass segregation arguments, confirmed by recent observations \citep{FR06, AT09, ME00, HA18}, predict the presence of a cusp of stellar black holes (SBHs) around the central SMBH.
At Galactocentric radii smaller than $0.05$\,pc,  the GC hosts the so called S-stars, young B-stars with ages between $10$ and $50$\,Myr and relatively high orbital eccentricities ($0.3<e<0.95$). 
The spatial and kinematic distribution of the S-stars is approximately isotropic, and not associated with any disc structure \citep{GI09}. There is not yet consensus on the origin of these stars, which in the two most prominent scenarios might have formed and migrated inward in a gaseous disc or be captured during the disruption of binary stars operated by the SMBH \citep{MO93,PG09, PG10}. 
Further from Sgr A$^*$, the central 0.5\,pc of the Galaxy harbour a clockwise disc of young and massive stars with ages between $2$ and $7$\,Myr and total mass of few $10^3\,M_\odot$ to $10^4\,M_\odot$ \citep{LE03, PA06, LU09, BA09, GE10, YG10, PF11}. { The presence of a second counter-clockwise and approximately coeval disc is still debated \citep{GS03, PA06, BA09, LU09, LU13}. }
As shown by hydrodynamical simulations, these stars might have formed locally inside a self-gravitating gaseous disc \cite[see e.g.][]{LE03, PA06,  NC05, HN09, M12} which fragmented in the central $0.5$\,pc of the Galaxy \citep[see e.g.][]{FN15}. { The mass function of the young stars has been claimed to be top-heavy when observing  only or mainly Wolf-Rayet and O-type stars \citep[$\propto m^{-0.85}$ or $\propto m^{-0.45}$][]{PA06, BA10}. A top-heavy mass function similar to the observed one has been reproduced in simulations under particular conditions on the temperature of the gas \citep{M12}. Using the same data and Bayesian approach adopted by \cite{DO13}, \cite{LU13} found a steeper mass function ($\propto m^{-1.7}$). We note that these are estimates for the current mass function and, as such, they are based on the observation of a small sample ($\sim 100$) of luminous stars (mostly O B and WR stars).}
The presence of the young stellar disc as well as the existence of  \ion{H}{ii}  regions and young
stars within the central $100$\,pc of the Milky Way \citep{FR04} suggest that the GC went through recent star formation processes in this region \citep{GE10}.
However, while the majority of current studies focused on the understanding of the build-up of NSCs through the infall of stellar clusters \citep{TR75, CD93, AN12, AS14, PMB14,MBP14, GN14, AS15, AN15, AS17b, AS17, TMB17, AMB18}, very few studies so far explore the long-term effects and evolution of NSCs with in-situ star formation \citep{AH15, GER16, BA18}. In particular the potential in-situ star formation producing stellar discs such as the one observed in the GC had not been investigated. 
By means of direct $N$-body and hybrid self-consistent field method models, \cite{PMB18} illustrated how spiral arms can arise in the current young stellar disc together with 
a warped disc structure. These instabilities, caused by collective effects introduced by a temporally aspherical NSC and strengthened by the presence of a SBH cusp, can explain the observations including the presence of overdensities in the disc, without invoking  external effects, like the presence of an intermediate mass black hole. { The warping of the disc due to the resonant relaxation happening between the disc and the surrounding old NSC could also be the responsible for the detection of the second counter-clockwise and  coeval second disc \citep{KT11}. In addition, \cite{US13} suggested the possibility that the second disc is the result of a self-interacting single disc breaking up into two components.}\\
Observations suggest that stellar discs at the GC formed at intervals of approximately 100\,Myr \citep[][]{HN09}.
Those discs must have co-evolved leaving observable chemo-dynamical signatures on the structure of the different stellar populations.
Here we focus on such multiple star formation epochs in stellar discs and explore the evolution and build-up of NSCs through these processes.    
We make use of $N$-body simulations in which we mimic the events of star formation at the GC by introducing several stellar discs one after the other, while we consider a smooth potential for the background stellar component of the central cusp of the NSC and include the discrete component of the potential by simulating a SBH cusp similar to the one expected from theory and observations.
 These simulations provide clues to the effects and observational signatures for the in-situ disc star formation processes. In particular, they can point towards long-lasting observable dynamical signatures of initial disc-like configurations. They can also reveal the effects of a new stellar disc on previously formed stars and older stellar discs as well as the mutual interactions between different stellar discs. Together these may shed new light on the physical and dynamical processes in galactic nuclei and their implications for our understanding of the build-up of NSCs, their star formation histories and observational implications. \\
This paper is organised as follows: in Section \ref{sec:mm} we describe the initial conditions adopted and the simulations.
 In Section \ref{sec:res} we illustrate the results obtained and in Section \ref{sec:con} we draw our conclusions.
\begin{figure}
\centering
\includegraphics[width=0.45\textwidth]{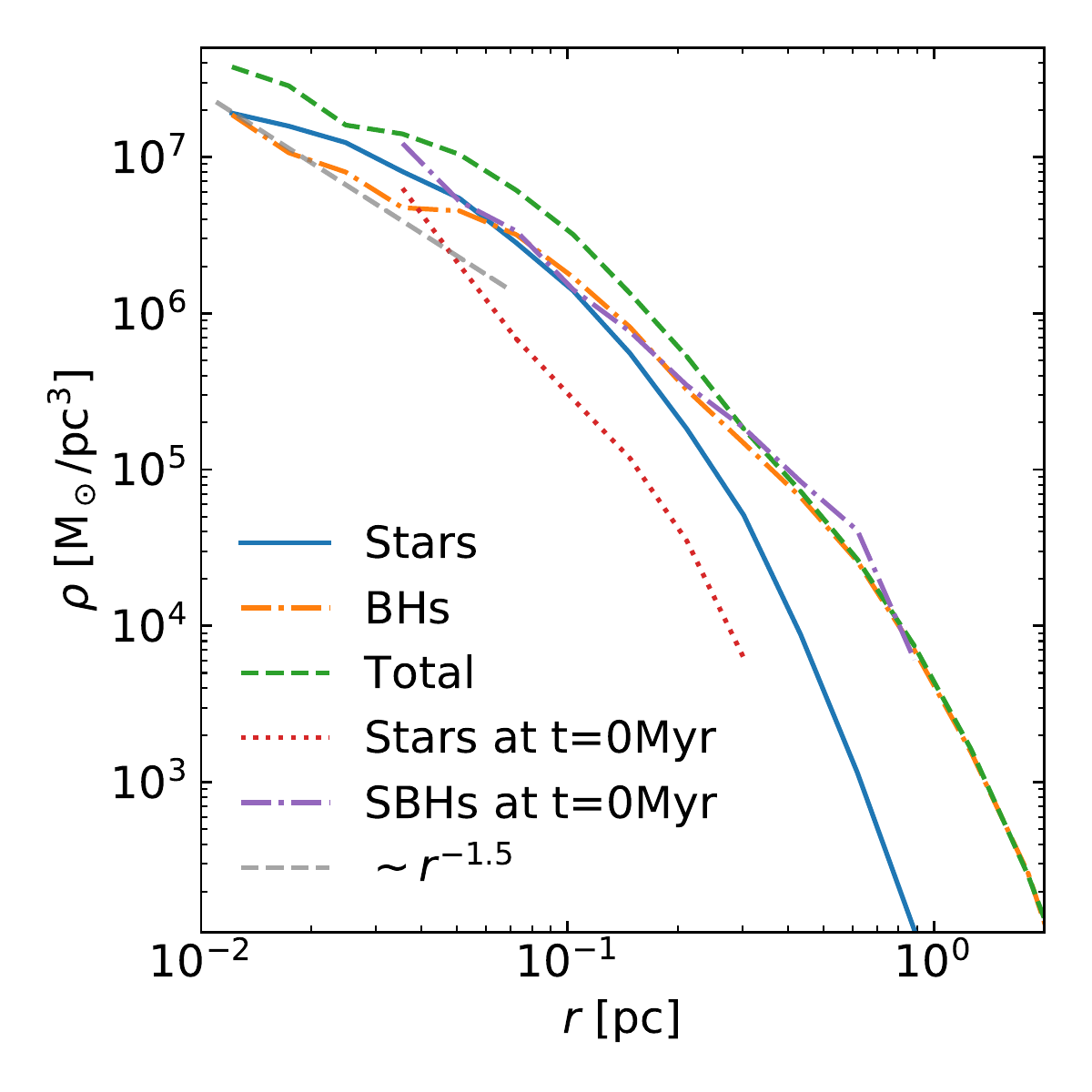}
\caption{Density profile of the stars in the five discs (solid light blue line),  the black hole cusp (dot-dashed orange line) and  the combination of the two populations (dashed green line) at the end of { the evolution, in the case of the first realisation}. The stars are characterised by a central shallower cusp compared to the initial profile of a single disc (dotted red line). The SBHs diffuse inwards and outwards and show a density profile similar to the initial one in the intermediate region (dot-dashed purple line) and a profile $\propto r^{-1.5}$ in the internal region (dashed grey line).}\label{fig:density}
\end{figure}
\begin{figure}
\centering
\includegraphics[width=0.45\textwidth]{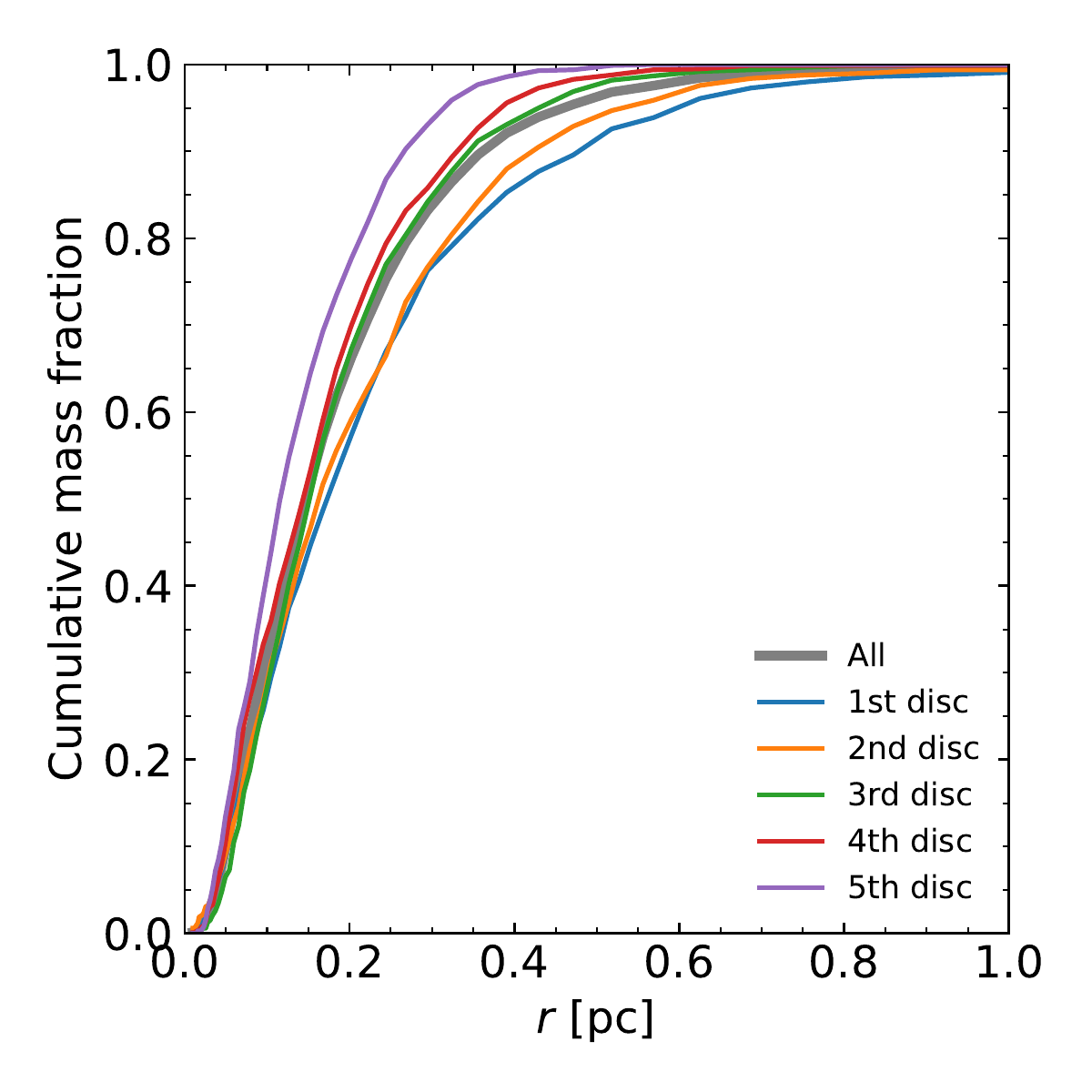}
\caption{Final cumulative mass fraction for each disc and for the whole stellar component { in the first realisation}. The distribution is shallower and more extended for older discs compared to younger discs, which are still more centrally concentrated with respect to older populations.}\label{fig:cumulative_mass}
\end{figure}
\begin{figure}
\centering
\includegraphics[width=0.45\textwidth]{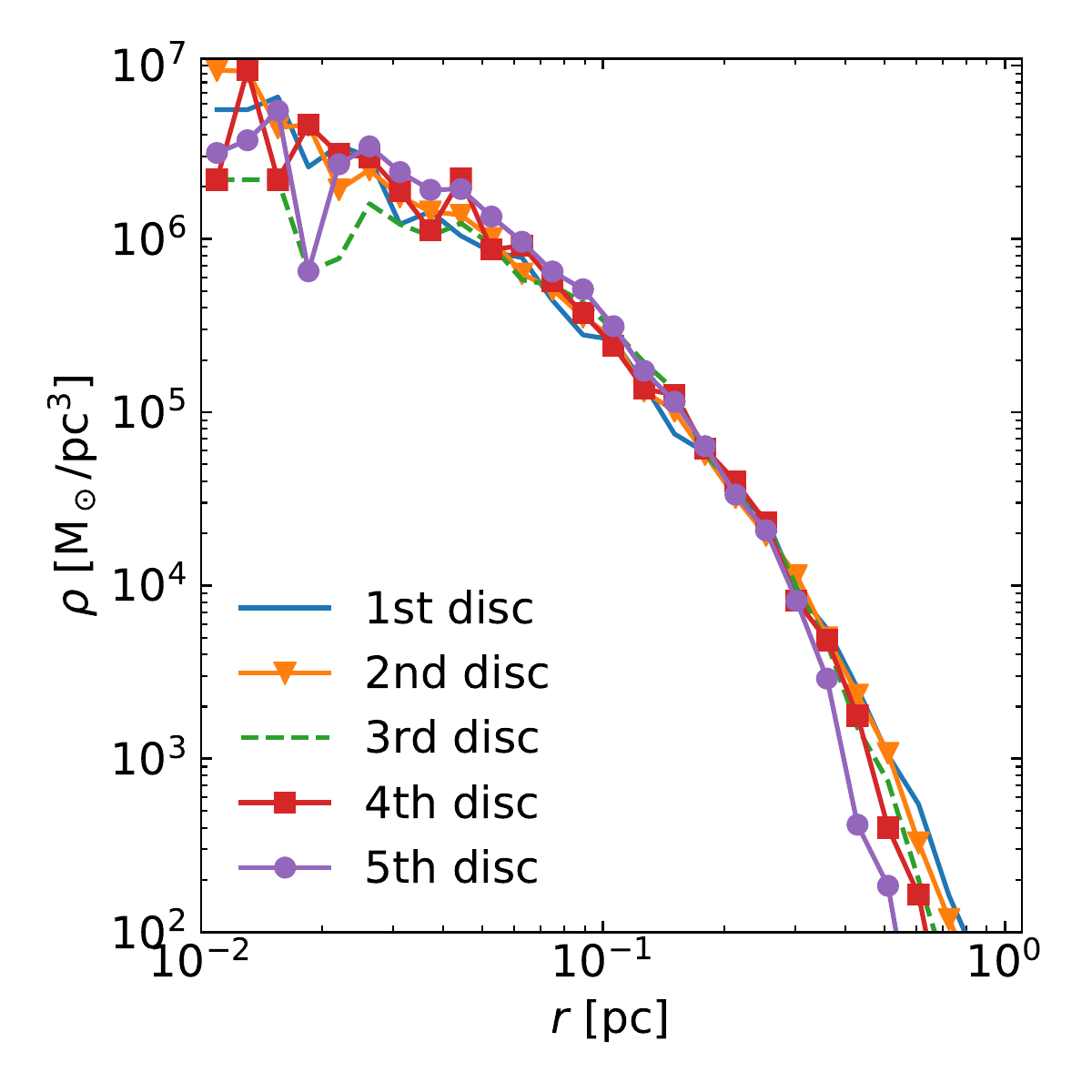}\\
\includegraphics[width=0.45\textwidth]{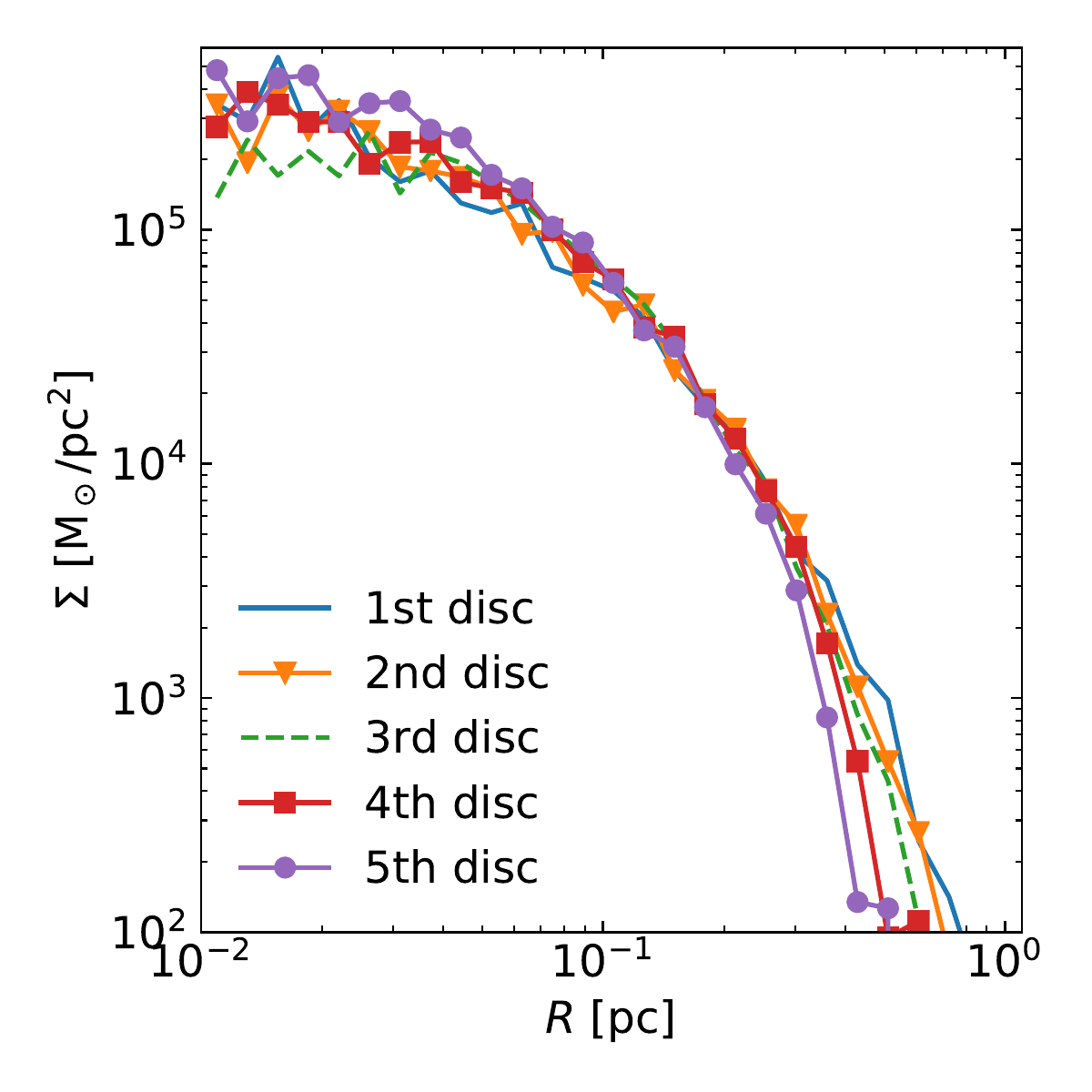}
\caption{Spatial (upper panel) and projected (bottom panel) radial density profiles of each disc { in the first realisation} at the end of the simulation.}\label{fig:density_sep}
\end{figure}
\begin{figure}
\centering
\includegraphics[width=0.45\textwidth]{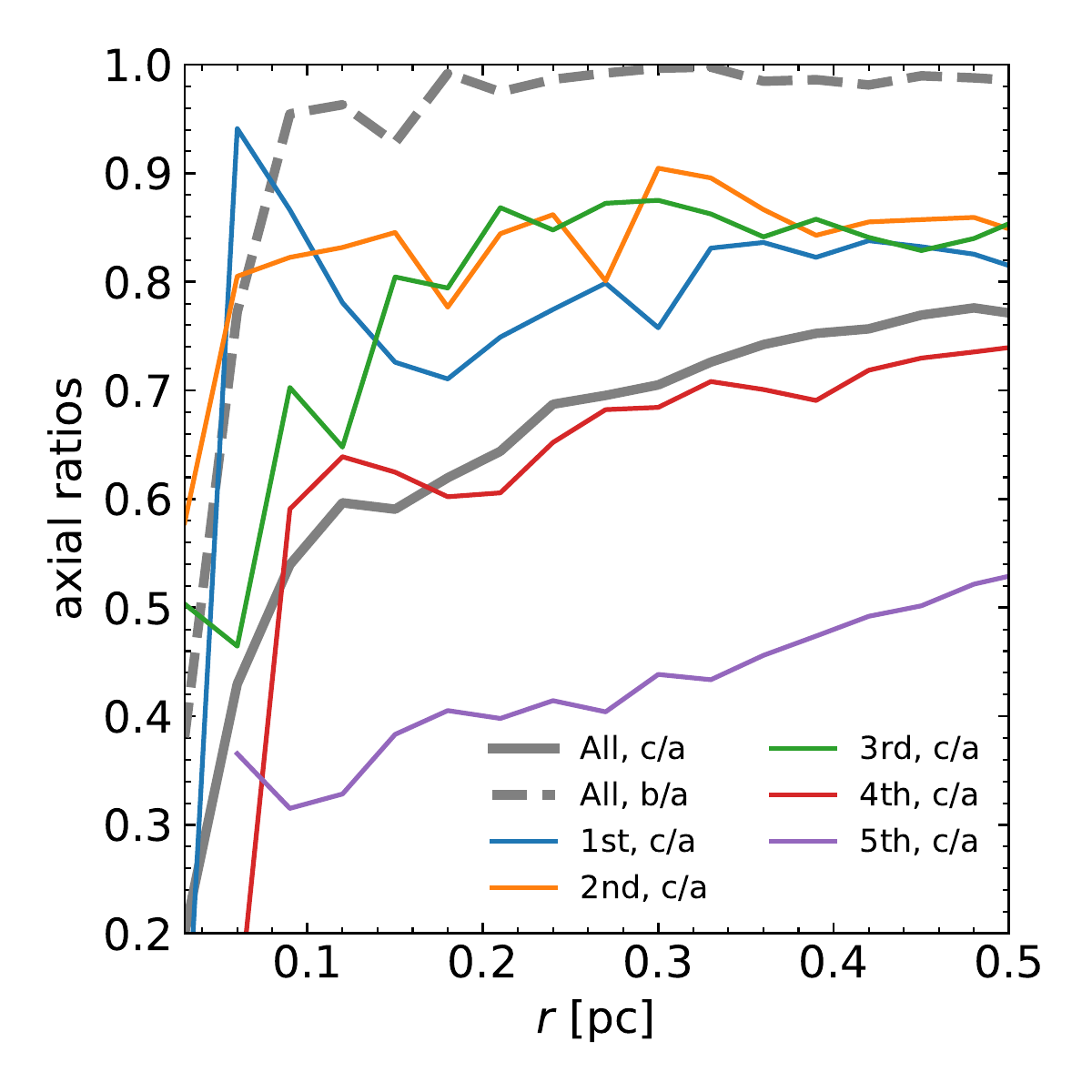}
\includegraphics[width=0.45\textwidth]{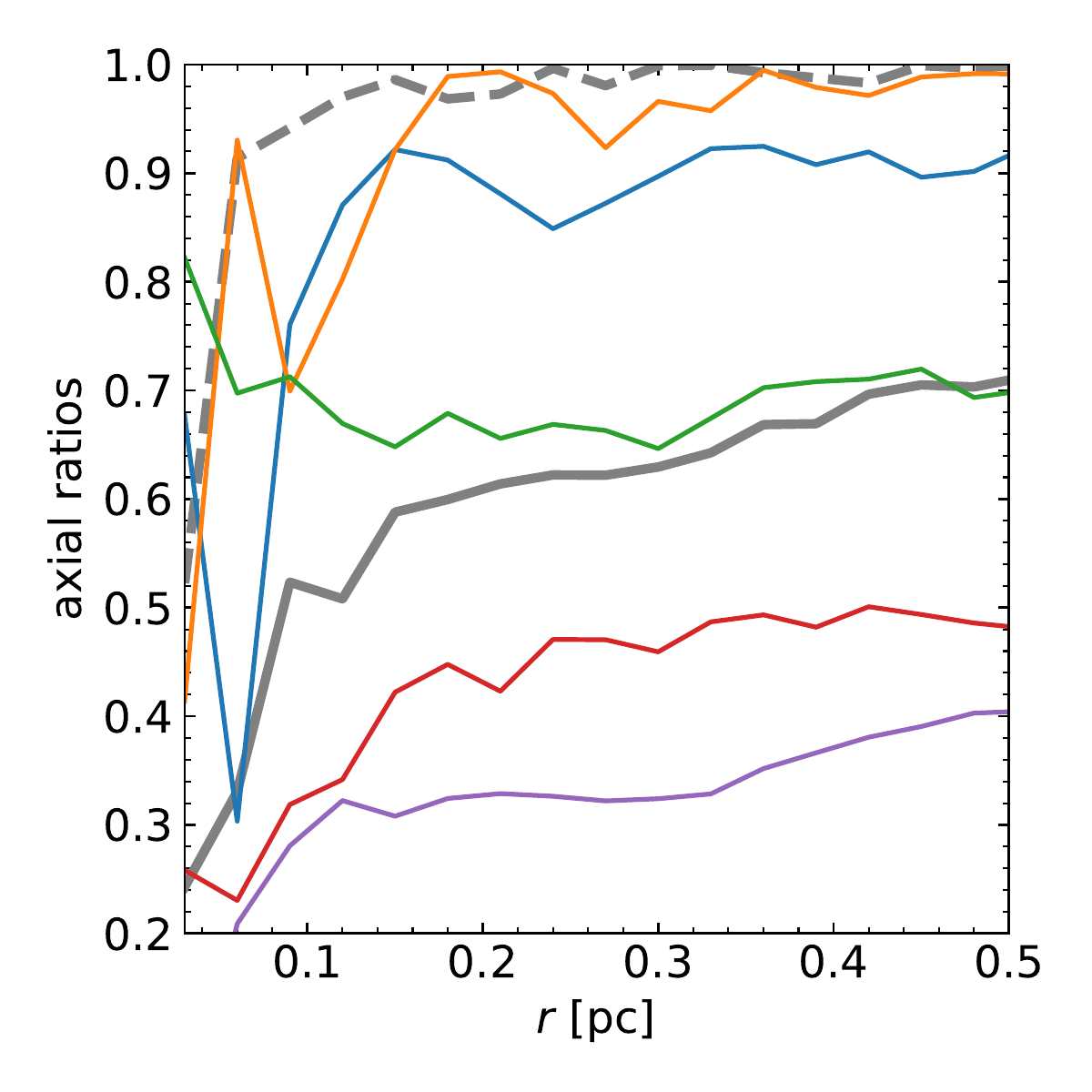}
\caption{Axial ratios of the total system (solid and dashed grey lines). The minor axial ratio $c/a$ ratio
is shown for the different discs separately. The top panel is for the first realisation while the bottom panel is for the second one.}\label{fig:ax_ratios}
\end{figure}
\begin{figure*}
\includegraphics[width=0.33\textwidth]{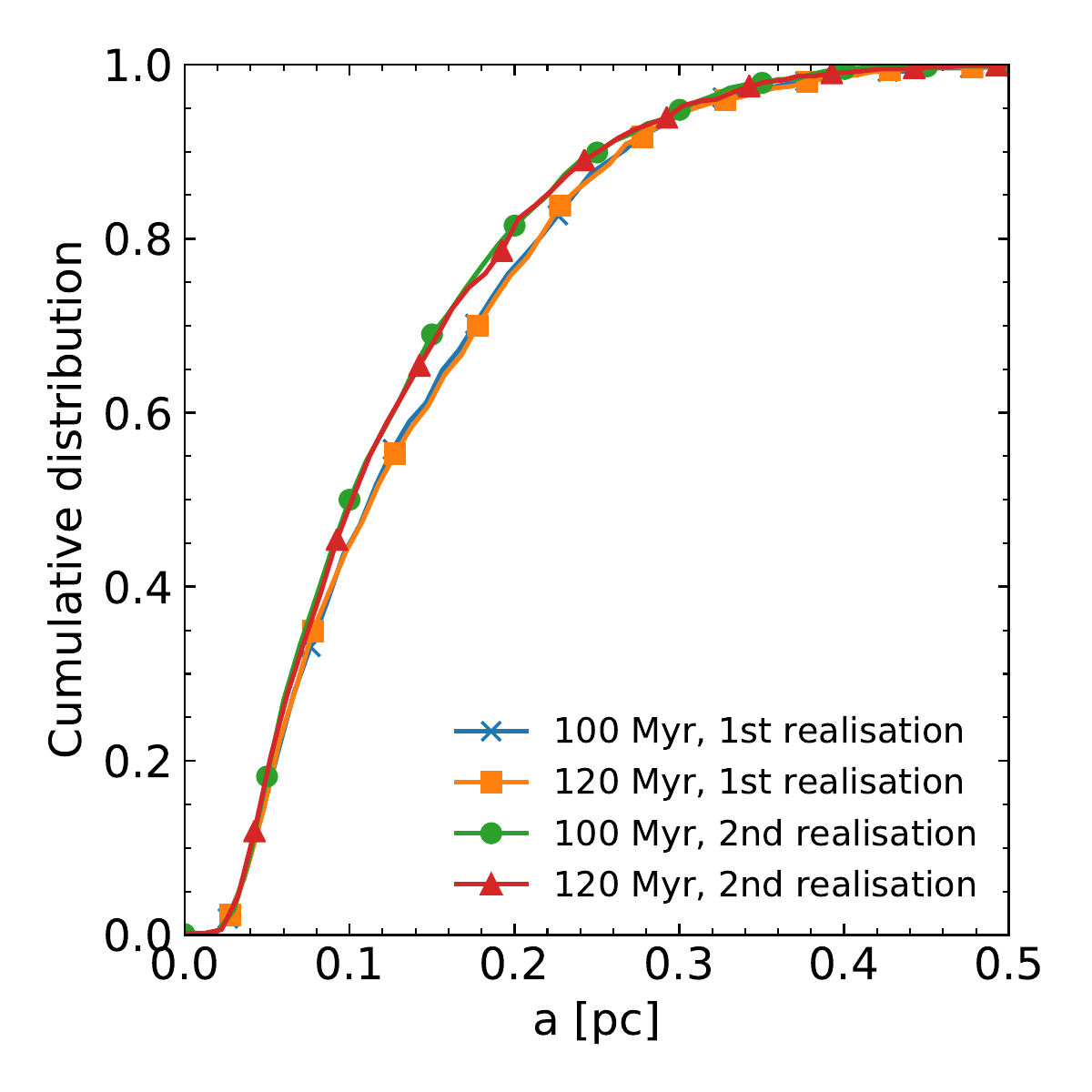}~\includegraphics[width=0.33\textwidth]{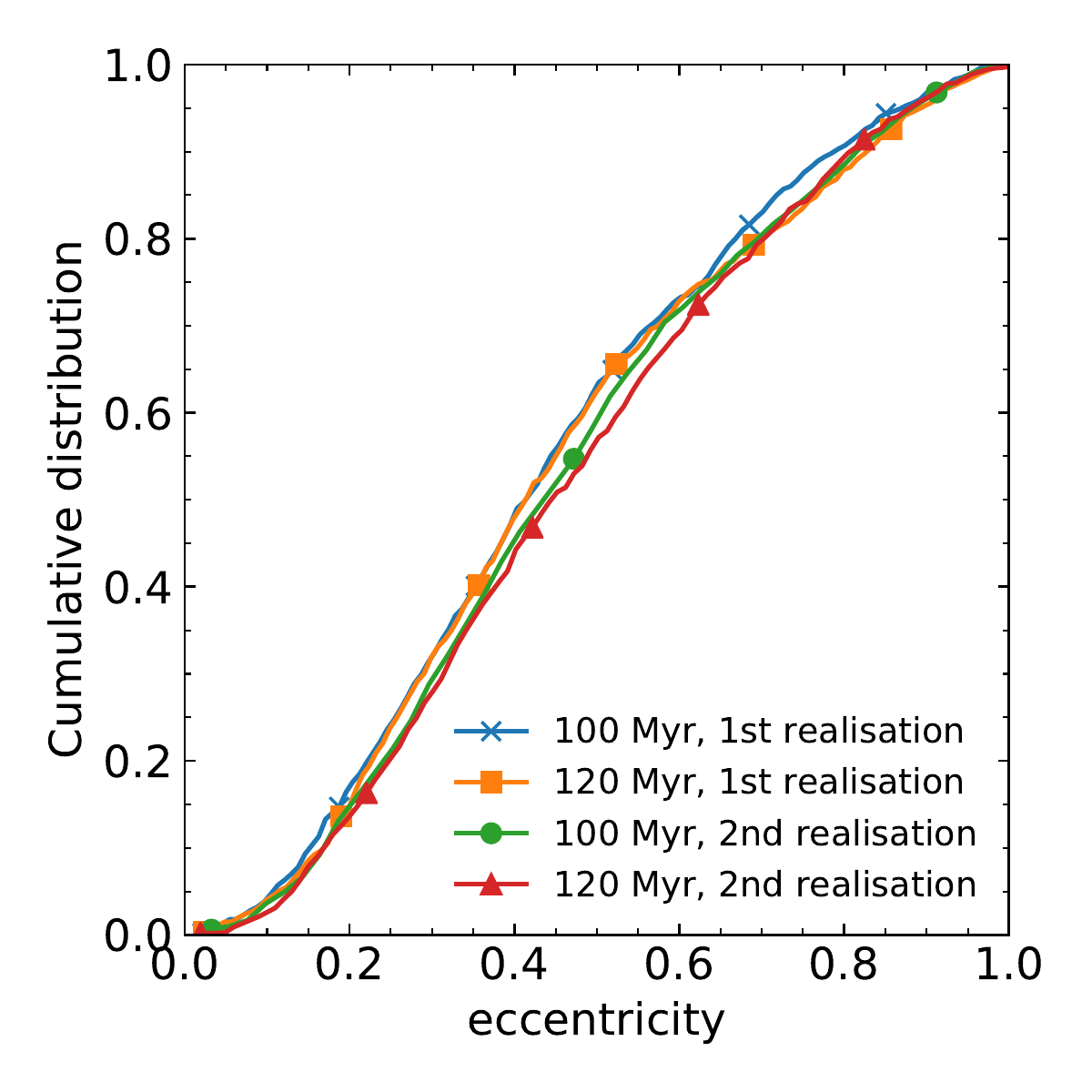}~\includegraphics[width=0.33\textwidth]{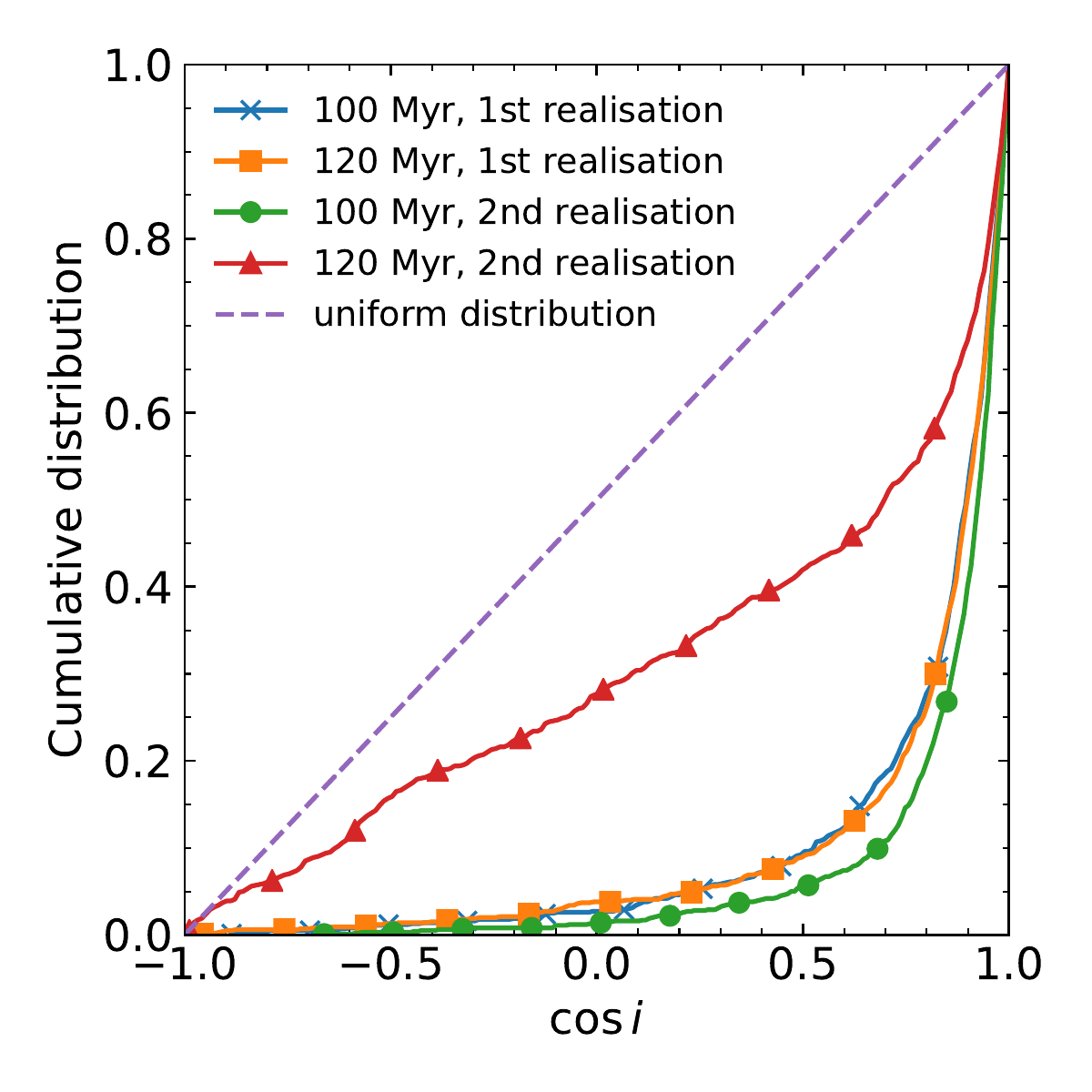}
\caption{The cumulative distribution of the semi-major axis (left panel), the eccentricities (middle panel) and the inclinations given as $\cos i$ (right panel) are shown for the fist disc after 100\,Myr and 120\,Myr of evolution for both the simulations described in the text. The counterrotation of the second disc in the case of the second realisation leads to a quick evolution of the disc towards a uniform distribution of the inclinations.}\label{fig:first_disc_second}
\end{figure*}
\begin{figure*}
\centering
  \begin{tabular}{@{}ccccc@{}}
\includegraphics[width=0.19\textwidth]{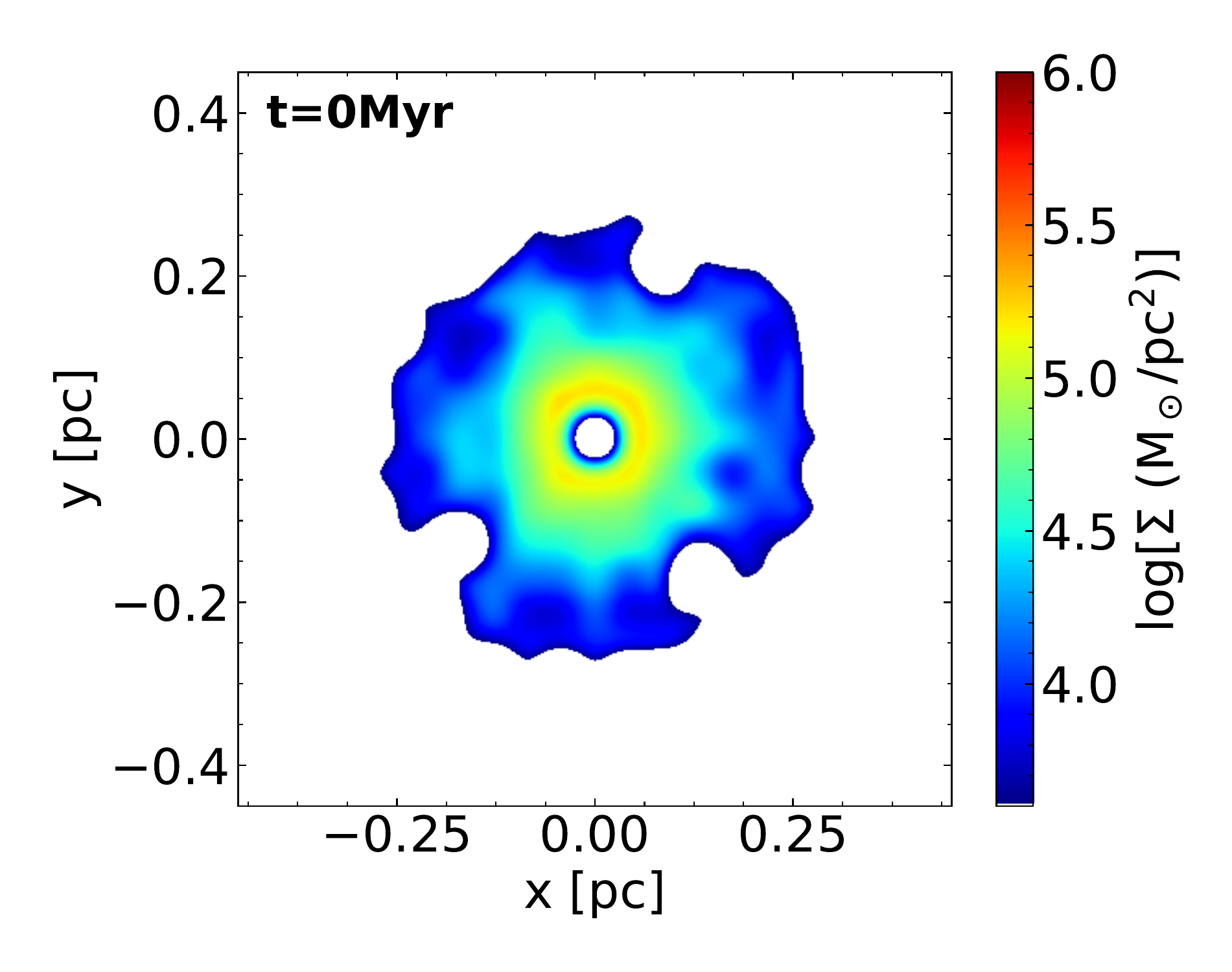}
\includegraphics[width=0.19\textwidth]{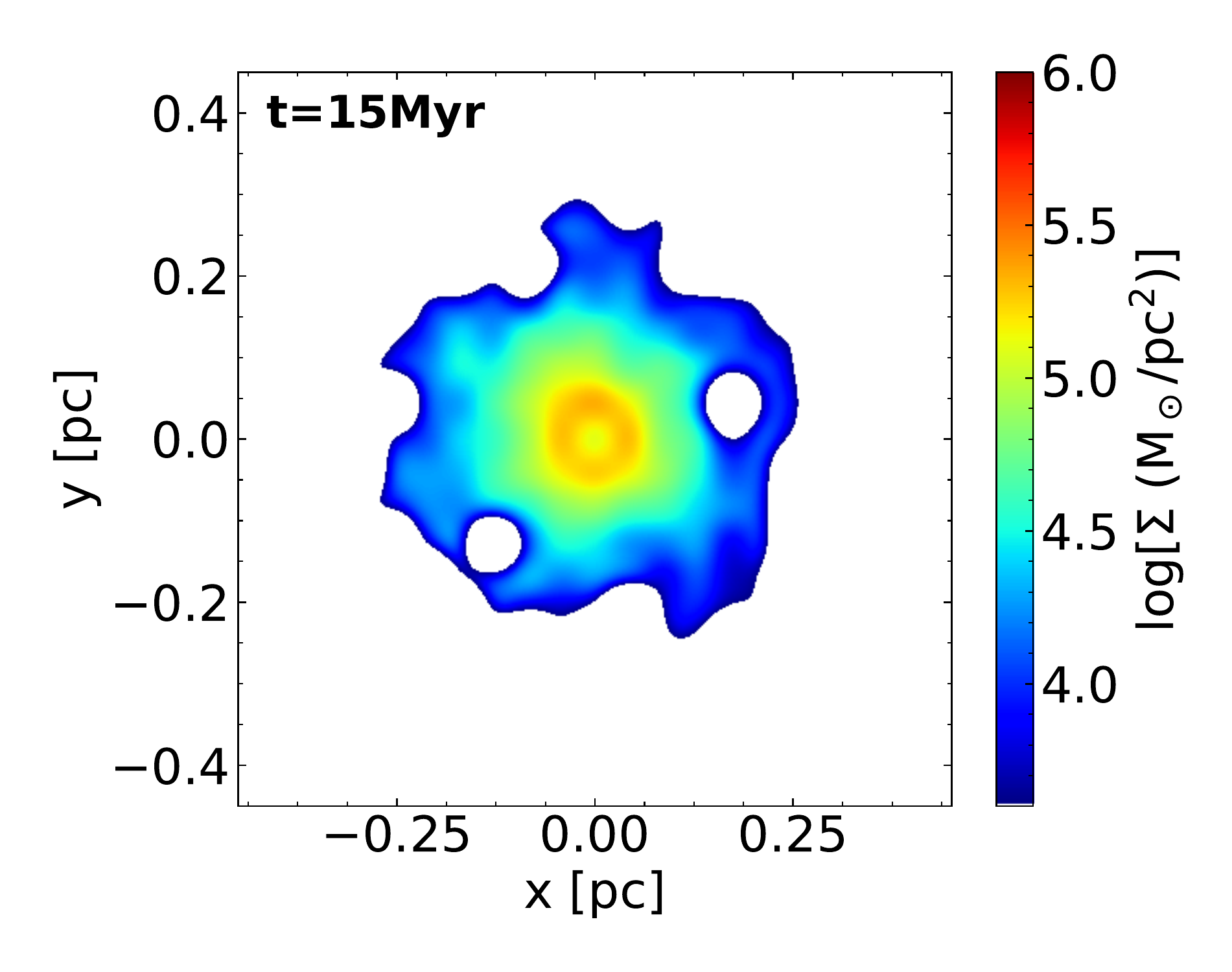}
\includegraphics[width=0.19\textwidth]{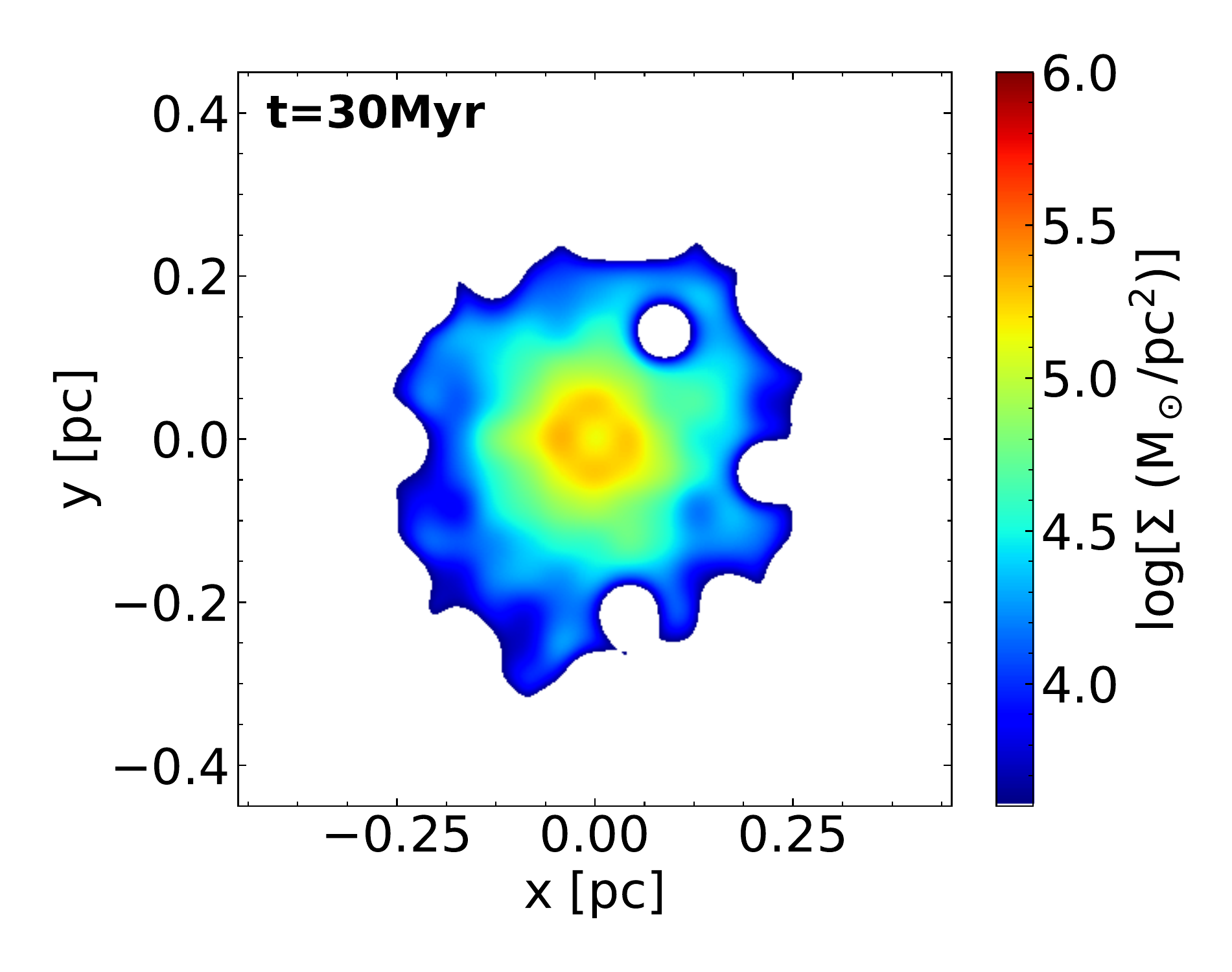}
\includegraphics[width=0.19\textwidth]{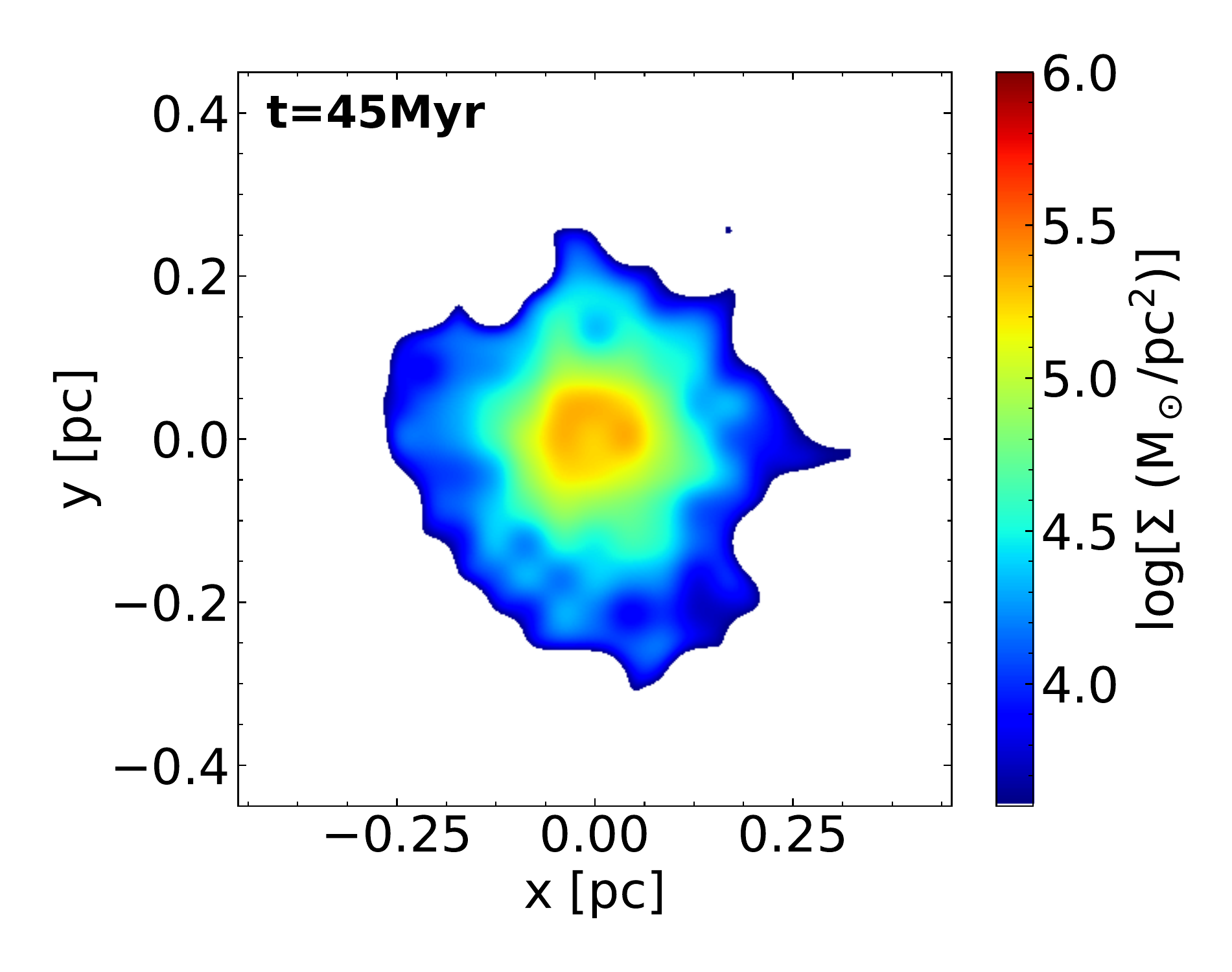}
\includegraphics[width=0.19\textwidth]{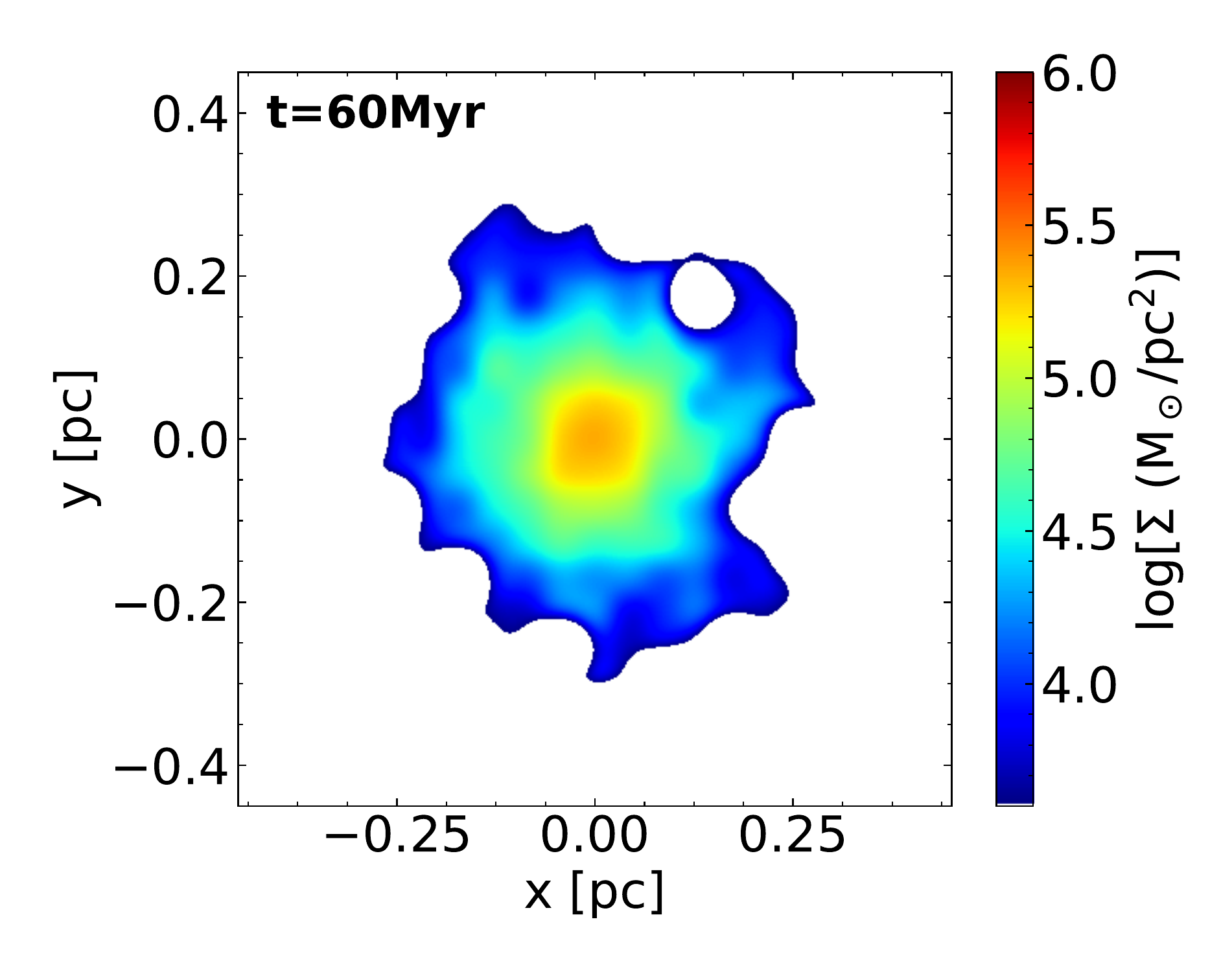}\\
\includegraphics[width=0.19\textwidth]{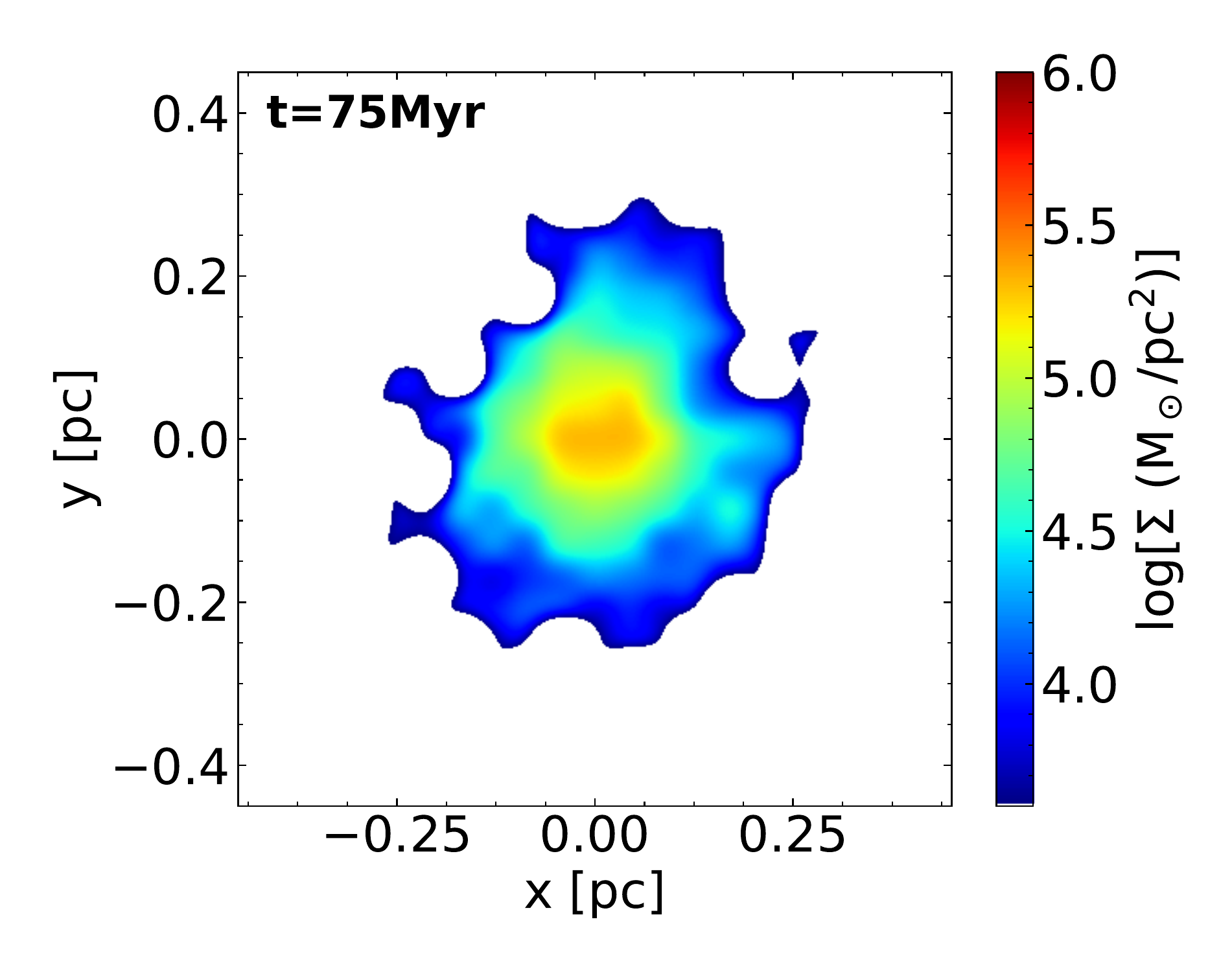}
\includegraphics[width=0.19\textwidth]{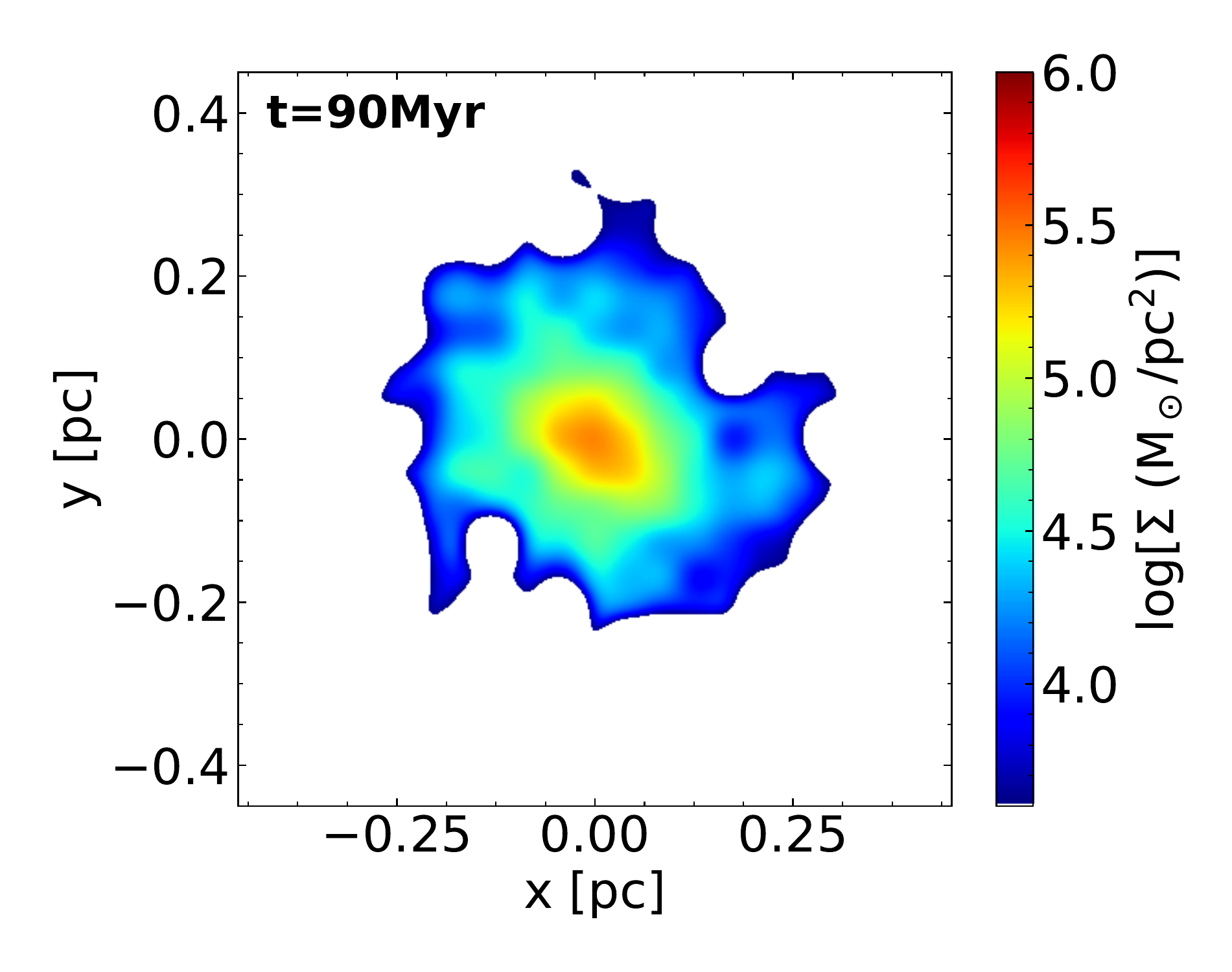}
\includegraphics[width=0.19\textwidth]{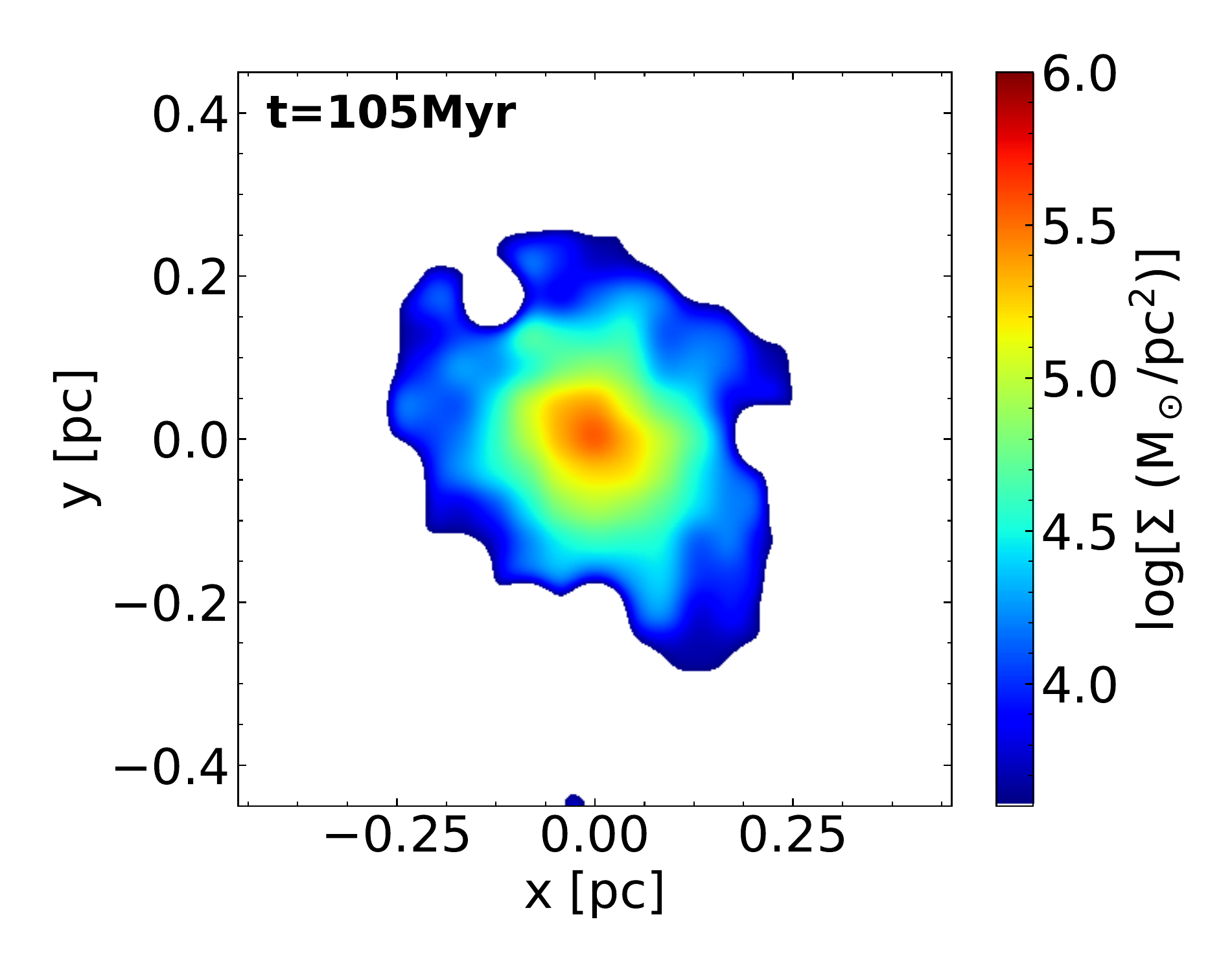}
\includegraphics[width=0.19\textwidth]{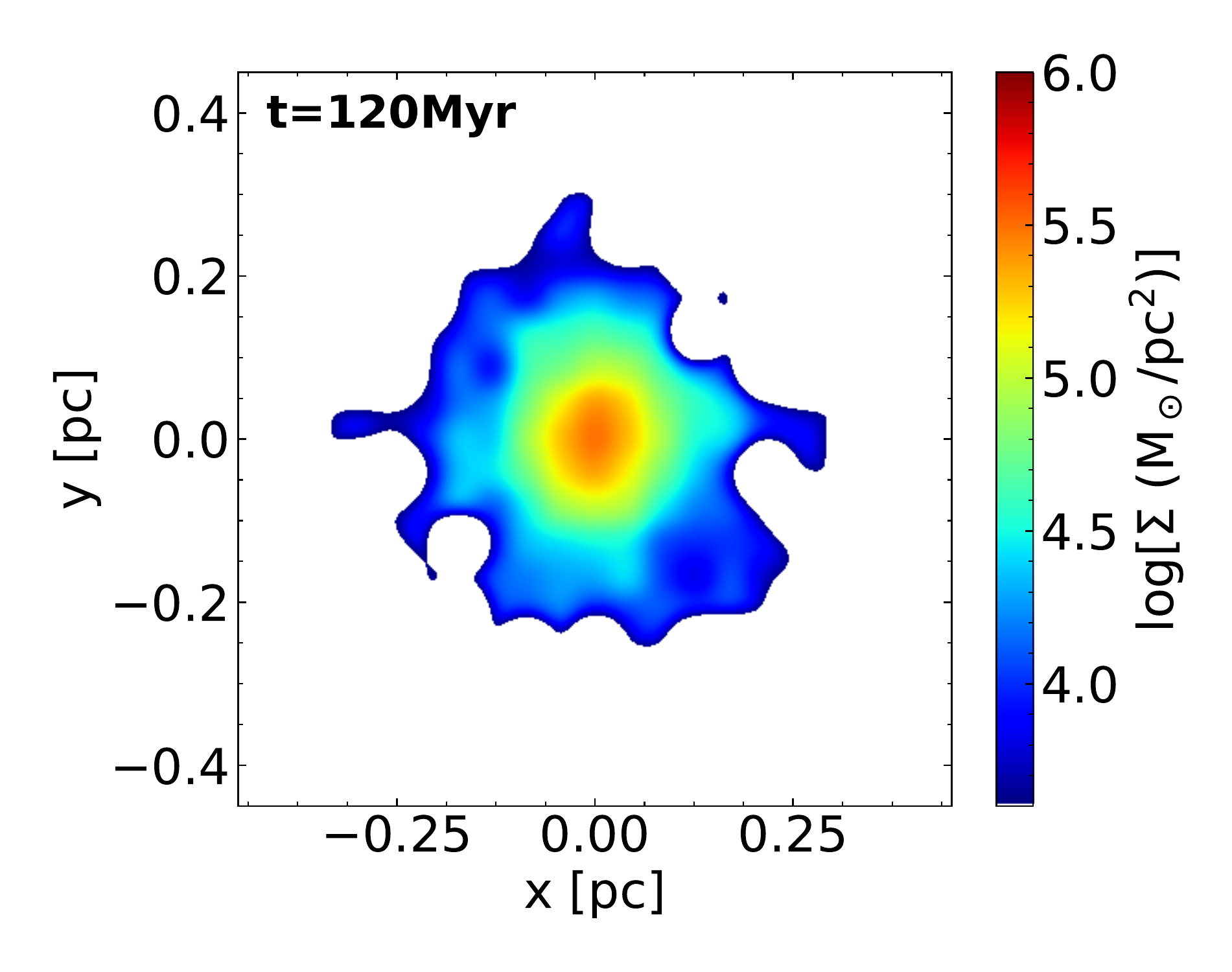}
\includegraphics[width=0.19\textwidth]{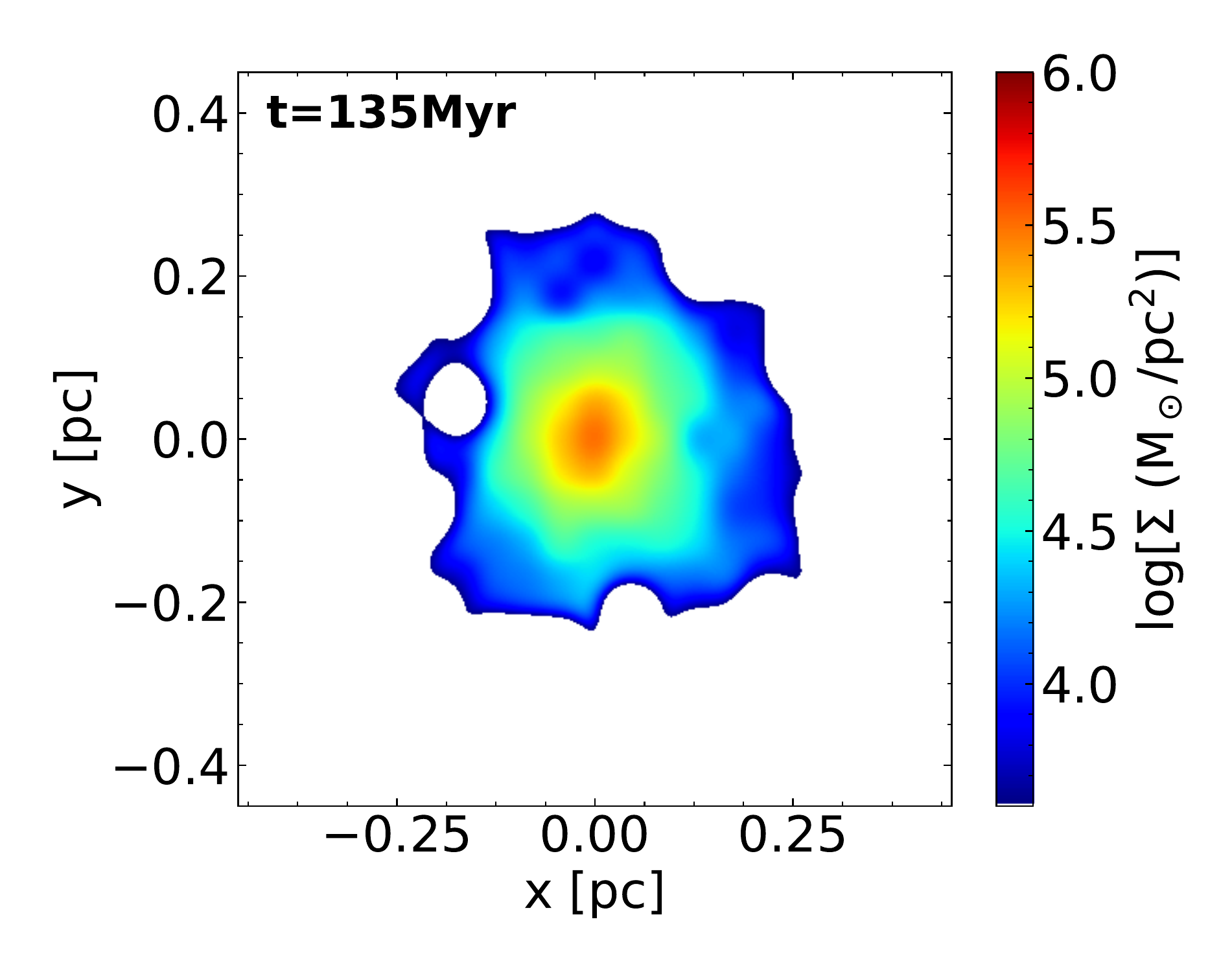}
 \end{tabular}
\caption{Face-on evolution of the first disc in the second realisation. Snapshots are taken every 15\,Myr. The second disc appears after $100$\,Myr. The second panel in the second row is for a snapshot taken at 90\,Myr, just before the second disc appears. The third panel in the same row corresponds to a snapshot taken at 105\,Myr, right after the second disc has appeared.}\label{fig:fmaps_fo}
\end{figure*}
\begin{figure*}
\centering
  \begin{tabular}{@{}ccccc@{}}
\includegraphics[width=0.19\textwidth]{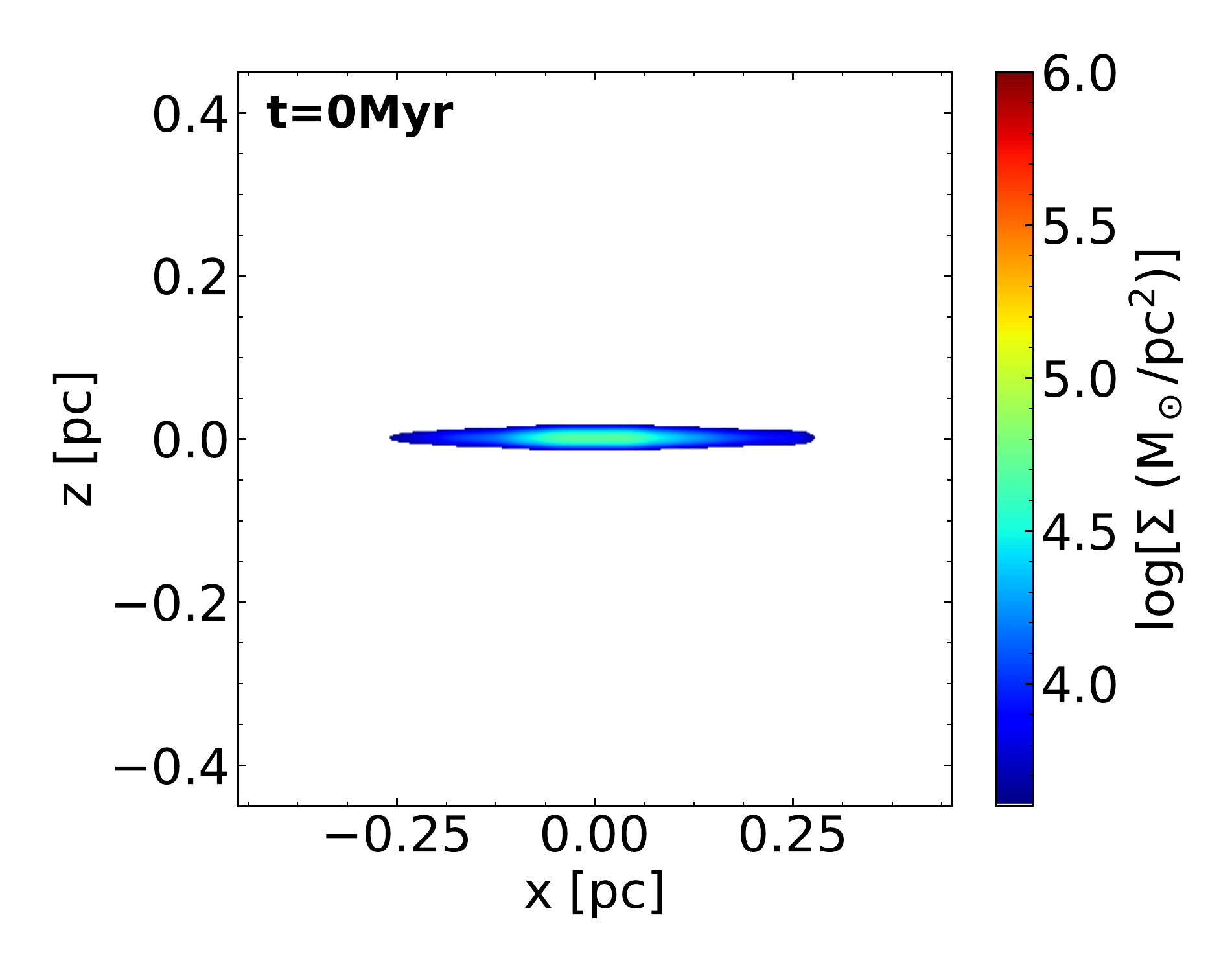}
\includegraphics[width=0.19\textwidth]{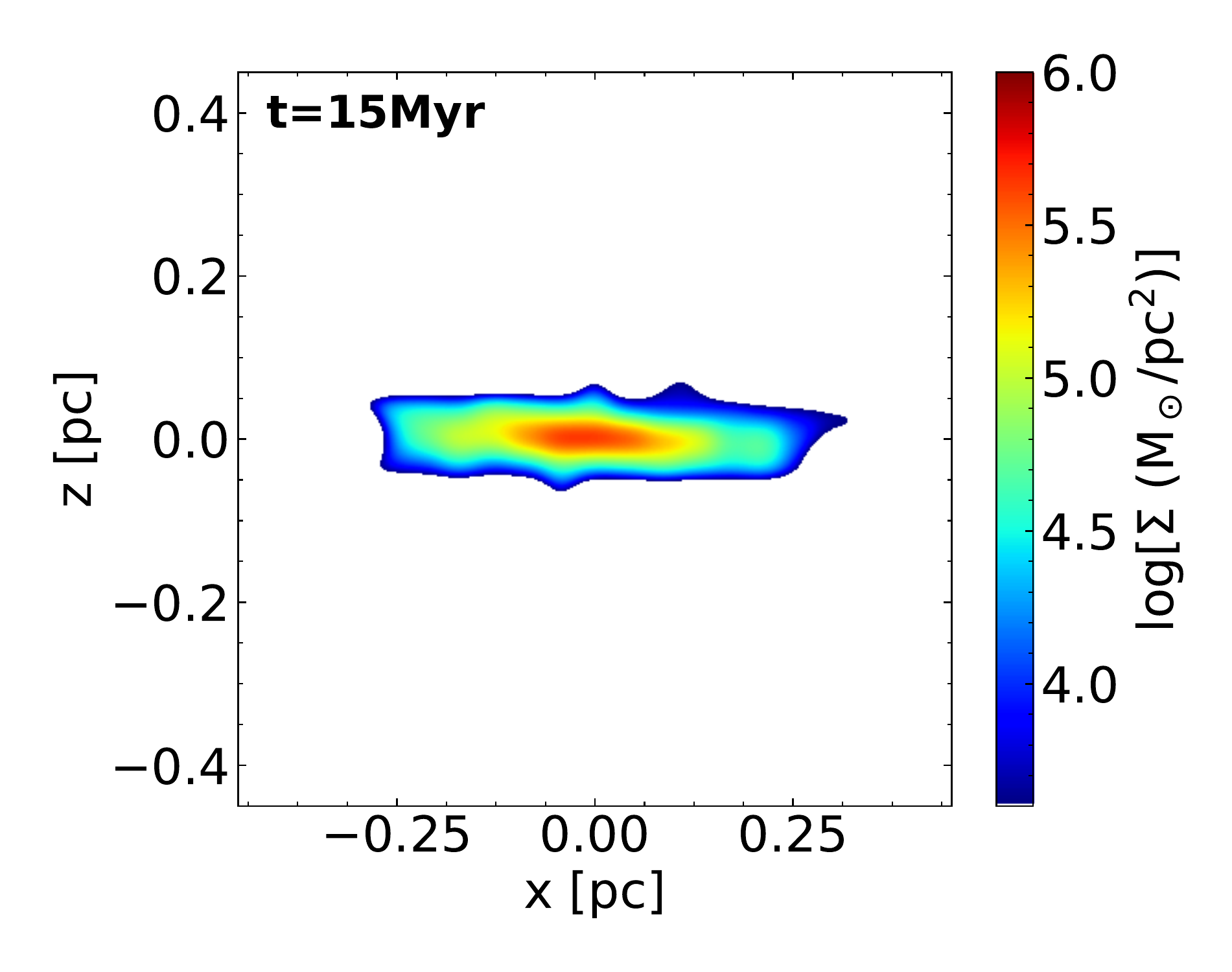}
\includegraphics[width=0.19\textwidth]{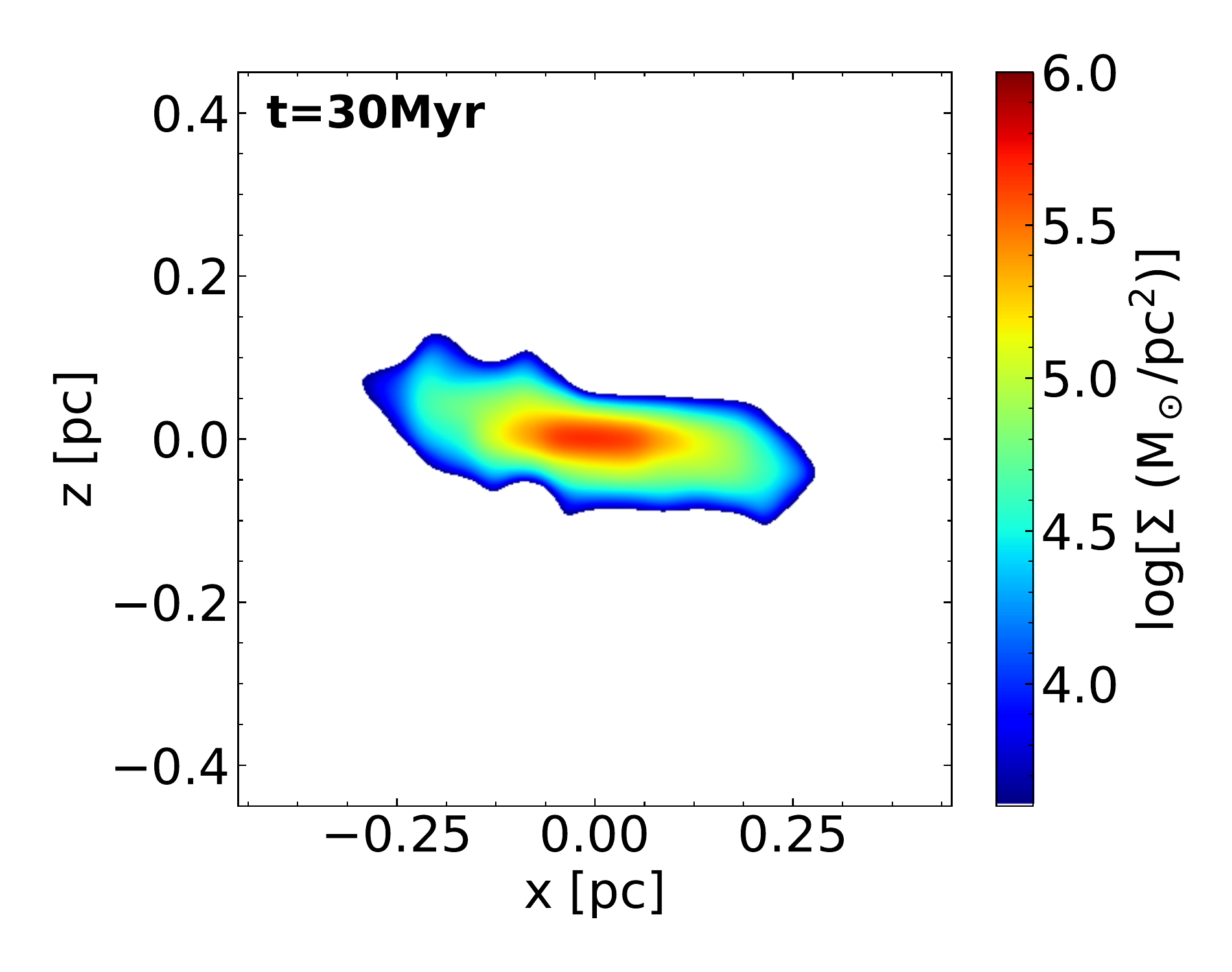} 
\includegraphics[width=0.19\textwidth]{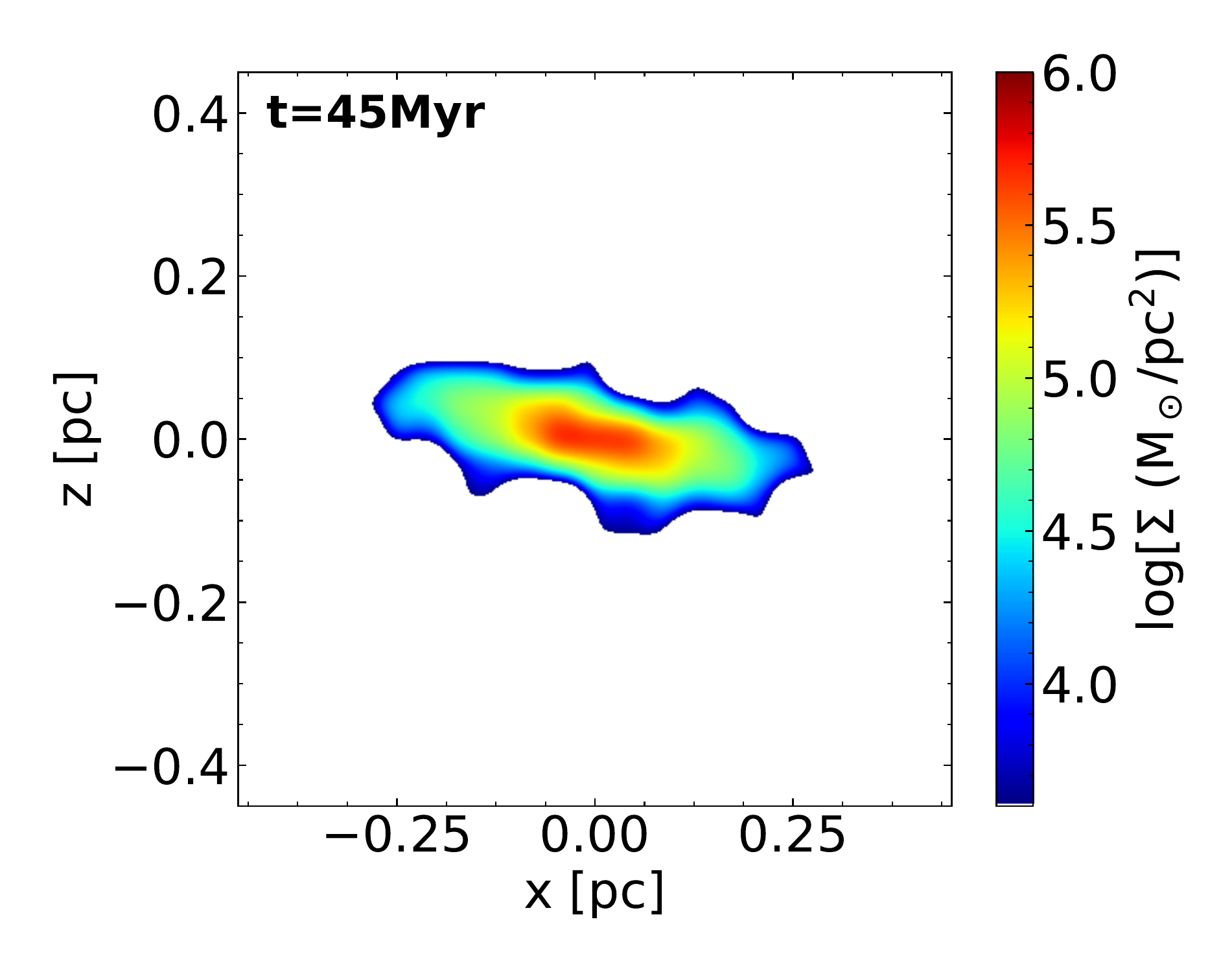}
\includegraphics[width=0.19\textwidth]{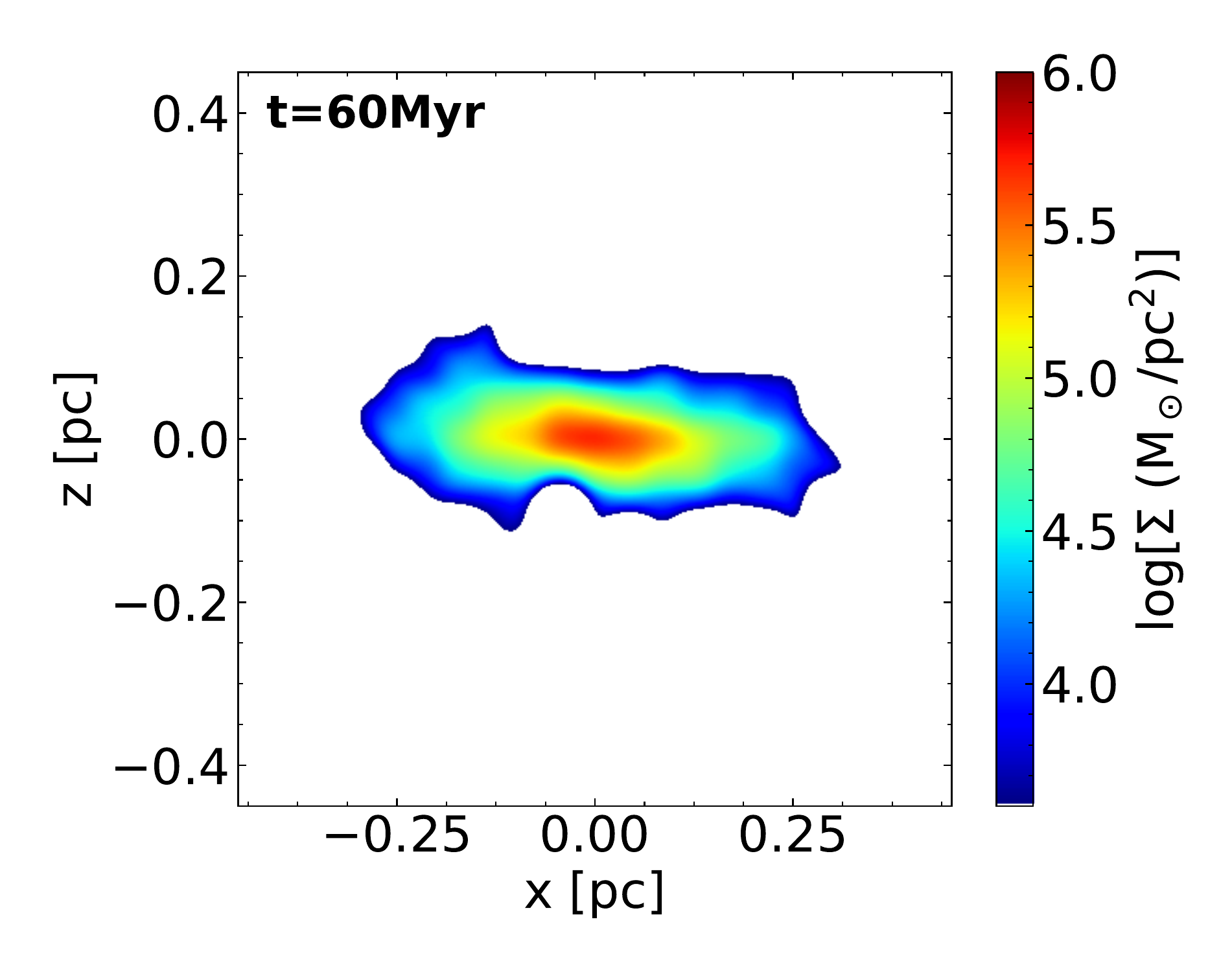}\\
\includegraphics[width=0.19\textwidth]{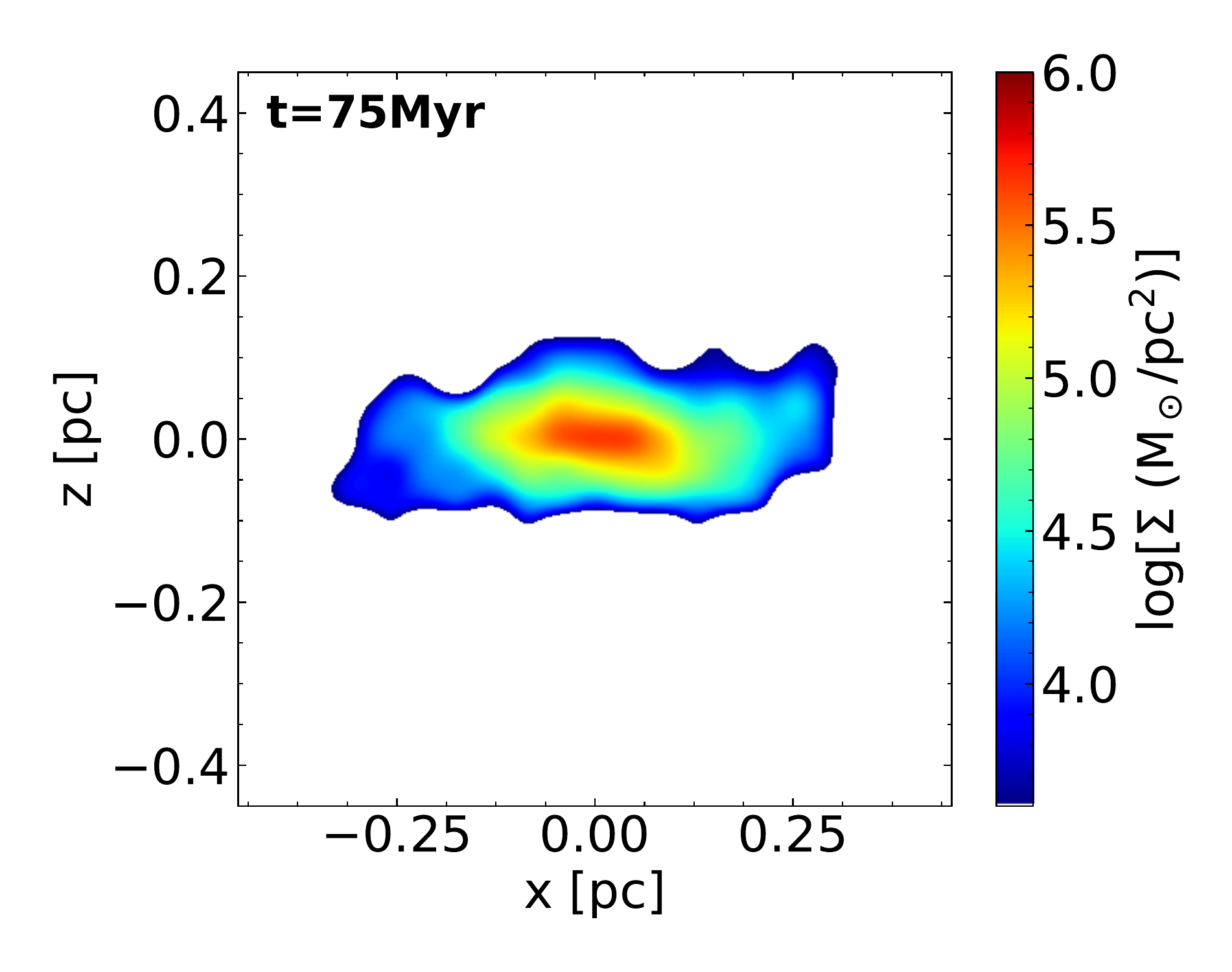}
\includegraphics[width=0.19\textwidth]{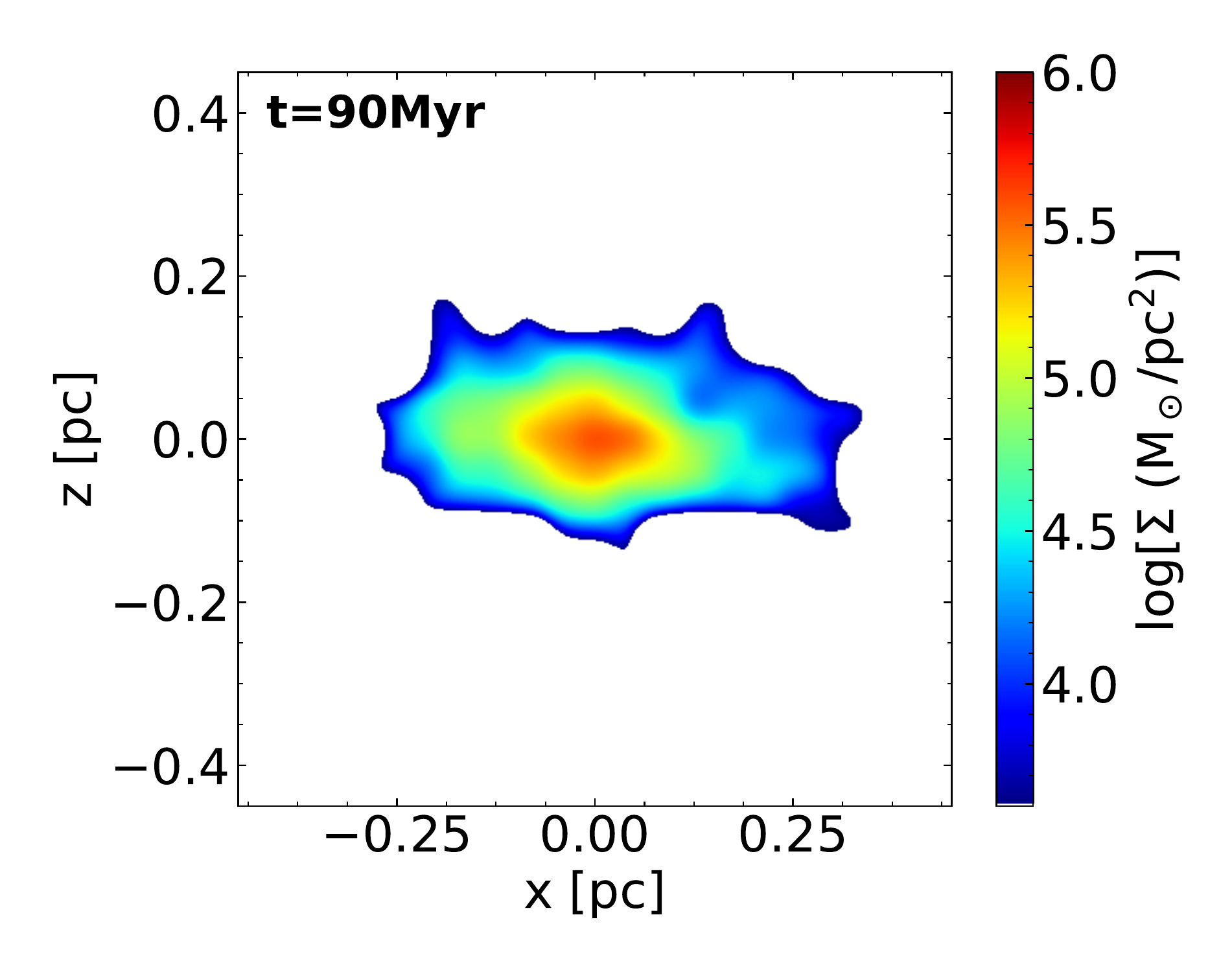}
\includegraphics[width=0.19\textwidth]{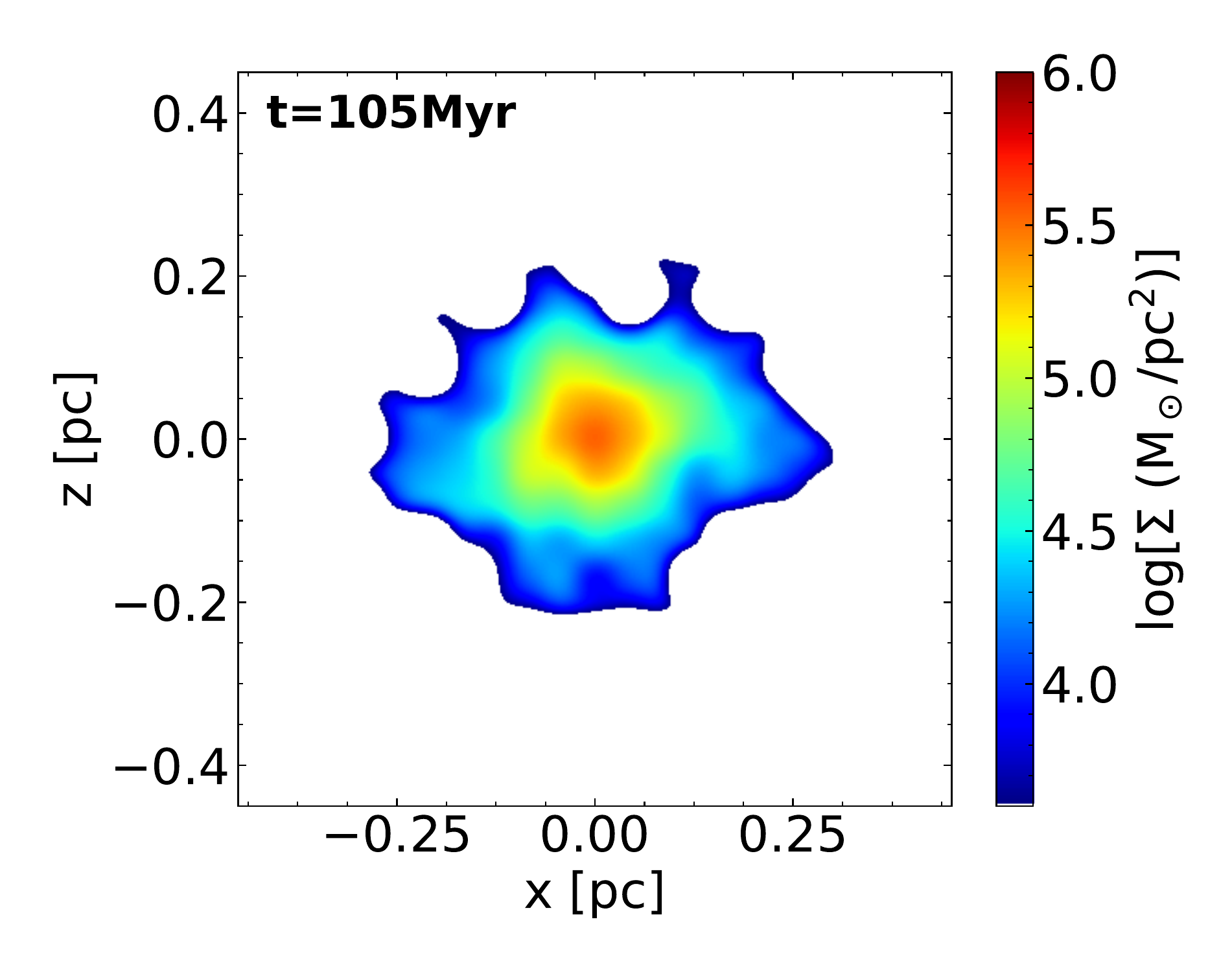}
\includegraphics[width=0.19\textwidth]{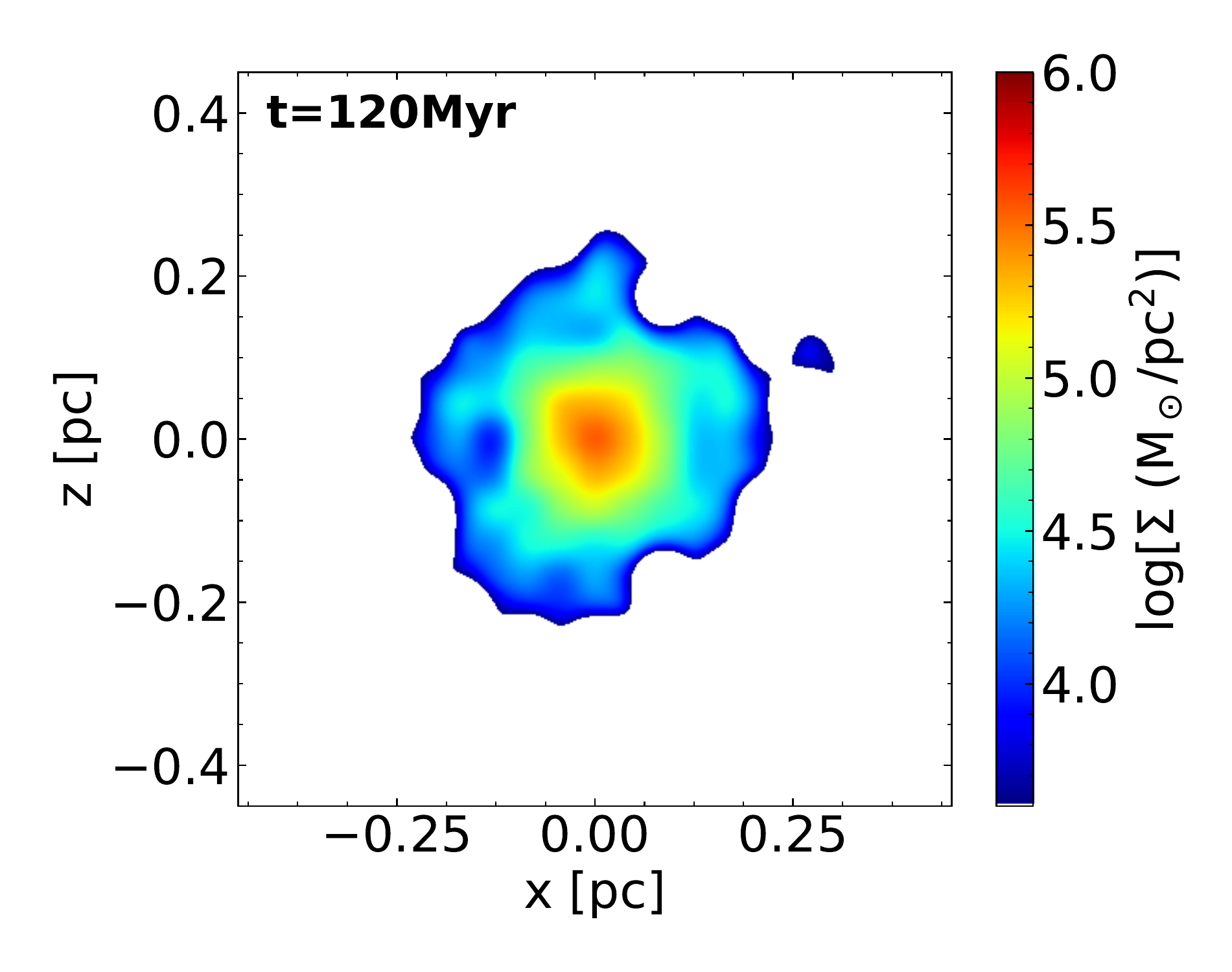}
\includegraphics[width=0.19\textwidth]{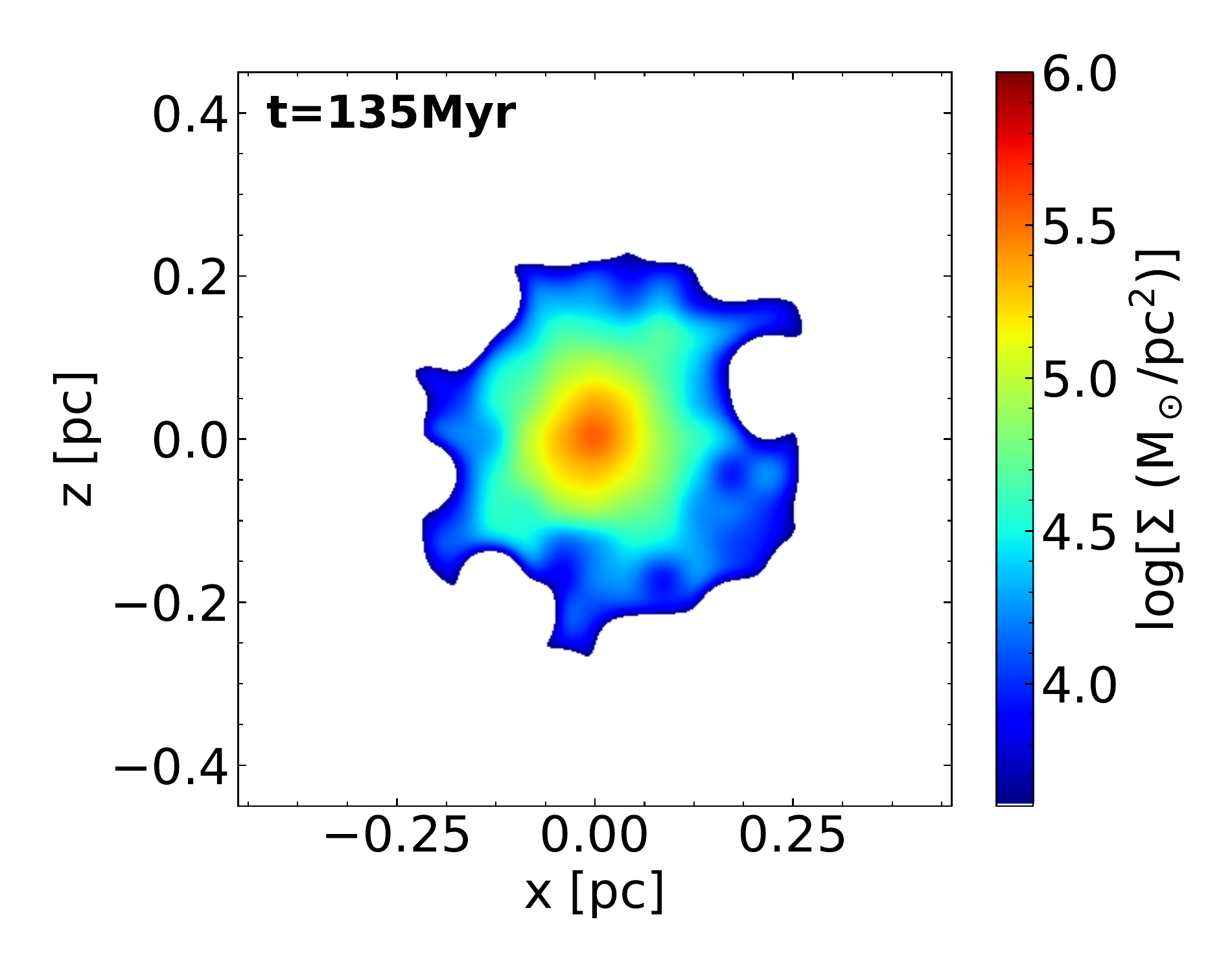}
 \end{tabular}
\caption{Edge-on evolution of the first disc in the second realisation. Snapshots are taken every 15\,Myr. The second disc appears after $100$\,Myr. The second panel in the second row is for a snapshot taken at 90\,Myr, just before the second disc appears. The third panel in the same row corresponds to a snapshot taken at 105\,Myr, right after the second disc has appeared.}\label{fig:fmaps_eo}
\end{figure*}
\section{Models and methods}\label{sec:mm}
To study the interaction of multiple stellar populations born one after the other at the GC, we  followed the evolution of five stellar discs by means of direct $N$-body simulations. { The number of simulated discs is constrained by the high computational demand and uncertainties on the knowledge of star formation rate at the GC. As such, our simulations are not able to reproduce the exact conditions at the GC, but rather provide observational constraints on the effects of multiple episodes of star formation in stellar discs.}
We introduced each disc after the previous one had evolved for 100\,Myr to mimic the star formation events. This time-scale choice is justified by the missing evidence for star formation episodes in the  $100$ Myr prior to the episode that lead to the formation of the observed young stellar disc \citep{HN09}. The detailed models of the discs and NSC as well as the adopted initial conditions and the simulations are described in the following paragraphs. 

\subsection{{\it N}-body simulations}
We performed our $N$-body simulations using the direct $N$-body code phiGRAPE \citep{HA07},
adapted to run on NVIDIA graphic processing units (GPUs) through the Sapporo library \citep{GA09}.
The basic version of phiGRAPE has been modified to include an external analytic potential in the form of a stellar cusp and to consider the 
central SMBH as a fixed central potential \citep[see][]{PG09, PMB18}. This latter choice is justified by the necessity of reducing spurious Brownian motion of the SMBH due to low particle numbers in the model \citep{BG16}. Simulations with a fixed and free SMBH show a comparable behaviour \citep[see][]{PMB18}.
The simulations have been run both on the cluster Tamnun at the Technion and on the cluster BwFor at the Computing Center of Heidelberg University (URZ).
To avoid close encounters between particles, we adopted a softening length of $3\times10^{-4}$ pc.  
The relative change in energy is $(E_{f}-E_i)/E_i\leq0.1$, where $E_i$ and $E_{f}$ are the values of the energy at the beginning and at the end of the evolution of each of the discs.
\begin{figure}
\centering
\includegraphics[width=0.4\textwidth]{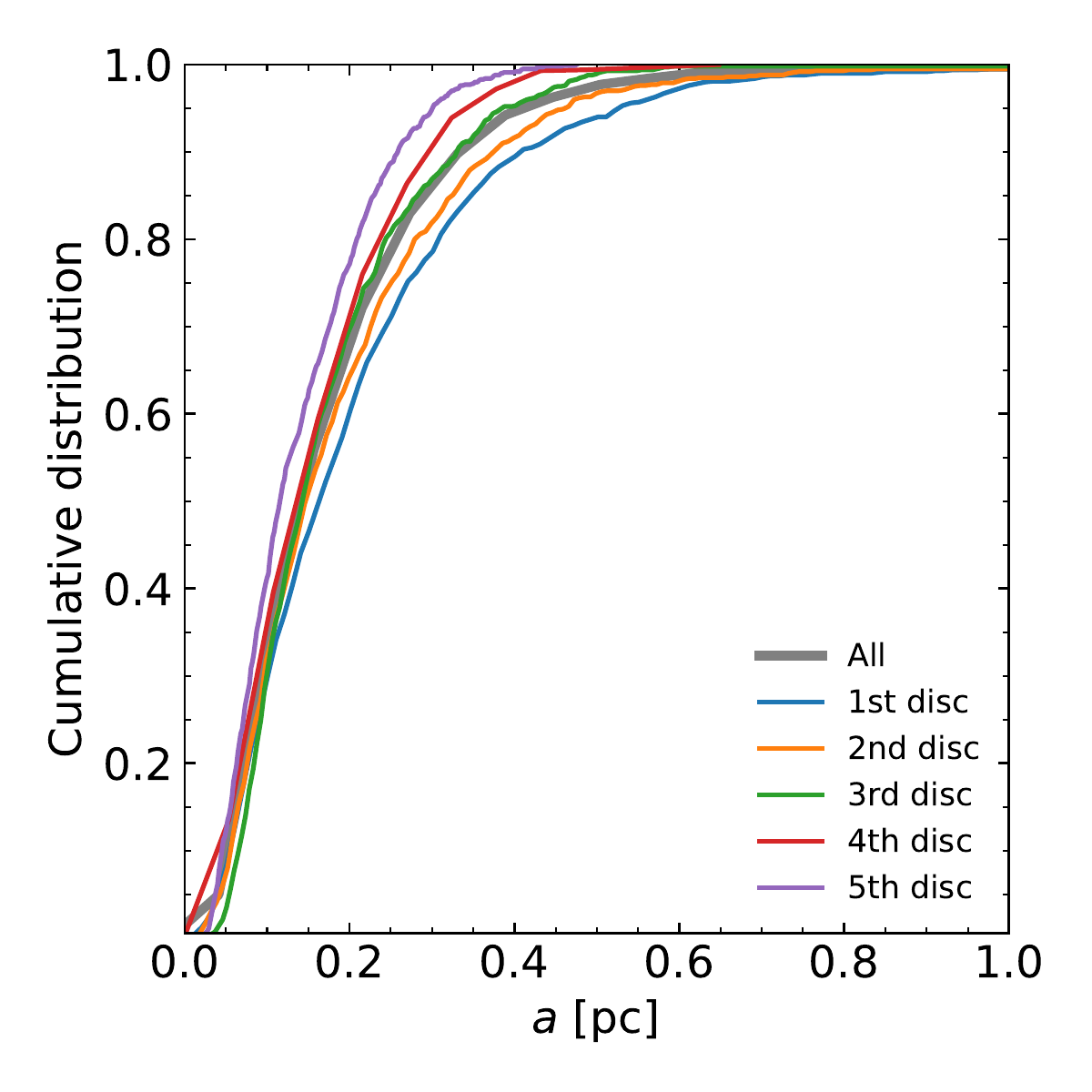}
\includegraphics[width=0.4\textwidth]{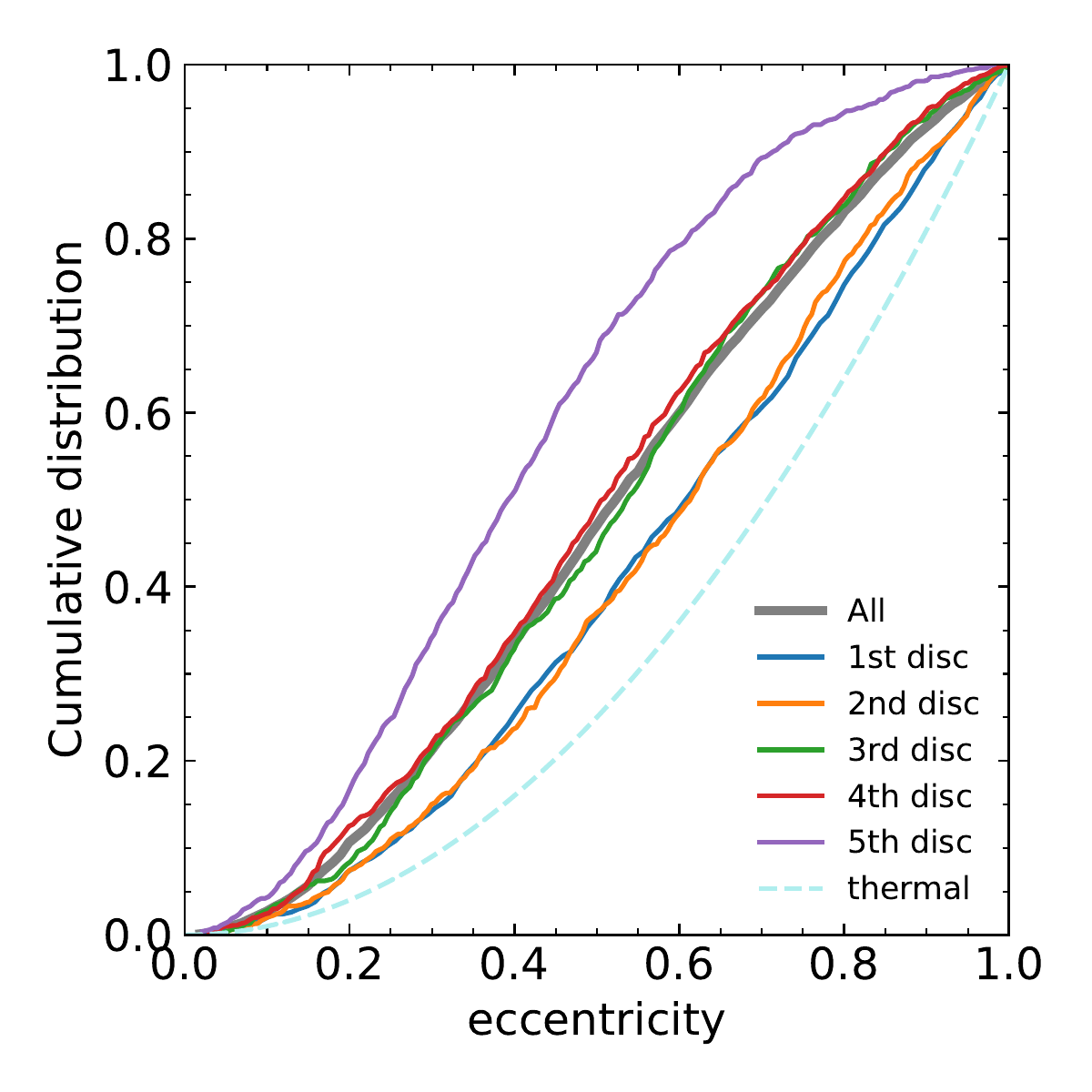}
\includegraphics[width=0.4\textwidth]{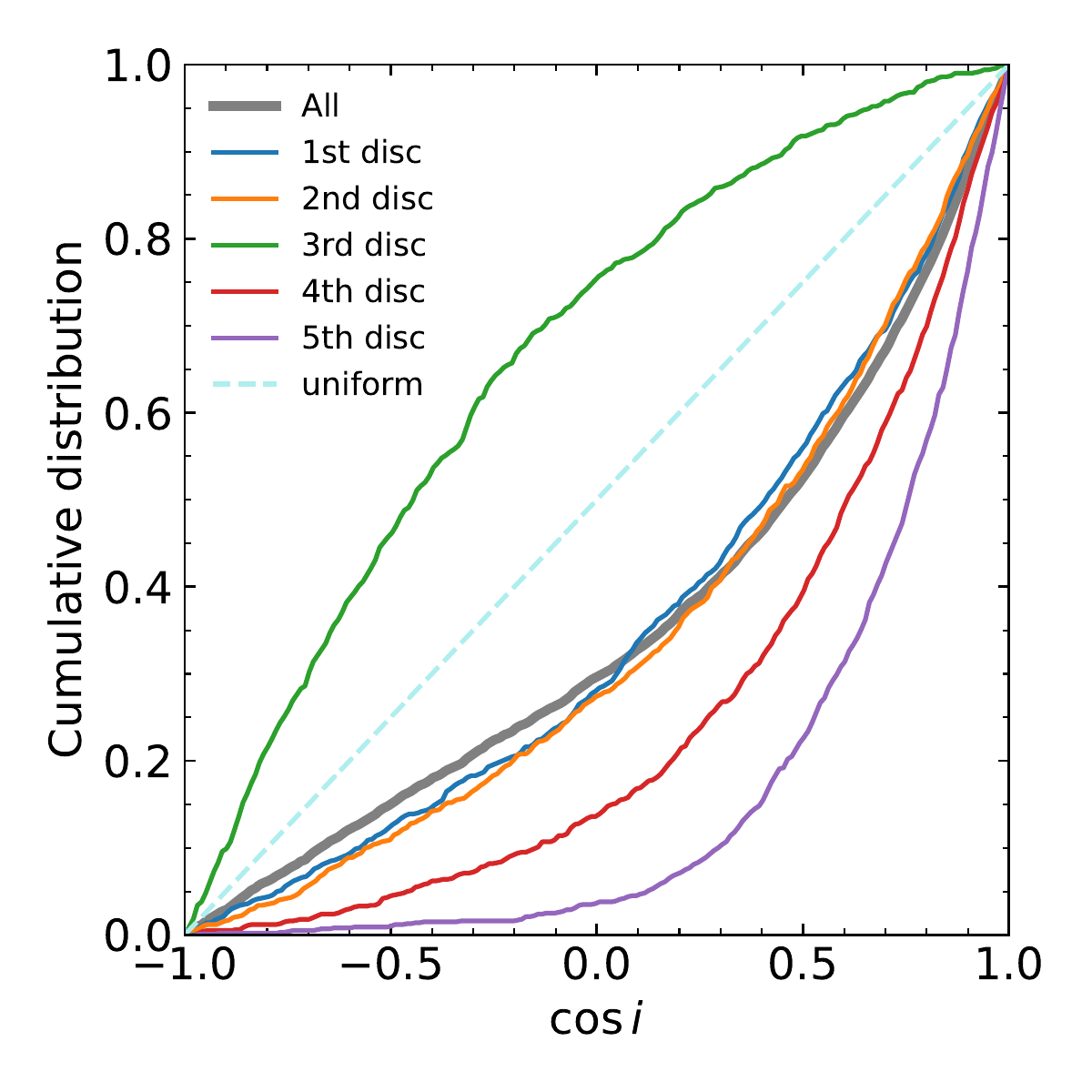}
\caption{Final cumulative distribution of the orbital parameters for the stars belonging to each disc and for the whole stellar component { in the first realisation}. The semi-major axis distribution is shown in the top panel, the eccentricity distribution in the middle panel and the inclination distribution in the bottom panel. The green (third disc) and grey (all stars) curves overlap in the top and middle panels.}\label{fig:par_orb_fin}
\end{figure}

\begin{figure}
\centering
\includegraphics[width=0.4\textwidth]{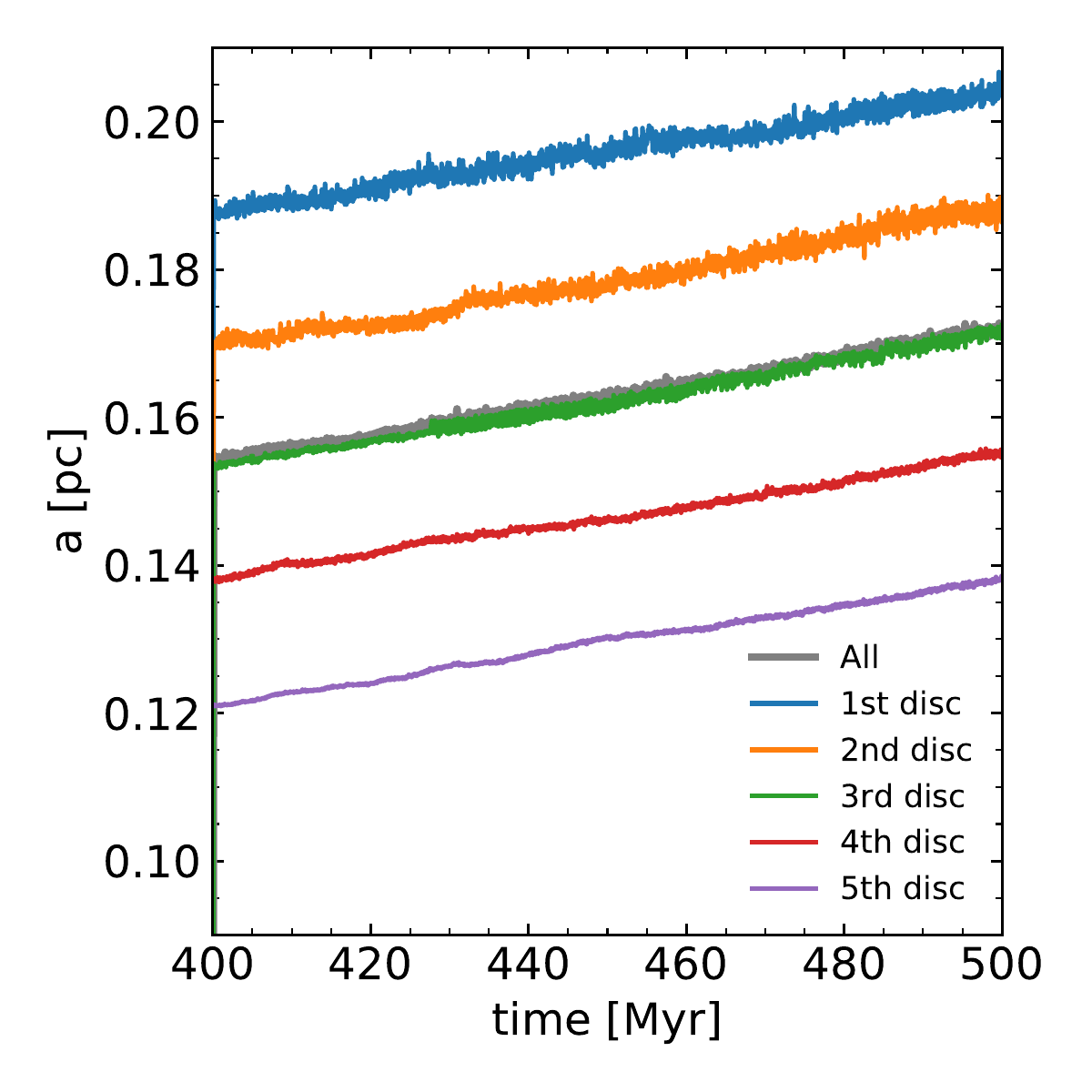}
\includegraphics[width=0.4\textwidth]{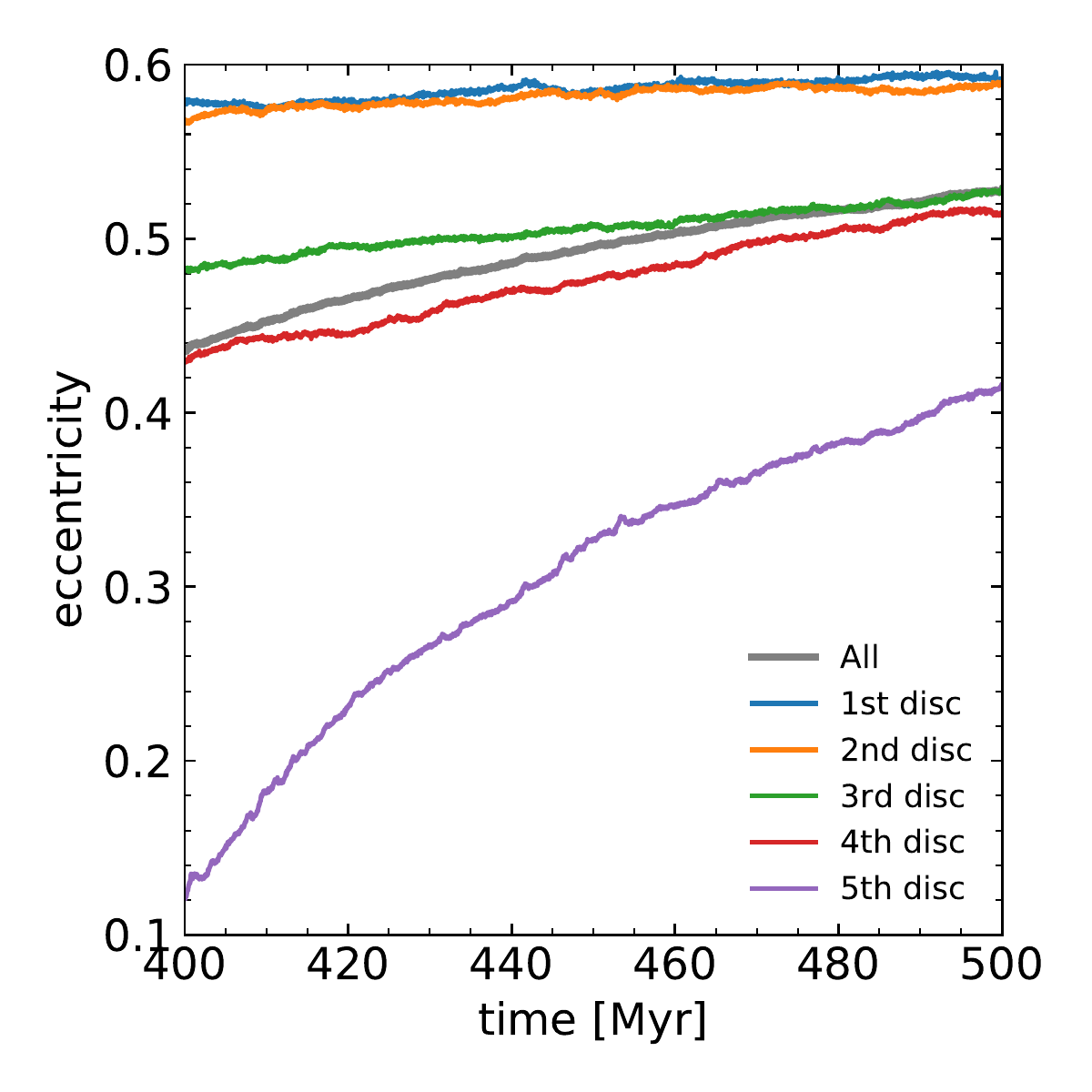}
\includegraphics[width=0.4\textwidth]{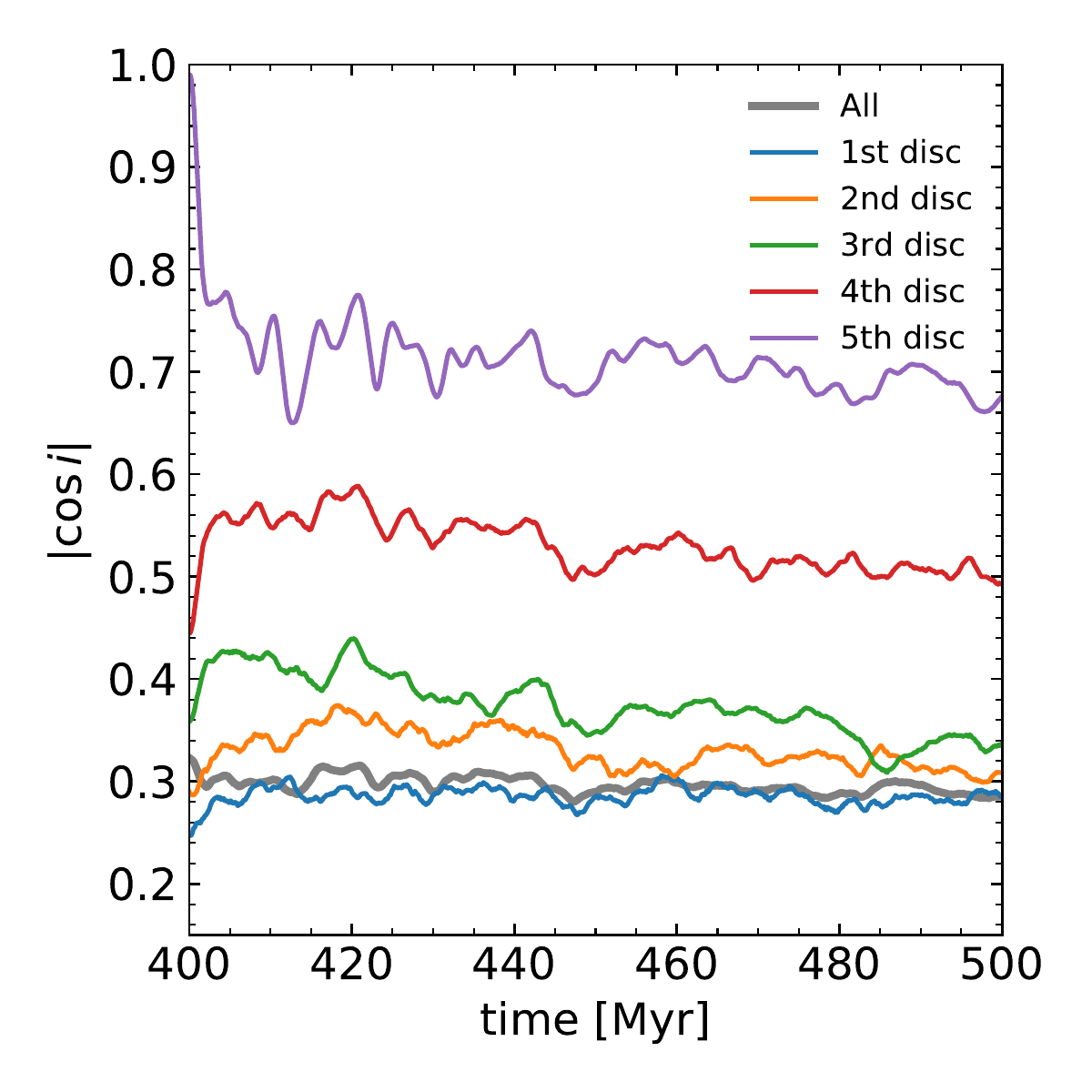}
\caption{Evolution of the orbital parameters for the different populations during the last 100\,Myr of the { first} simulation. The average semi-major axis is shown in the top panel, the average eccentricity in the middle panel and the average inclination in the bottom panel.  The green (third disc) and grey (all stars) curves overlap in the top panel.}\label{fig:par_orb_ev_100}
\end{figure}

\begin{figure*}
\centering
\includegraphics[width=0.45\textwidth]{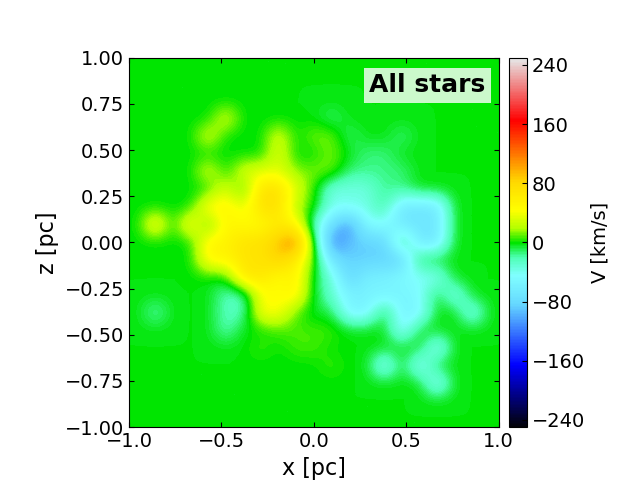}~
\includegraphics[width=0.45\textwidth]{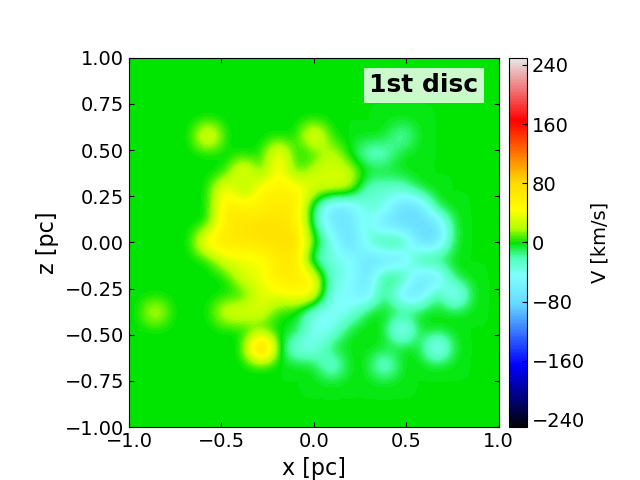}
\includegraphics[width=0.45\textwidth]{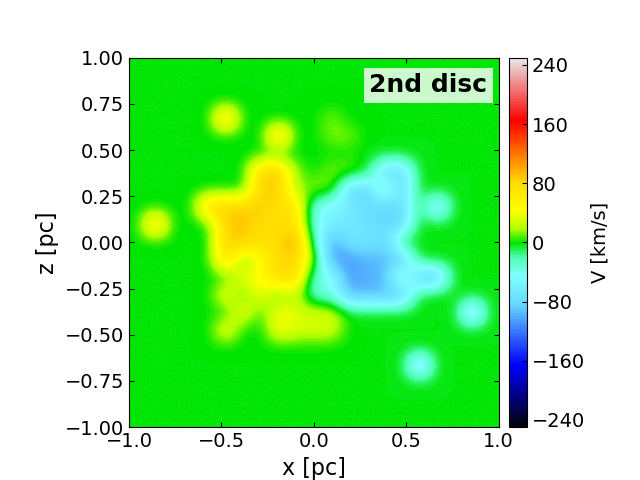}~
\includegraphics[width=0.45\textwidth]{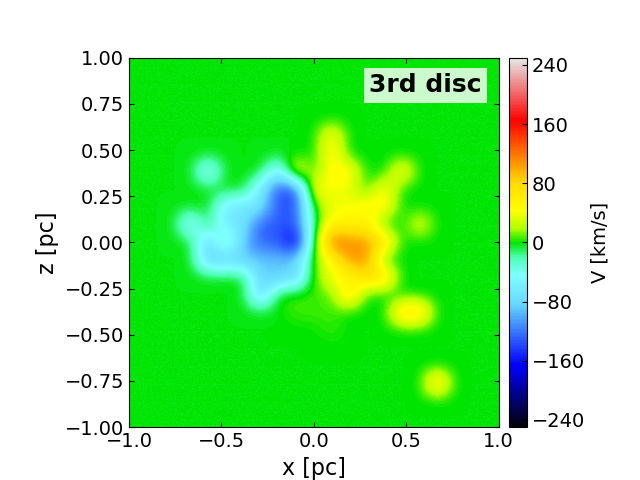}
\includegraphics[width=0.45\textwidth]{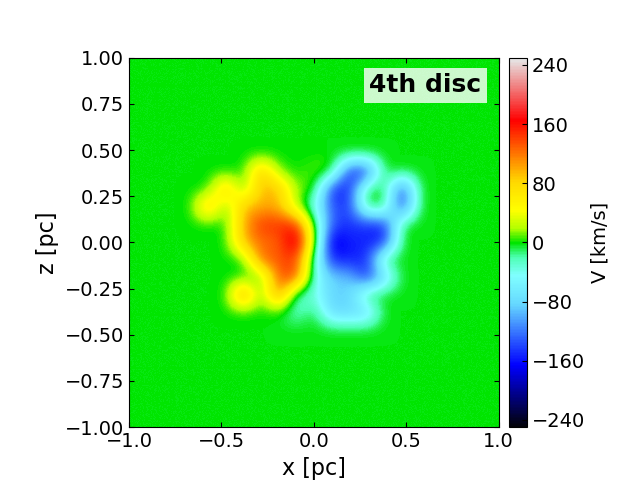}
\includegraphics[width=0.45\textwidth]{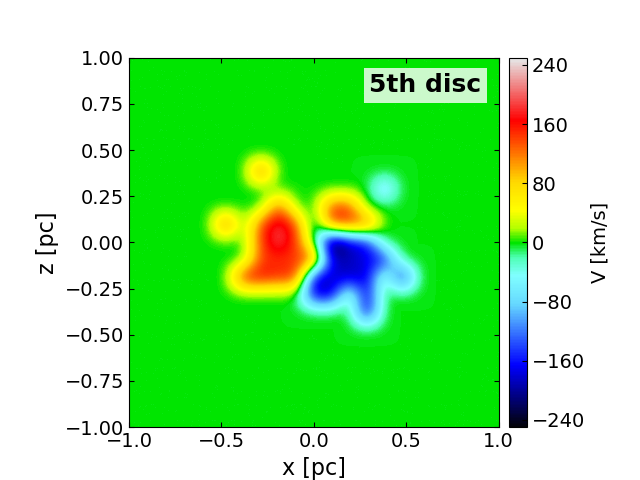}
\caption{Line of sight velocity maps for all the discs considered together and separately { in the first realisation}. The line of sight is perpendicular to the total angular momentum of the discs considered together.}\label{fig:rot}
\end{figure*}
\begin{figure*}
\centering
\includegraphics[width=0.45\textwidth]{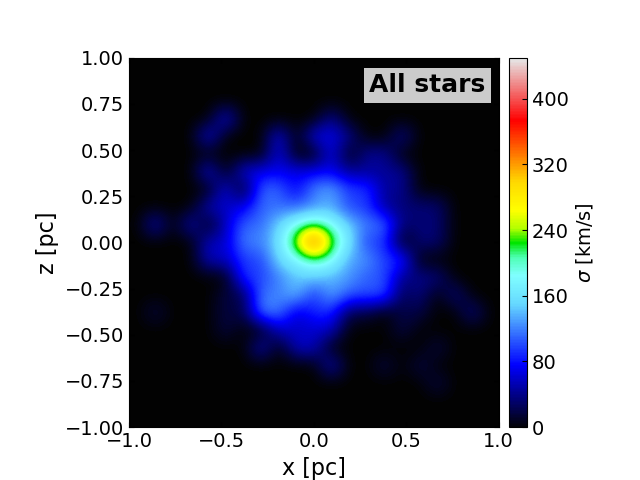}~
\includegraphics[width=0.45\textwidth]{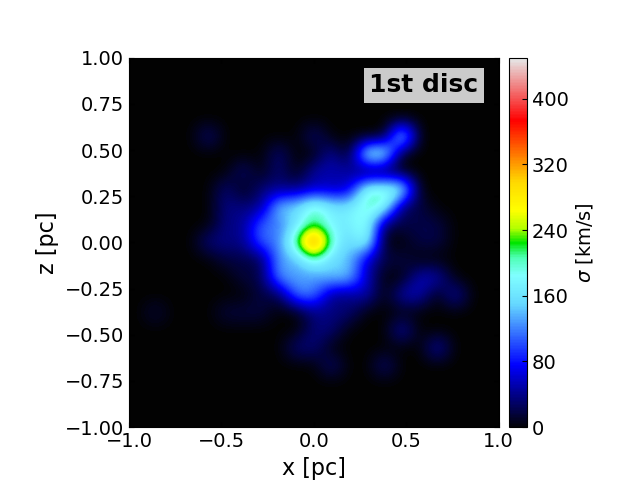}
\includegraphics[width=0.45\textwidth]{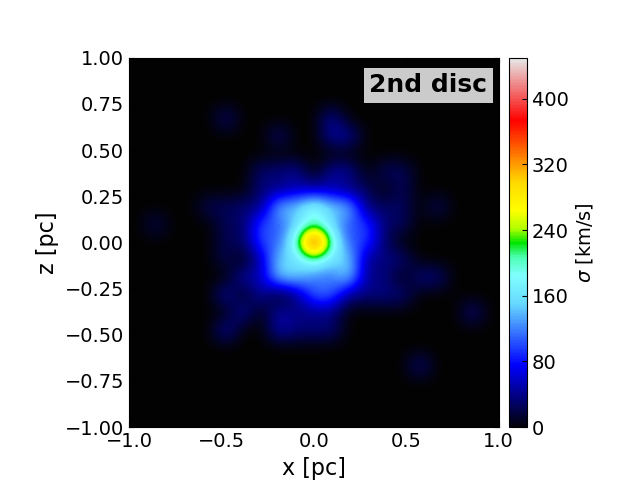}~
\includegraphics[width=0.45\textwidth]{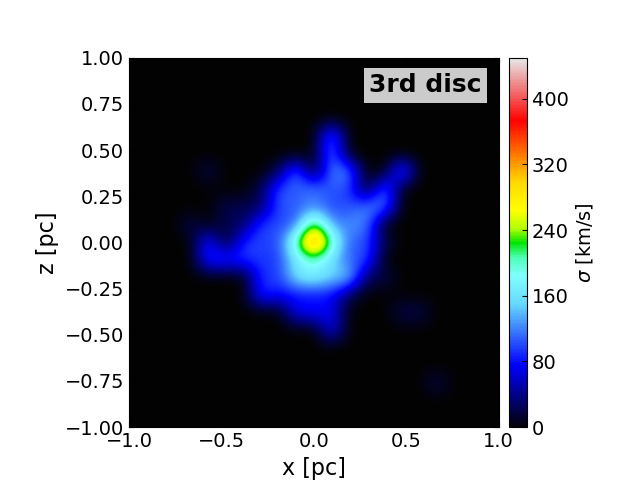}
\includegraphics[width=0.45\textwidth]{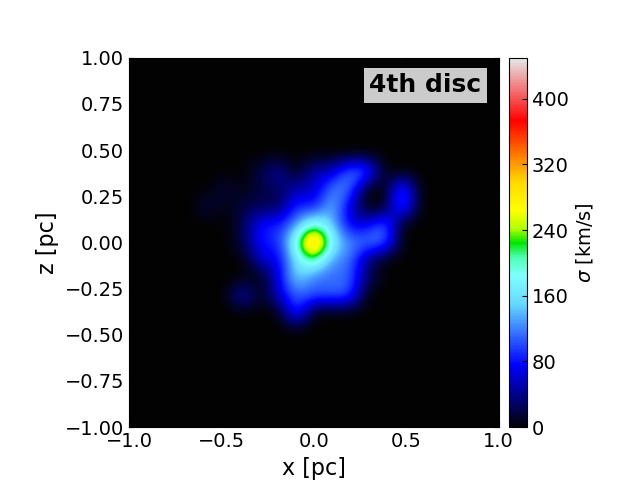}
\includegraphics[width=0.45\textwidth]{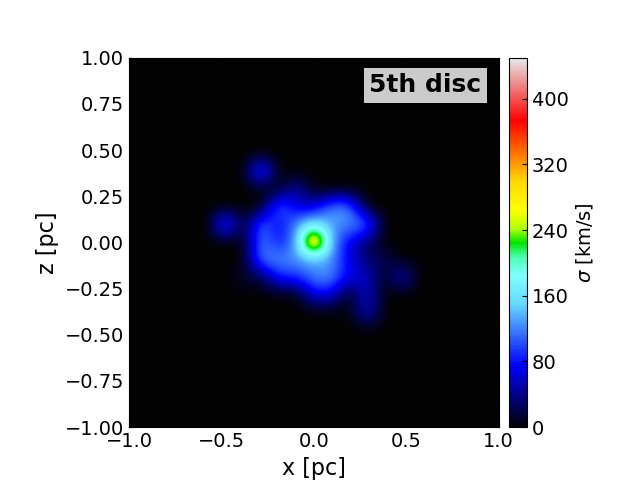}
\caption{Velocity dispersion maps for all the discs considered together and separately { in the first realisation}. The line of sight is perpendicular to the total angular momentum of the discs considered together.}\label{fig:sigma}
\end{figure*}

\begin{table}
\centering
\begin{tabular}{lccccc}
\hline 
Component & $M$ ($M_\odot$) &  $N$ & $\gamma$ & $r_{\rm min}$ (pc) & $r_{\rm max}$ (pc)
 \tabularnewline
\hline 
NSC & $4.3\times10^6$ & $-$  &$-1.75$ & 0  & $3$ \tabularnewline
SMBH & $4.3\times10^6$ & $1$ &  $-$ & $-$ & $-$ \tabularnewline
SBHs & $1.6\times10^5$ & $3200$ & $-2$ & $0.03$ & $0.8$ \tabularnewline
Each disc & $9\times10^3$ & 1000 & $-2$ & 0.03 & 0.3 \tabularnewline
\hline 
\end{tabular}
\caption{Summary of the parameters used for the $N$-body realisation of each component considered in the simulations. $M$ is the total mass, $N$ is the number of particles adopted to model the component, $\gamma$ is the slope of the cusp or of the surface density profile in the case of the discs, $r_{\rm min}$ and $r_{\rm max}$ are the internal and external limiting radii of the component. The old stellar NSC is modelled as a smooth analytic potential.}\label{tab1}
\end{table}
\begin{table}
\centering
\begin{tabular}{ccccc}
\hline 
 & \multicolumn{2}{c}{Simulation 1} & \multicolumn{2}{c}{Simulation 2}\tabularnewline
\#ID & $i \,(\deg)$ & $\Omega\,(\deg)$ & $i \,(\deg)$ & $\Omega \,(\deg)$\tabularnewline
\hline 
d1 & 0 & 0 & 0 & 0 \tabularnewline
d2 & $4.8$ & $196$ & $103$ & $108$ \tabularnewline
d3 & $173$ & $17$  & 66  & $200$ \tabularnewline
d4 & $82$ & $114$ &  $40$ & $123$ \tabularnewline
d5 & $2.9$ & $17$ & $26$  &$40$ \tabularnewline
\hline 
\end{tabular}
\caption{Initial orbital parameters for the five simulated discs in both realisations. The longitude
of the ascending node, $\Omega$, and the inclination, $i$, 
calculated with respect to the simulation reference frame.}\label{tab2}
\end{table}

\subsection{Initial conditions: discs, stellar black holes and stellar cusp}
The presence of the old stellar component of the NSC is taken into account through an analytic potential representing a cusp with slope 
$\gamma=-1.75$, i.e. the slope of the \cite{BW76} cusp, extending between $0$ and $3$ pc (similar to the influence radius of Sgr A$^*$). The central SMBH has a mass of $4.3\times10^6\,M_\odot$ and the cusp is as massive as the central SMBH. 
Stellar evolution and dynamics arguments predict the presence of $\sim 20\,000$ SBHs at the GC \citep{MG00,MO93, FR06}; 
the existence of these objects has been recently confirmed by \cite{HA18} through observations of X-ray binaries within $1$ pc from Sgr A$^*$.
Most of the SBHs, remnants of the evolution of massive ($\geq 20$\,M$\sun$) stars, have a mass of $\sim 7$\,M$_\odot$ (\citealt{BA98}; see \citealt{AP16} for a discussion on the presence and influence of more massive SBHs).
We modelled this component, whose effect is crucial for the dynamical evolution of the stars, assuming a population of $1.6\times 10^4$  SBHs with mass of $10$\,$M_\odot$  distributed isotropically between $0.03$ and $0.8$ pc  according to a density profile $\propto r^{-2}$ \citep[see][]{FR06, HA06, AT09, PAS10, AP16, PMB18}.
The young stellar discs, which represent the stellar populations forming at the GC, extend between $0.03$ pc and $0.3$ pc from the central SMBH and have a mass of  
$\sim9\times10^3$ M$_\odot$ each \citep[][]{GE10}. Their surface density profile is described by a power-law with index $\gamma=-2$ as found in observations \citep[see][for the structural properties of the young stellar disc observed at the GC]{GE10}. The stars in the discs are initially on circular orbits around the SMBH.
The discs are initially thin as suggested by \cite{NA07} and their scale height is set to be 5\% of their radial extension ($0.015$ pc). { Stars in each disc are initially on circular orbits around the SMBH, which dominates the potential within its radius of influence \citep[$\sim 2-3$\,pc,][]{MD10}. The thickness of each disc is generated by adding a 10\% randomly oriented perturbation to the velocities of the stars.}\\
For computational reasons, and in particular to speed up the simulation,
we sampled each disc using $1000$ single mass stars with $\sim9$ M$_\odot$ each. This corresponds to five times the average stellar mass of a population with masses distributed according to a \citep{SA55} initial mass function (IMF) sampled between $0.6$ M$_\odot$ and $60$ M$_\odot$.  { We adopted a Salpeter mass function for simplicity, because of the uncertainties on the IMF at the GC.}
To have the same mass ratio between the simulated and real system both in the SBH cusp and in the disc, we simulated $3200$ SBHs of $50$\,M$_\odot$, multiplying their mass by a factor $5$, as done for the stars. {With the adopted number of particles each 100\,Myr of evolution required a wall-clock time of up to 4 weeks\footnote{The particle by particle simulation run by \cite{PMB18} required ~4 weeks to follow 6Myr of evolution with 21000 particles.}.} The parameters used for each component are summarised in Table \ref{tab1}. \\
To readjust the simulation time to the real evolutionary time, we rescaled the times using the ratio between the relaxation times of the real and simulated systems as done in \cite{MBP13}. 
In particular, we calculated 
\begin{equation}
\frac{t_{\rm sim}}{t_{\rm real}}=\frac{n_{\rm real}}{n_{\rm sim}}\left(\frac{m_{\rm sim}}{m_{\rm real}}\right)^2=\frac{1}{5}
\end{equation} 
where $n_{\rm real}=5n_{\rm sim}$ and $n_{\rm real}$,  $m_{\rm real}$ and $n_{\rm sim}$, $m_{\rm sim}$
are the number densities and particle masses in the real and simulated system.\\
We simulated each disc after the previous one has evolved for $100$\,Myr, hence the total simulation time is $500$\,Myr. The inclination of each disc is chosen by randomly picking the direction of its total angular momentum.
We ran two simulations with similar initial set-ups but different random realisations of the initial conditions for the discs (see Table \ref{tab2} and Figure \ref{fig:snaps} for the disc initial orbital parameters and for the snapshots of the initial conditions for each disc in both realisations). The third disc in the first realisation and the second disc in the second realisation are retrograde.
Given the similarity between the results of the two simulations, which supports the robustness of our approach, in the next sections
we mostly focus on the first realisation.

\section{Results}\label{sec:res}
As shown in the next sections, the evolution of the discs  leaves  long-lasting morphological and kinematical signatures
that, if observed, can shed light on the star formation processes happening at the GC. 
\subsection{Morphology and axial ratios}
Figure \ref{fig:maps} shows the density maps in the face-on (top panel) and edge-on (bottom two panels) projections of the final stellar system.
The stellar density peaks at the GC and quickly drops going outwards. After 500\,Myr of evolution, the stars are still in a significantly flattened configuration, however they have diffused beyond 0.3\,pc which was their initial maximum distance from the SMBH. 
The stellar density profile evolves from a $\propto r^{-2}$ power law to a shallow cusp in the central region, followed by an external steep decline of the density (see Fig. \ref{fig:density}). 
The SBHs diffuse inwards as well as outwards, and after $500$\,Myr, their density profile follows a power law with index $\sim -1.5$ at radii smaller than $0.03$\,pc (i.e. within their initial minimum distance from the centre, see grey dashed line in Fig. \ref{fig:density}). { This is close to the slope ($-1.75$) of a stable \cite{BW76} equilibrium cusp towards which the SBHs are gradually evolving while diffusing. } The SBH density profile shows a core between $0.03$ and $0.05$\,pc while, at radii larger than $0.05$\,pc, it is comparable to the initial profile. 
The density profile with slope $-2$ is stable in the intermediate regions, as predicted by dynamical studies \citep{FR06, HA06, AT09, PAS10}. However, we expect the presence of SBHs both inside 0.03\,pc and outside 0.08\,pc. Tidal disruption events and gravitational wave emission events due to SBH mergers are expected in a wide radial range around the SMBH.\\
Figure \ref{fig:cumulative_mass} shows the radial cumulative mass fraction for all the stars and for each of the five discs considered separately.
While the stars in the fifth disc are still mostly distributed within their initial radius, with the majority of the stellar mass still enclosed within $0.3$\,pc, the stars belonging
to more evolved discs have diffused to larger distances from the SMBH. A significant fraction (20\%) of the stars initially belonging to the first disc are found beyond a radial distance of $0.3$\,pc.
The three-dimensional and projected density profiles of the five discs slightly differ in the central and intermediate regions ($<0.1$\,pc, see Figure \ref{fig:density_sep}). Hence, we expect a larger density of old stars  beyond 0.3\,pc. Interestingly, the third disc, which in the first realisation is counterrotating with respect to the other discs (see Table \ref{tab1}), has the lowest projected density profile and shows a flat central core, suggesting that its stars were scattered at larger radii more efficiently than stars belonging to prograde discs. 
Figure \ref{fig:ax_ratios} shows the axial ratios for the discs considered together and separately { in the case of the first (top panel) and second (bottom panel) realisation. In both realisations, the intermediate axial
ratio, $b/a$, for the whole stellar component -- and for each disc -- is larger than $0.9$ at radii smaller than $0.1$\,pc, and approximately $1.0$ at larger radii.
The overall minor axial ratio, $c/a$, is, at any radius, smaller than 0.8 for the first realisation and smaller than 0.7 for the second one. The first and second discs in the second realisation are significantly less flattened than in the first realisation. This is due to the effect of the second counterrotating disc in the second realisation (see Sect. \ref{sec:coev} for the details). In both simulations, the systems are oblate and the flattening decreases with increasing distance from the centre. Interestingly, stars escaped beyond $0.3$\,pc show a significant degree of flattening, depending on the epoch of formation.}
As expected, the first disc is the most relaxed one, but it is still considerably flattened with $c/a$ ranging between 0.6 and 0.75.

The fifth disc is still significantly flattened, with $c/a\approx 0.3$ in the most central regions and $c/a\approx 0.5$ in the outskirts of the disc. 
The second disc which, in the first realisation, is followed by the formation of the third counterrotating disc (see Table \ref{tab1}) is slightly more spherical than the first disc, suggesting that counterrotating discs have a significant dynamical impact on the dynamical evolution of stellar populations (see Section \ref{sec:coev}).
\subsection{The dynamical impact of young populations on old ones}\label{sec:coev}
To assess if and how the birth of a new stellar disc alters the structure of older stellar populations we analysed the first 150\,Myr of evolution of the first disc in both our realisations. 
Figure \ref{fig:first_disc_second} shows the cumulative distribution of the semi-major axes, eccentricities and inclinations at 100\,Myr, i.e. immediately before the birth of the second disc, and after additional 20\,Myr of co-evolution.
In the first realisation, the second disc is prograde with respect to the first one and its presence does not affect the evolution of the orbital parameters of the stars belonging to the first disc. In the second realisation, where the second disc is counterrotating with respect to the previous population, the evolution of the semi-major axes and eccentricities of the stars in the first disc is slightly affected by the formation of the second disc.
The counterrotation of the second disc mostly affects the distribution of inclinations of the stars belonging to the first disc, i.e. their spatial distribution. As seen in the right panel of Figure \ref{fig:first_disc_second}, the first disc becomes immediately more uniform due to the interaction with the second disc. Figures \ref{fig:fmaps_fo} and \ref{fig:fmaps_eo} show the face-on and edge-on evolution of the first disc in the second realisation, plotted at intervals of 15\,Myr from 0 to 150\,Myr. While the face-on disc does not modify its structure significantly, the edge-on view shows that after 45\,Myr of evolution, the disc starts warping and, as soon as the second disc appears, its vertical size increases and the disc becomes quickly more spherical as suggested by the distribution of the inclinations. 
In conclusion, while prograde discs do not affect each other's evolution significantly, retrograde discs  sensibly modify  the evolution 
of older populations, accelerating their relaxation toward a more spherical distribution.
\subsection{Distribution and evolution of the orbital parameters}
We calculated the cumulative distribution of the final orbital parameters considering the five discs separately and all together.
The top panel of Figure \ref{fig:par_orb_fin} shows the cumulative distributions of the semi-major axis. As already seen in
Figure \ref{fig:cumulative_mass}, 
stars in the first two discs have significantly diffused both towards the GC and 
beyond $0.3$\,pc. Few percent of the stars are found in a more extended and isotropic configuration (see Figure \ref{fig:ax_ratios}).
Consistent with observations of young stars, which are preferentially found closer to the GC \citep[see e.g.][]{FN15}, stars belonging to younger populations are more centrally concentrated than older stars.
In addition, the stars in the fifth disc are distributed on lower eccentricity orbits compared to earlier discs (see middle panel of Figure \ref{fig:par_orb_fin}). After $500$\,Myr of evolution, the first and second discs have approximately settled on a thermal distribution for the eccentricities;  younger discs are farther from reaching this point than older stars. 
In the case of the first realisation all the discs except the third one are co-rotating (see bottom panel of Figure \ref{fig:par_orb_fin} and Table \ref{tab1}) and the discs relax approaching an isotropic distribution of the inclinations. Stars in the first and second discs are significantly closer to a uniform spatial distribution than the ones belonging to the remaining discs.  
Figure \ref{fig:par_orb_ev_100} shows the evolution of the average semi-major axis (top panel), eccentricity (middle panel) and inclination ($|\cos i|$, bottom panel) for each disc and for all the stars taken together during the last $100$\,Myr of evolution. While the first disc increased its radius significantly, the  other discs are still centrally concentrated. { The linear evolution of the semi-major axes, with each new disc reaching the same size that the previous one had 100\,Myr earlier, suggests that every disc relaxes and diffuses its stars at a similar rate as well as that their mutual gravitational interactions are negligible}. The fifth disc is the least relaxed one and shows the smallest average semi-major axis for its stars. The average semi-major axis increases at the same rate for all the discs. The first and second discs are the closest to a thermal distribution of the eccentricities and they show the largest average orbital eccentricity. The stars belonging to the fifth disc are on more circular orbits than the stars belonging to older discs and show the lowest average orbital eccentricity.
While the first four discs have significantly decreased their initial average $|\cos i|$ moving toward a more uniform distribution, the fifth disc is still significantly flattened and shows the smallest average orbital inclination (i.e. largest flattening).
\subsection{Kinematics of the discs}
The discs, which are initially rotationally supported, are expected to relax becoming more spherical, losing part of their angular momentum and redistributing it among the SBHs. 
Depending on the time-scale of this latter process, a signature of the initial rotational velocity could still be observable today in the different populations.
Figure \ref{fig:rot} shows the line of sight velocity maps for all the discs considered together and separately,  observed perpendicular to
the total angular momentum of the system. The maps are obtained  by applying the Voronoi binning procedure described by \cite{Cap03} with a fixed S/N ratio of 5 in each bin\footnote{We assume Poissonian noise; the signal to noise is $\sqrt{N}$, where $N$ is the number of particles in the bin.}.
The whole stellar system and each disc show different maximum rotation velocities.
Since they are less relaxed, the stars in the fifth disc rotate faster than the stars in the older discs and are  more spatially concentrated. 
The velocity dispersion peaks at the centre and it is higher for the earlier discs (see Figure
\ref{fig:sigma}). 
The clumps and irregularities in the line of sight velocity and velocity dispersion observed at radii larger than $0.5$\,pc both for the single discs  and for the full stellar component are due to stars diffusing farther from the GC. 
The angular momentum lost by the relaxing discs is acquired by the SBHs. However, no regular rotation is observed in this component. 

\section{Discussion and summary}\label{sec:con}
We have explored the co-evolution of five young stellar populations born consecutively in discs at the GC,
where a population of stellar black holes is expected to reside in a cusp \citep{ME00, FR06, AT09, HA18}.
Observations of the GC suggest that the nucleus of the Milky Way experienced at least one recent star formation episode that produced the currently observed young stellar disc orbiting the central $0.5$\,pc of the Galaxy \citep{LE03, PA06, LU09, BA09, GE10, PF11, FN15}. In-situ star formation can play an important role in the build-up of  NSCs \citep{LKT82,AH15,BA18}. Gas infalling from the circumnuclear regions can produce a gaseous disc that subsequently fragments and forms a new stellar population showing a disc-like configuration. We show that the co-evolution of multiple stellar populations gives rise to  dynamical and morphological signatures that, if observed, could trace the star formation processes that generated the discs.\\
In this work we modelled equally time-spaced star formation episodes in stellar discs,  constraints from observations \citep{HN09}. Each of the modelled discs has a mass and initial distribution of the orbital parameters drawn randomly from the same function. However, they are randomly inclined with respect to each other.
We simulated the discs for a total time of 500\,Myr using the direct $N$-body code phiGRAPE. We performed two different realisations of the same simulation using two sets of randomly picked initial conditions. Both realisations provide comparable results and in this paper we mainly focused primarily on the first realisation.

The discs are born at intervals of 100\,Myr and they slowly evolve towards spatial and kinematical isotropy. As illustrated by \cite{PMB18}, the evolution of a stellar disc is strongly affected by the temporal asphericity of the NSC, in this case introduced by the presence of a discrete SBH cusp.  { The same disc evolved in a smooth NSC potential, without SBHs, keeps its initial configuration over long time-scales. The presence of massive O-stars, which are not taken into consideration in our simulations, could have affected the initial evolutionary phases of the discs. However, disentangling their effect from that of the more numerous and more anisotropically distributed SBHs would be challenging.} 
The density profile of the final stellar component is significantly different from the initial disc profile. It shows a central shallow cusp followed by a sharp decline and stars are present both inside and outside the initial radial limits of the discs. The SBH spatial distribution does not change significantly over $500$\,Myr and a small fraction of SBHs diffuse inwards and outwards. 

At the end of the simulation (i.e. after a $500$\,Myr of evolution), only the last disc is still considerably flattened ($c/a\approx0.3$) while the previous discs have had time to relax and achieve flattening ratios between $0.25$ and $0.4$. The second disc, which in the first realisation is followed by a counterrotating disc, is less flattened than the first disc. A significant fraction of stars diffuse beyond the initial radial limit, and a steady growth of the average semi-major axis is observed for all the discs. 
We notice that, in our second realisation, the relaxation of the first disc is accelerated by the formation of the following counterrotating disc, which quickly randomises the inclinations of the stars belonging to the older population. In this realisation, the first and second discs are indeed less flattened than in the first realisation.

The five stellar populations, after $500$\,Myr, have distinct orbital parameter distributions. Younger discs are still significantly more centrally concentrated and flattened than the older ones. We expect a growing fraction of old stars moving at larger distances from the GC. 
The first and second discs have comparable cumulative eccentricity distributions, both close to a thermal configuration.
The birth of the third counterrotating disc creates a gap between the evolution of the first two and the last two discs. 
 The stars in the fifth disc move on more circular orbits than the stars in the other discs. The inclination distributions of the first and second discs are almost uniform after 500\,Myr of evolution, while the other discs are still far from becoming isotropic. The fifth disc is the only one still showing low inclinations at the end of the simulation. We clearly observe age segregation in the outskirts of the system. The discs rotate at different speeds, with the older discs rotating slower than the younger ones. Clumps and irregularities due to diffusing stars are observed in the velocity dispersion maps outside the central $0.3$\,pc. These overdensities could help to identify different kinematical populations in a less crowded area. The SBHs acquire the angular momentum lost by the discs, however they do not show any significant coherent rotation signal.

Due to the complexity of the problem presented in this work, i.e. the need to evolve the analytic potential of the GC in combination with the population of SBHs as well as the individual discs, we have purposely refrained from introducing stellar evolution as an additional ingredient. This is crucial to be able to disentangle the effects of the SBHs and the discs onto each other as a first step. We note, that the lifetime of the massive stars in our simulated discs can be shorter than the duration of our simulations. For this reason, a more detailed analysis of this problem including stellar evolution is foreseen in the near future. However, given the steep nature of the Salpeter IMF, only a small fraction of the stars would have been affected by severe mass loss within the first $500$\,Myr of evolution.

Our results suggest that star formation in multiple discs gives rise to a complex dynamical evolution, with the discs mutually affecting each other. { The relative inclination of the discs is an important parameter} determining the specific type of evolution. Generally, the introduction of new counterrotating discs can perturb the dynamics of older stellar populations leading to accelerated relaxation. We find that the long-term evolution of the discs leaves behind unique observable morphologic and kinematic signatures whose strength depends on the age of the stellar population. These signatures, which include flattening, rotation and spatial distribution, might be observable in our own GC and other galactic nuclei, proving important clues on the assembly histories of NSCs.

\section*{Acknowledgements}
We thank the referee for the helpful comments.
AMB, NN and AS acknowledge support by Sonderforschungsbereich
(SFB) 881 ``The Milky Way System'' of the German Research Foundation
(DFG). 
The simulations have been carried out on the Tamnun (Technion, Israel) and on the BwFor (Computing Center of Heidelberg University, Germany) clusters.




\bibliographystyle{mnras}
\bibliography{mdnsc} 

\begin{thebibliography}{}
\makeatletter
\relax
\def\mn@urlcharsother{\let\do\@makeother \do\$\do\&\do\#\do\^\do\_\do\%\do\~}
\def\mn@doi{\begingroup\mn@urlcharsother \@ifnextchar [ {\mn@doi@}
  {\mn@doi@[]}}
\def\mn@doi@[#1]#2{\def\@tempa{#1}\ifx\@tempa\@empty \href
  {http://dx.doi.org/#2} {doi:#2}\else \href {http://dx.doi.org/#2} {#1}\fi
  \endgroup}
\def\mn@eprint#1#2{\mn@eprint@#1:#2::\@nil}
\def\mn@eprint@arXiv#1{\href {http://arxiv.org/abs/#1} {{\tt arXiv:#1}}}
\def\mn@eprint@dblp#1{\href {http://dblp.uni-trier.de/rec/bibtex/#1.xml}
  {dblp:#1}}
\def\mn@eprint@#1:#2:#3:#4\@nil{\def\@tempa {#1}\def\@tempb {#2}\def\@tempc
  {#3}\ifx \@tempc \@empty \let \@tempc \@tempb \let \@tempb \@tempa \fi \ifx
  \@tempb \@empty \def\@tempb {arXiv}\fi \@ifundefined
  {mn@eprint@\@tempb}{\@tempb:\@tempc}{\expandafter \expandafter \csname
  mn@eprint@\@tempb\endcsname \expandafter{\@tempc}}}

\bibitem[\protect\citeauthoryear{{Abbate}, {Mastrobuono-Battisti}, {Colpi},
  {Possenti}, {Sippel}  \& {Dotti}}{{Abbate} et~al.}{2018}]{AMB18}
{Abbate} F.,  {Mastrobuono-Battisti} A.,  {Colpi} M.,  {Possenti} A.,  {Sippel}
  A.~C.,   {Dotti} M.,  2018, \mn@doi [\mnras] {10.1093/mnras/stx2364}, \href
  {http://adsabs.harvard.edu/abs/2018MNRAS.473..927A} {473, 927}

\bibitem[\protect\citeauthoryear{{Aharon} \& {Perets}}{{Aharon} \&
  {Perets}}{2015}]{AH15}
{Aharon} D.,  {Perets} H.~B.,  2015, \mn@doi [\apj]
  {10.1088/0004-637X/799/2/185}, \href
  {http://adsabs.harvard.edu/abs/2015ApJ...799..185A} {799, 185}

\bibitem[\protect\citeauthoryear{{Aharon} \& {Perets}}{{Aharon} \&
  {Perets}}{2016}]{AP16}
{Aharon} D.,  {Perets} H.~B.,  2016, \mn@doi [\apjl]
  {10.3847/2041-8205/830/1/L1}, \href
  {http://adsabs.harvard.edu/abs/2016ApJ...830L...1A} {830, L1}

\bibitem[\protect\citeauthoryear{{Alexander} \& {Hopman}}{{Alexander} \&
  {Hopman}}{2009}]{AT09}
{Alexander} T.,  {Hopman} C.,  2009, \mn@doi [\apj]
  {10.1088/0004-637X/697/2/1861}, \href
  {http://adsabs.harvard.edu/abs/2009ApJ...697.1861A} {697, 1861}

\bibitem[\protect\citeauthoryear{{Antonini}, {Capuzzo-Dolcetta},
  {Mastrobuono-Battisti}  \& {Merritt}}{{Antonini} et~al.}{2012}]{AN12}
{Antonini} F.,  {Capuzzo-Dolcetta} R.,  {Mastrobuono-Battisti} A.,   {Merritt}
  D.,  2012, \mn@doi [\apj] {10.1088/0004-637X/750/2/111}, \href
  {http://adsabs.harvard.edu/abs/2012ApJ...750..111A} {750, 111}

\bibitem[\protect\citeauthoryear{{Antonini}, {Barausse}  \& {Silk}}{{Antonini}
  et~al.}{2015}]{AN15}
{Antonini} F.,  {Barausse} E.,   {Silk} J.,  2015, \mn@doi [\apj]
  {10.1088/0004-637X/812/1/72}, \href
  {http://adsabs.harvard.edu/abs/2015ApJ...812...72A} {812, 72}

\bibitem[\protect\citeauthoryear{{Arca-Sedda} \&
  {Capuzzo-Dolcetta}}{{Arca-Sedda} \& {Capuzzo-Dolcetta}}{2014}]{AS14}
{Arca-Sedda} M.,  {Capuzzo-Dolcetta} R.,  2014, \mn@doi [\mnras]
  {10.1093/mnras/stu1683}, \href
  {http://adsabs.harvard.edu/abs/2014MNRAS.444.3738A} {444, 3738}

\bibitem[\protect\citeauthoryear{{Arca-Sedda} \&
  {Capuzzo-Dolcetta}}{{Arca-Sedda} \& {Capuzzo-Dolcetta}}{2017a}]{AS17b}
{Arca-Sedda} M.,  {Capuzzo-Dolcetta} R.,  2017a, \mn@doi [\mnras]
  {10.1093/mnras/stw2483}, \href
  {http://adsabs.harvard.edu/abs/2017MNRAS.464.3060A} {464, 3060}

\bibitem[\protect\citeauthoryear{{Arca-Sedda} \&
  {Capuzzo-Dolcetta}}{{Arca-Sedda} \& {Capuzzo-Dolcetta}}{2017b}]{AS17}
{Arca-Sedda} M.,  {Capuzzo-Dolcetta} R.,  2017b, \mn@doi [\mnras]
  {10.1093/mnras/stx1586}, \href
  {http://adsabs.harvard.edu/abs/2017MNRAS.471..478A} {471, 478}

\bibitem[\protect\citeauthoryear{{Arca-Sedda}, {Capuzzo-Dolcetta}, {Antonini}
  \& {Seth}}{{Arca-Sedda} et~al.}{2015}]{AS15}
{Arca-Sedda} M.,  {Capuzzo-Dolcetta} R.,  {Antonini} F.,   {Seth} A.,  2015,
  \mn@doi [\apj] {10.1088/0004-637X/806/2/220}, \href
  {http://adsabs.harvard.edu/abs/2015ApJ...806..220A} {806, 220}

\bibitem[\protect\citeauthoryear{{Bahcall} \& {Wolf}}{{Bahcall} \&
  {Wolf}}{1976}]{BW76}
{Bahcall} J.~N.,  {Wolf} R.~A.,  1976, \mn@doi [\apj] {10.1086/154711}, \href
  {https://ui.adsabs.harvard.edu/abs/1976ApJ...209..214B} {209, 214}

\bibitem[\protect\citeauthoryear{{Bailyn}, {Jain}, {Coppi}  \&
  {Orosz}}{{Bailyn} et~al.}{1998}]{BA98}
{Bailyn} C.~D.,  {Jain} R.~K.,  {Coppi} P.,   {Orosz} J.~A.,  1998, \mn@doi
  [\apj] {10.1086/305614}, \href
  {http://adsabs.harvard.edu/abs/1998ApJ...499..367B} {499, 367}

\bibitem[\protect\citeauthoryear{{Bartko} et~al.,}{{Bartko}
  et~al.}{2009}]{BA09}
{Bartko} H.,  et~al., 2009, \mn@doi [\apj] {10.1088/0004-637X/697/2/1741},
  \href {http://adsabs.harvard.edu/abs/2009ApJ...697.1741B} {697, 1741}

\bibitem[\protect\citeauthoryear{{Bartko} et~al.,}{{Bartko}
  et~al.}{2010}]{BA10}
{Bartko} H.,  et~al., 2010, \mn@doi [\apj] {10.1088/0004-637X/708/1/834}, \href
  {https://ui.adsabs.harvard.edu/abs/2010ApJ...708..834B} {708, 834}

\bibitem[\protect\citeauthoryear{{Baumgardt}, {Amaro-Seoane}  \&
  {Sch{\"o}del}}{{Baumgardt} et~al.}{2018}]{BA18}
{Baumgardt} H.,  {Amaro-Seoane} P.,   {Sch{\"o}del} R.,  2018, \mn@doi [\aap]
  {10.1051/0004-6361/201730462}, \href
  {http://adsabs.harvard.edu/abs/2018A\%26A...609A..28B} {609, A28}

\bibitem[\protect\citeauthoryear{{Boehle} et~al.,}{{Boehle}
  et~al.}{2016}]{BGS16}
{Boehle} A.,  et~al., 2016, \mn@doi [\apj] {10.3847/0004-637X/830/1/17}, \href
  {http://adsabs.harvard.edu/abs/2016ApJ...830...17B} {830, 17}

\bibitem[\protect\citeauthoryear{{Bortolas}, {Gualandris}, {Dotti}, {Spera}  \&
  {Mapelli}}{{Bortolas} et~al.}{2016}]{BG16}
{Bortolas} E.,  {Gualandris} A.,  {Dotti} M.,  {Spera} M.,   {Mapelli} M.,
  2016, \mn@doi [\mnras] {10.1093/mnras/stw1372}, \href
  {http://adsabs.harvard.edu/abs/2016MNRAS.461.1023B} {461, 1023}

\bibitem[\protect\citeauthoryear{{Cappellari} \& {Copin}}{{Cappellari} \&
  {Copin}}{2003}]{Cap03}
{Cappellari} M.,  {Copin} Y.,  2003, \mn@doi [\mnras]
  {10.1046/j.1365-8711.2003.06541.x}, \href
  {http://adsabs.harvard.edu/abs/2003MNRAS.342..345C} {342, 345}

\bibitem[\protect\citeauthoryear{{Capuzzo-Dolcetta}}{{Capuzzo-Dolcetta}}{1993}]{CD93}
{Capuzzo-Dolcetta} R.,  1993, \mn@doi [\apj] {10.1086/173189}, \href
  {http://adsabs.harvard.edu/abs/1993ApJ...415..616C} {415, 616}

\bibitem[\protect\citeauthoryear{{Do}, {Lu}, {Ghez}, {Morris}, {Yelda},
  {Martinez}, {Wright}  \& {Matthews}}{{Do} et~al.}{2013}]{DO13}
{Do} T.,  {Lu} J.~R.,  {Ghez} A.~M.,  {Morris} M.~R.,  {Yelda} S.,  {Martinez}
  G.~D.,  {Wright} S.~A.,   {Matthews} K.,  2013, \mn@doi [\apj]
  {10.1088/0004-637X/764/2/154}, \href
  {https://ui.adsabs.harvard.edu/abs/2013ApJ...764..154D} {764, 154}

\bibitem[\protect\citeauthoryear{{Eisenhauer} et~al.}{{Eisenhauer}
  et~al.}{2005}]{EI05}
{Eisenhauer} F.,  et~al., 2005, \mn@doi [\apj] {10.1086/430667}, \href
  {http://adsabs.harvard.edu/cgi-bin/nph-bib_query?bibcode=2005ApJ...628..246E&db_key=AST}
  {628, 246}

\bibitem[\protect\citeauthoryear{{Feldmeier-Krause} et~al.,}{{Feldmeier-Krause}
  et~al.}{2015}]{FN15}
{Feldmeier-Krause} A.,  et~al., 2015, \mn@doi [\aap]
  {10.1051/0004-6361/201526336}, \href
  {http://adsabs.harvard.edu/abs/2015A\%26A...584A...2F} {584, A2}

\bibitem[\protect\citeauthoryear{{Feldmeier} et~al.,}{{Feldmeier}
  et~al.}{2014}]{FN14}
{Feldmeier} A.,  et~al., 2014, \mn@doi [\aap] {10.1051/0004-6361/201423777},
  \href {http://adsabs.harvard.edu/abs/2014A%26A...570A...2F} {570, A2}

\bibitem[\protect\citeauthoryear{{Figer}, {Rich}, {Kim}, {Morris}  \&
  {Serabyn}}{{Figer} et~al.}{2004}]{FR04}
{Figer} D.~F.,  {Rich} R.~M.,  {Kim} S.~S.,  {Morris} M.,   {Serabyn} E.,
  2004, \mn@doi [\apj] {10.1086/380392}, \href
  {http://adsabs.harvard.edu/abs/2004ApJ...601..319F} {601, 319}

\bibitem[\protect\citeauthoryear{{Freitag}, {Amaro-Seoane}  \&
  {Kalogera}}{{Freitag} et~al.}{2006}]{FR06}
{Freitag} M.,  {Amaro-Seoane} P.,   {Kalogera} V.,  2006, \mn@doi [\apj]
  {10.1086/506193}, \href {http://adsabs.harvard.edu/abs/2006ApJ...649...91F}
  {649, 91}

\bibitem[\protect\citeauthoryear{{Gaburov}, {Harfst}  \& {Portegies
  Zwart}}{{Gaburov} et~al.}{2009}]{GA09}
{Gaburov} E.,  {Harfst} S.,   {Portegies Zwart} S.,  2009, \mn@doi [\na]
  {10.1016/j.newast.2009.03.002}, \href
  {http://adsabs.harvard.edu/abs/2009NewA...14..630G} {14, 630}

\bibitem[\protect\citeauthoryear{{Genzel} et~al.,}{{Genzel}
  et~al.}{2003}]{GS03}
{Genzel} R.,  et~al., 2003, \mn@doi [\apj] {10.1086/377127}, \href
  {https://ui.adsabs.harvard.edu/abs/2003ApJ...594..812G} {594, 812}

\bibitem[\protect\citeauthoryear{{Genzel}, {Eisenhauer}  \&
  {Gillessen}}{{Genzel} et~al.}{2010}]{GE10}
{Genzel} R.,  {Eisenhauer} F.,   {Gillessen} S.,  2010, \mn@doi [Reviews of
  Modern Physics] {10.1103/RevModPhys.82.3121}, \href
  {http://adsabs.harvard.edu/abs/2010RvMP...82.3121G} {82, 3121}

\bibitem[\protect\citeauthoryear{{Ghez}, {Klein}, {Morris}  \&
  {Becklin}}{{Ghez} et~al.}{1998}]{GH98}
{Ghez} A.~M.,  {Klein} B.~L.,  {Morris} M.,   {Becklin} E.~E.,  1998, \apj,
  \href
  {http://cdsads.u-strasbg.fr/cgi-bin/nph-bib_query?bibcode=1998ApJ...509..678G&amp;db_key=AST}
  {509, 678}

\bibitem[\protect\citeauthoryear{{Gillessen}, {Eisenhauer}, {Trippe},
  {Alexander}, {Genzel}, {Martins}  \& {Ott}}{{Gillessen} et~al.}{2009}]{GI09}
{Gillessen} S.,  {Eisenhauer} F.,  {Trippe} S.,  {Alexander} T.,  {Genzel} R.,
  {Martins} F.,   {Ott} T.,  2009, \mn@doi [\apj]
  {10.1088/0004-637X/692/2/1075}, \href
  {http://adsabs.harvard.edu/abs/2009ApJ...692.1075G} {692, 1075}

\bibitem[\protect\citeauthoryear{{Gillessen} et~al.,}{{Gillessen}
  et~al.}{2017}]{GP17}
{Gillessen} S.,  et~al., 2017, \mn@doi [\apj] {10.3847/1538-4357/aa5c41}, \href
  {http://adsabs.harvard.edu/abs/2017ApJ...837...30G} {837, 30}

\bibitem[\protect\citeauthoryear{{Gnedin}, {Ostriker}  \& {Tremaine}}{{Gnedin}
  et~al.}{2014}]{GN14}
{Gnedin} O.~Y.,  {Ostriker} J.~P.,   {Tremaine} S.,  2014, \mn@doi [\apj]
  {10.1088/0004-637X/785/1/71}, \href
  {http://adsabs.harvard.edu/abs/2014ApJ...785...71G} {785, 71}

\bibitem[\protect\citeauthoryear{{Guillard}, {Emsellem}  \&
  {Renaud}}{{Guillard} et~al.}{2016}]{GER16}
{Guillard} N.,  {Emsellem} E.,   {Renaud} F.,  2016, \mn@doi [\mnras]
  {10.1093/mnras/stw1570}, \href
  {http://adsabs.harvard.edu/abs/2016MNRAS.461.3620G} {461, 3620}

\bibitem[\protect\citeauthoryear{{Hailey}, {Mori}, {Bauer}, {Berkowitz}, {Hong}
   \& {Hord}}{{Hailey} et~al.}{2018}]{HA18}
{Hailey} C.~J.,  {Mori} K.,  {Bauer} F.~E.,  {Berkowitz} M.~E.,  {Hong} J.,
  {Hord} B.~J.,  2018, \mn@doi [\nat] {10.1038/nature25029}, \href
  {http://adsabs.harvard.edu/abs/2018Natur.556...70H} {556, 70}

\bibitem[\protect\citeauthoryear{{Harfst}, {Gualandris}, {Merritt}, {Spurzem},
  {Zwart}  \& {Berczik}}{{Harfst} et~al.}{2007}]{HA07}
{Harfst} S.,  {Gualandris} A.,  {Merritt} D.,  {Spurzem} R.,  {Zwart} S.~P.,
  {Berczik} P.,  2007, \mn@doi [New Astronomy] {10.1016/j.newast.2006.11.003},
  \href {http://adsabs.harvard.edu/abs/2007NewA...12..357H} {12, 357}

\bibitem[\protect\citeauthoryear{{Hobbs} \& {Nayakshin}}{{Hobbs} \&
  {Nayakshin}}{2009}]{HN09}
{Hobbs} A.,  {Nayakshin} S.,  2009, \mn@doi [\mnras]
  {10.1111/j.1365-2966.2008.14359.x}, \href
  {http://adsabs.harvard.edu/abs/2009MNRAS.394..191H} {394, 191}

\bibitem[\protect\citeauthoryear{{Hopman} \& {Alexander}}{{Hopman} \&
  {Alexander}}{2006}]{HA06}
{Hopman} C.,  {Alexander} T.,  2006, \mn@doi [\apjl] {10.1086/506273}, \href
  {https://ui.adsabs.harvard.edu/abs/2006ApJ...645L.133H} {645, L133}

\bibitem[\protect\citeauthoryear{{Kocsis} \& {Tremaine}}{{Kocsis} \&
  {Tremaine}}{2011}]{KT11}
{Kocsis} B.,  {Tremaine} S.,  2011, \mn@doi [\mnras]
  {10.1111/j.1365-2966.2010.17897.x}, \href
  {https://ui.adsabs.harvard.edu/abs/2011MNRAS.412..187K} {412, 187}

\bibitem[\protect\citeauthoryear{{Levin} \& {Beloborodov}}{{Levin} \&
  {Beloborodov}}{2003}]{LE03}
{Levin} Y.,  {Beloborodov} A.~M.,  2003, \apjl, \href
  {http://adsabs.harvard.edu/cgi-bin/nph-bib_query?bibcode=2003ApJ...590L..33L&amp;db_key=AST}
  {590, L33}

\bibitem[\protect\citeauthoryear{{Loose}, {Kruegel}  \& {Tutukov}}{{Loose}
  et~al.}{1982}]{LKT82}
{Loose} H.~H.,  {Kruegel} E.,   {Tutukov} A.,  1982, \aap, \href
  {http://adsabs.harvard.edu/abs/1982A%26A...105..342L} {105, 342}

\bibitem[\protect\citeauthoryear{{Lu}, {Ghez}, {Hornstein}, {Morris}, {Becklin}
   \& {Matthews}}{{Lu} et~al.}{2009}]{LU09}
{Lu} J.~R.,  {Ghez} A.~M.,  {Hornstein} S.~D.,  {Morris} M.~R.,  {Becklin}
  E.~E.,   {Matthews} K.,  2009, \mn@doi [\apj] {10.1088/0004-637X/690/2/1463},
  \href {http://adsabs.harvard.edu/abs/2009ApJ...690.1463L} {690, 1463}

\bibitem[\protect\citeauthoryear{{Lu}, {Do}, {Ghez}, {Morris}, {Yelda}  \&
  {Matthews}}{{Lu} et~al.}{2013}]{LU13}
{Lu} J.~R.,  {Do} T.,  {Ghez} A.~M.,  {Morris} M.~R.,  {Yelda} S.,   {Matthews}
  K.,  2013, \mn@doi [\apj] {10.1088/0004-637X/764/2/155}, \href
  {https://ui.adsabs.harvard.edu/abs/2013ApJ...764..155L} {764, 155}

\bibitem[\protect\citeauthoryear{{Mapelli}, {Hayfield}, {Mayer}  \&
  {Wadsley}}{{Mapelli} et~al.}{2012}]{M12}
{Mapelli} M.,  {Hayfield} T.,  {Mayer} L.,   {Wadsley} J.,  2012, \mn@doi
  [\apj] {10.1088/0004-637X/749/2/168}, \href
  {https://ui.adsabs.harvard.edu/abs/2012ApJ...749..168M} {749, 168}

\bibitem[\protect\citeauthoryear{{Mastrobuono-Battisti} \&
  {Perets}}{{Mastrobuono-Battisti} \& {Perets}}{2013}]{MBP13}
{Mastrobuono-Battisti} A.,  {Perets} H.~B.,  2013, \mn@doi [\apj]
  {10.1088/0004-637X/779/1/85}, \href
  {http://adsabs.harvard.edu/abs/2013ApJ...779...85M} {779, 85}

\bibitem[\protect\citeauthoryear{{Mastrobuono-Battisti}, {Perets}  \&
  {Loeb}}{{Mastrobuono-Battisti} et~al.}{2014}]{MBP14}
{Mastrobuono-Battisti} A.,  {Perets} H.~B.,   {Loeb} A.,  2014, \mn@doi [\apj]
  {10.1088/0004-637X/796/1/40}, \href
  {http://adsabs.harvard.edu/abs/2014ApJ...796...40M} {796, 40}

\bibitem[\protect\citeauthoryear{{Merritt}}{{Merritt}}{2010}]{MD10}
{Merritt} D.,  2010, \mn@doi [\apj] {10.1088/0004-637X/718/2/739}, \href
  {https://ui.adsabs.harvard.edu/abs/2010ApJ...718..739M} {718, 739}

\bibitem[\protect\citeauthoryear{{Miralda-Escud{\'e}} \&
  {Gould}}{{Miralda-Escud{\'e}} \& {Gould}}{2000a}]{ME00}
{Miralda-Escud{\'e}} J.,  {Gould} A.,  2000a, \mn@doi [\apj] {10.1086/317837},
  \href {http://adsabs.harvard.edu/abs/2000ApJ...545..847M} {545, 847}

\bibitem[\protect\citeauthoryear{{Miralda-Escud{\'e}} \&
  {Gould}}{{Miralda-Escud{\'e}} \& {Gould}}{2000b}]{MG00}
{Miralda-Escud{\'e}} J.,  {Gould} A.,  2000b, \mn@doi [\apj] {10.1086/317837},
  \href {http://adsabs.harvard.edu/abs/2000ApJ...545..847M} {545, 847}

\bibitem[\protect\citeauthoryear{{Morris}}{{Morris}}{1993}]{MO93}
{Morris} M.,  1993, \mn@doi [\apj] {10.1086/172607}, \href
  {http://adsabs.harvard.edu/abs/1993ApJ...408..496M} {408, 496}

\bibitem[\protect\citeauthoryear{{Nayakshin} \& {Cuadra}}{{Nayakshin} \&
  {Cuadra}}{2005}]{NC05}
{Nayakshin} S.,  {Cuadra} J.,  2005, \aap, \href
  {http://adsabs.harvard.edu/cgi-bin/nph-bib_query?bibcode=2005A\%26A...437..437N&amp;db_key=AST}
  {437, 437}

\bibitem[\protect\citeauthoryear{{Nayakshin}, {Cuadra}  \&
  {Springel}}{{Nayakshin} et~al.}{2007}]{NA07}
{Nayakshin} S.,  {Cuadra} J.,   {Springel} V.,  2007, \mn@doi [\mnras]
  {10.1111/j.1365-2966.2007.11938.x}, \href
  {http://adsabs.harvard.edu/abs/2007MNRAS.379...21N} {379, 21}

\bibitem[\protect\citeauthoryear{{Paumard} et~al.}{{Paumard}
  et~al.}{2006}]{PA06}
{Paumard} T.,  et~al., 2006, \mn@doi [\apj] {10.1086/503273}, \href
  {http://adsabs.harvard.edu/cgi-bin/nph-bib_query?bibcode=2006ApJ...643.1011P&db_key=AST}
  {643, 1011}

\bibitem[\protect\citeauthoryear{{Perets} \& {Gualandris}}{{Perets} \&
  {Gualandris}}{2010}]{PG10}
{Perets} H.~B.,  {Gualandris} A.,  2010, \mn@doi [\apj]
  {10.1088/0004-637X/719/1/220}, \href
  {http://adsabs.harvard.edu/abs/2010ApJ...719..220P} {719, 220}

\bibitem[\protect\citeauthoryear{{Perets} \& {Mastrobuono-Battisti}}{{Perets}
  \& {Mastrobuono-Battisti}}{2014}]{PMB14}
{Perets} H.~B.,  {Mastrobuono-Battisti} A.,  2014, \mn@doi [\apjl]
  {10.1088/2041-8205/784/2/L44}, \href
  {http://adsabs.harvard.edu/abs/2014ApJ...784L..44P} {784, L44}

\bibitem[\protect\citeauthoryear{{Perets}, {Gualandris}, {Kupi}, {Merritt}  \&
  {Alexander}}{{Perets} et~al.}{2009}]{PG09}
{Perets} H.~B.,  {Gualandris} A.,  {Kupi} G.,  {Merritt} D.,   {Alexander} T.,
  2009, \mn@doi [\apj] {10.1088/0004-637X/702/2/884}, \href
  {http://adsabs.harvard.edu/abs/2009ApJ...702..884P} {702, 884}

\bibitem[\protect\citeauthoryear{{Perets}, {Mastrobuono-Battisti}, {Meiron}  \&
  {Gualandris}}{{Perets} et~al.}{2018}]{PMB18}
{Perets} H.~B.,  {Mastrobuono-Battisti} A.,  {Meiron} Y.,   {Gualandris} A.,
  2018, preprint, \href {http://adsabs.harvard.edu/abs/2018arXiv180200012P} {}
  (\mn@eprint {arXiv} {1802.00012})

\bibitem[\protect\citeauthoryear{{Pfuhl} et~al.,}{{Pfuhl} et~al.}{2011}]{PF11}
{Pfuhl} O.,  et~al., 2011, \mn@doi [\apj] {10.1088/0004-637X/741/2/108}, \href
  {http://adsabs.harvard.edu/abs/2011ApJ...741..108P} {741, 108}

\bibitem[\protect\citeauthoryear{{Preto} \& {Amaro-Seoane}}{{Preto} \&
  {Amaro-Seoane}}{2010}]{PAS10}
{Preto} M.,  {Amaro-Seoane} P.,  2010, \mn@doi [\apjl]
  {10.1088/2041-8205/708/1/L42}, \href
  {https://ui.adsabs.harvard.edu/abs/2010ApJ...708L..42P} {708, L42}

\bibitem[\protect\citeauthoryear{{Salpeter}}{{Salpeter}}{1955}]{SA55}
{Salpeter} E.~E.,  1955, \mn@doi [\apj] {10.1086/145971}, \href
  {http://adsabs.harvard.edu/abs/1955ApJ...121..161S} {121, 161}

\bibitem[\protect\citeauthoryear{{Sch{\"o}del}, {Feldmeier}, {Neumayer},
  {Meyer}  \& {Yelda}}{{Sch{\"o}del} et~al.}{2014}]{S14}
{Sch{\"o}del} R.,  {Feldmeier} A.,  {Neumayer} N.,  {Meyer} L.,   {Yelda} S.,
  2014, \mn@doi [Classical and Quantum Gravity]
  {10.1088/0264-9381/31/24/244007}, \href
  {http://adsabs.harvard.edu/abs/2014CQGra..31x4007S} {31, 244007}

\bibitem[\protect\citeauthoryear{{Tremaine}, {Ostriker}  \&
  {Spitzer}}{{Tremaine} et~al.}{1975}]{TR75}
{Tremaine} S.~D.,  {Ostriker} J.~P.,   {Spitzer} Jr. L.,  1975, \mn@doi [\apj]
  {10.1086/153422}, \href {http://adsabs.harvard.edu/abs/1975ApJ...196..407T}
  {196, 407}

\bibitem[\protect\citeauthoryear{{Tsatsi}, {Mastrobuono-Battisti}, {van de
  Ven}, {Perets}, {Bianchini}  \& {Neumayer}}{{Tsatsi} et~al.}{2017}]{TMB17}
{Tsatsi} A.,  {Mastrobuono-Battisti} A.,  {van de Ven} G.,  {Perets} H.~B.,
  {Bianchini} P.,   {Neumayer} N.,  2017, \mn@doi [\mnras]
  {10.1093/mnras/stw2593}, \href
  {http://adsabs.harvard.edu/abs/2017MNRAS.464.3720T} {464, 3720}

\bibitem[\protect\citeauthoryear{{Ulubay-Siddiki}, {Bartko}  \&
  {Gerhard}}{{Ulubay-Siddiki} et~al.}{2013}]{US13}
{Ulubay-Siddiki} A.,  {Bartko} H.,   {Gerhard} O.,  2013, \mn@doi [\mnras]
  {10.1093/mnras/sts167}, \href
  {https://ui.adsabs.harvard.edu/abs/2013MNRAS.428.1986U} {428, 1986}

\bibitem[\protect\citeauthoryear{{Yelda}, {Ghez}, {Lu}, {Clarkson}, {Anderson}
  \& {Matthews}}{{Yelda} et~al.}{2010}]{YG10}
{Yelda} S.,  {Ghez} A.,  {Lu} J.,  {Clarkson} W.,  {Anderson} J.,   {Matthews}
  K.,  2010, in American Astronomical Society Meeting Abstracts \#215. p.~322

\makeatother
\end{thebibliography}

\bsp	
\label{lastpage}
\end{document}